\RequirePackage[orthodox,l2tabu]{nag}

\documentclass[a4paper,11pt]{article}
\usepackage[T1]{fontenc}        %
\usepackage[utf8]{inputenc}
\usepackage[english]{babel}

\usepackage[left=2.7cm,
            right=2.7cm,
            top=2.7cm,
            bottom=3.2cm
            ]{geometry}        %

\usepackage[protrusion,
            expansion
            ]{microtype} %

\usepackage{booktabs}    %

\usepackage{caption}
\usepackage{subcaption}

\usepackage{enumitem}

\usepackage{authblk}    %
\usepackage{appendix}
\usepackage{fancyhdr}
\usepackage{cite}

\usepackage{braket}
\usepackage{xspace}
\usepackage{mathdots}
\usepackage{stackrel}
\usepackage{xcolor}
\usepackage{mathtools}
\usepackage{graphicx}
\usepackage{amsmath,          %
            amsfonts,
            amssymb,          %
            amsthm}          %
            
\usepackage{xfrac}

\usepackage[colorlinks,
            linkcolor=blue,
            citecolor=blue,
            bookmarks,
            bookmarksnumbered
            ]{hyperref}    %

\allowdisplaybreaks

\theoremstyle{definition}

\theoremstyle{remark}

\newcommand{\ZZ}{\mathbb{Z}}

\newcommand{\ii}{\mathrm{i}}

\newcommand{\dd}{\mathrm{d}}

\newcommand{\abs}[1]{\left\vert#1\right\vert}
\newcommand{\avg}[1]{\left \langle #1 \right \rangle}
\newcommand{\nord}[1]{:\! #1\! :}   %

\DeclareMathOperator{\Pf}{Pf}   %
\DeclareMathOperator{\Tr}{Tr}   %
\DeclareMathOperator{\Ei}{Ei}   %

\DeclareMathOperator*{\res}{Res}    %

\newcommand{\twist}{\mathcal{T}}
\newcommand{\antitwist}{\widetilde{\mathcal{T}}}
\newcommand{\Em}{\mathcal{M}}

\date{}
\title{\textbf{Entanglement entropy along a massless renormalisation flow: the tricritical to critical Ising crossover}}

\author[1]{Federico Rottoli}
\author[1]{Filiberto Ares}
\author[1,2]{Pasquale Calabrese}
\author[1,3]{D{\'a}vid X. Horv{\'a}th}

\affil[1]{\textit{SISSA and INFN Sezione di Trieste, via Bonomea 265, 34136 Trieste, Italy.}}
\affil[2]{\textit{International Centre for Theoretical Physics (ICTP), Strada Costiera 11, 34151 Trieste, Italy.}}
\affil[3]{\textit{Department of Mathematics, King’s College London,	Strand, London WC2R 2LS, U.K.}}

\begin{document}
\maketitle

\begin{abstract}
    We study the Rényi entanglement entropies along the massless renormalisation group flow that connects the tricritical and critical Ising field theories. 
    Similarly to the massive integrable field theories, we derive a set of bootstrap equations, from which we can analytically calculate the twist field form factors in a recursive way. Additionally,  we also obtain them as a non-trivial `roaming limit' of the sinh-Gordon theory. 
    Then the Rényi entanglement entropies are obtained as expansions in terms of the form factors of these branch point twist fields.  
    We find that the form factor expansion of the entanglement entropy along the flow organises in two different kind of terms. Those that couple particles with the same chirality, and reproduce the entropy of the infrared Ising theory, and those that couple particles with different chirality, which provide the ultraviolet contributions. 
    The massless flow under study possesses a global $\mathbb{Z}_2$ spin-flip symmetry. We further consider the composite twist fields associated to this group, which enter in the study of the symmetry resolution of the entanglement. We derive analytical expressions for their form factors both from the bootstrap equations and from the roaming limit of the sinh-Gordon theory. 
    
\end{abstract}

\newpage

\tableofcontents
 
\section{Introduction}\label{sec:intro}

Our understanding of many-body quantum systems both at and out 
equilibrium has been dramatically boosted in recent decades. An 
important part of this progress is due to the research carried 
out on entanglement in extended quantum systems. Being the 
fundamental feature of quantum mechanics, entanglement is responsible 
for a plethora of quantum phenomena, novel phases of matter, and 
collective effects~\cite{vNE1, vNE2, vNE3, vNE4}. Different quantities have been introduced to 
characterise and measure the amount of entanglement. The most 
prominent ones are the Rényi entanglement entropies. 
If we consider an extended quantum system in a pure state 
$\ket{\Psi}$ and we take a spatial bipartition into subsystems $A$ 
and $B$, then the Rényi entanglement entropies are defined as
\begin{equation}\label{eq:renyi_ent}
S_n(\rho_A)=\frac{1}{1-n}\log \Tr(\rho_A^n),
\end{equation}
where $\rho_A$ is the reduced density matrix that describes the state 
of subsystem $A$, which can be obtained by taking the partial trace 
to the complementary subsystem $B$, $\rho_A=\Tr_B(\ket{\Psi}\bra{\Psi})$. In the 
limit $n\to 1$, Eq.~\eqref{eq:renyi_ent} gives the von Neumann 
entanglement entropy $S(\rho_A)=-\Tr(\rho_A\log\rho_A)$.

The Rényi entanglement entropies are particularly interesting 
quantities to study in quantum field theories (QFT). In the ground 
state of two-dimensional conformal field theories (CFT), they grow 
logarithmically with the subsystem size, violating the area law, and 
they are proportional to the central charge of the theory~\cite{hlw-94, cc-04}. In general, using the path integral approach, the ground state Rényi entanglement entropies can be cast as partition functions of the 
field theory on a Riemann surface. Alternatively, they 
can also be computed as correlation functions of branch points twist 
fields inserted at the end-points of subsystem $A$, which are 
spinless primaries in CFTs~\cite{cc-04,cc-09}. 

The previous framework has been further extended to massive 
integrable and non-integrable QFTs in two dimensions~\cite{Ola, OlaBen, cd-09, doyon-09}. In this case, 
the Rényi entanglement entropies admit a spectral expansion in terms 
of the twist field form factors. In integrable QFTs, these form 
factors can be determined analytically via the bootstrap program~\cite{SmirnovBook, KarowskiU1}, 
which has been applied to calculate entanglement entropies in different theories and contexts~\cite{Ola, OlaBen, cd-09,cd-09b, cl-11, cdl-12, leviFFandVEV, lcd-13, bcd-16, Ola2, Ola1, Ola-c, Ola-c1, Ola-c2, Ola-c3, clsv-19, Ola3, SGRE}. In 
general, for large enough subsystems sizes, the behaviour of the 
entanglement entropy is well captured by the lower-particle form 
factors.

In the renormalisation group picture of QFTs as perturbed CFTs, an important result in two dimensions is the Zamolodchikov $c$-theorem~\cite{ZamoRoaming, CardyCentralCharge}, which describes the loss of information about the short-distance degrees of freedom along the flow.
In Refs.~\cite{cl-11, cdl-12}, it was recognised that the $\Delta$-sum~\cite{delta_theorem} for the branch point twist fields provides another quantity with the same qualitative behaviour as the Zamolodchikov $c$-function.
This $\Delta$-function monotonically decreases with the distance and it is equal to the 
scaling dimension of the twist fields at the IR and UV fixed points of the flow.
A different $c$-function can also be directly constructed from the entanglement entropy~\cite{ch-04}.

In this paper, we investigate the ground state 
Rényi entanglement entropies in the massless QFT associated to the 
renormalisation group that connects the tricritical and 
critical Ising theories by perturbing the former with a relevant
field. This theory is the simplest member of the well-known family of 
massless renormalisation group flows that have as UV and IR fixed points two 
consecutive $A$-series unitary conformal minimal models~\cite{Zamo-87-serie, cl-87-serie, AnZamo123, AnZamo2, AnZamo3}. The form factor 
bootstrap program has been successfully applied in Ref.~\cite{mds-95} 
to certain correlators along the tricritical-critical Ising flow. 
Here we extend it to the branch point twist fields and we obtain 
explicit expressions for the two and four-particle form factors. To 
this end, we follow the same strategy as in the massive case, we 
write the set of form factor bootstrap equations that take into 
account the particular exchange properties of the 
twist fields and we propose a general ansatz for their solution. 
Furthermore, we also derive the two and four-particle form factors 
along the massless flow from the roaming limit of the sinh-Gordon ones. By analytically continuing the scattering matrix of the 
sinh-Gordon model, one can find the Zamolodchikov's staircase model~\cite{RefZamo}, 
a two-dimensional integrable scattering theory that describes a 
renormalisation group flow which interpolates between 
the successive $A$-series unitary conformal minimal models. It has 
been shown~\cite{RefRoaming, RoamingFF} that the form factors of different fields in the 
tricritical-critical Ising model flow can be obtained as roaming 
limits of certain form factors of the sinh-Gordon theory. We show 
here that a similar strategy holds for the twist field form factors.
We also study the $\Delta$-function associated to the twist fields 
along the flow, finding that it is monotonic and correctly reproduces
their scaling dimension at the fixed points.

In the last years, one of the main research lines in the study of 
entanglement in extended systems has been its interplay with 
symmetries. If the system presents a global symmetry, the 
entanglement entropy can be further decomposed into the 
contribution of each symmetry sector~\cite{lr-14, gs-18, Equipartitioning}. The massless renormalisation 
flow between the tricritical and the critical Ising theories has a 
global $\ZZ_2$ symmetry. Let us denote by $Q$ the charge 
operator that generates the symmetry and assume that it is the sum of 
the charges in subsystems $A$ and $B$, $Q=Q_A+Q_B$. 
In that case, the reduced density matrix admits the following 
decomposition in symmetry sectors
\begin{equation}
\rho_A=\bigoplus_{q}p(q) \rho_{A, q},
\end{equation}
where $\rho_{A, q}=\Pi_q\rho_A\Pi_q/p(q)$, $\Pi_q$ is the projector 
onto the eigenspace of $Q_A$ with eigenvalue $q$, and $p(q)=\Tr(\rho_A\Pi_q)$ guarantees the correct normalisation of $\rho_{A, q}$, i.e., 
$\Tr(\rho_{A, q})=1$.

The symmetry-resolved entanglement entropies $S_n(\rho_{A, q})$ 
quantify the amount of entanglement in each charge sector. One usual 
way of calculating them is through the charged moments of $\rho_A$,
\begin{equation}\label{eq:charged_mom}
Z_n(\alpha)=\Tr(\rho_A^n e^{i\alpha Q_A}),
\end{equation}
by applying the Fourier representation of the projector $\Pi_q$. For a $\mathbb{Z}_N$ symmetry group, $q$ can only take values $q=0, \dots, N-1$ and
\begin{equation}
\mathcal{Z}_n(q) = \Tr(\rho_A^n\Pi_q)=\frac{1}{N}\sum_{k=0}^{N-1} Z_n\left(\frac{2\pi k}{N}\right)e^{\frac{i2\pi k q}{N}}.
\end{equation}
Therefore, one can write
\begin{equation}\label{eq:sree}
S_n(\rho_{A, q})=\frac{1}{1-n}\log \frac{\mathcal{Z}_n(q)}{\mathcal{Z}_1(q)^n}.
\end{equation}

The charged moments~\eqref{eq:charged_mom} were initially introduced in an independent 
way in Refs.~\cite{belin-13, caputa-13} in the context of holography. After Ref.~\cite{gs-18} pointed 
out its connection with the symmetry-resolved entropies~\eqref{eq:sree}, both quantities 
have been intensively studied in two-dimensional CFTs~\cite{gs-18, Equipartitioning, crc-20, Zn, Chen-21, cc-21, cdmWZW-21, mt-23, acdgm-22, fmc-23, dgmnsz-23, fac-23, kmop-23, mdc-23} as well 
as in free and integrable QFTs~\cite{mdc-20b, U(1)FreeFF, SGSRE, cm-23, Z2IsingShg, PottsSRE, IQFTSREExcitedStatesI, IQFTSREExcitedStatesII}. Symmetry resolved entanglement has 
also been investigated in other systems such as lattice and spin models~\cite{brc-19, fg-20, FreeF1, mdc-20, ss-dxh-22, bcckr-23, Topology, SSHDefect, mcp-22}, as well as in ion trap and cold atom experiments~\cite{Greiner, ncv-21, vecd-20, rath}. 
Similarly to the neutral moments $Z_n(0)$ of $\rho_A$, the charged 
ones can be expressed as partition functions of the field theory on a 
Riemann surface, but now in presence of an external magnetic flux. 
The charged moments $Z_n(\alpha)$ can also be obtained from a 
correlator of composite branch point twist fields, which also take 
into account the non-trivial monodromy between the sheets of 
the Riemann surface due to the magnetic flux. The form factor 
bootstrap techniques developed for the standard branch point twist 
fields in integrable QFTs have been extended 
to the composite ones to study $U(1)$ symmetries in free~\cite{U(1)FreeFF} and 
interacting QFTs such as sine-Gordon model~\cite{SGSRE}, the $\mathbb{Z}_2$ symmetry of 
massive Ising and sinh-Gordon models~\cite{Z2IsingShg, cm-23}, and the $\mathbb{Z}_3$ symmetry 
in the 3-state Potts model~\cite{PottsSRE}. In this work, we consider the composite 
branch point twist fields associated to the $\mathbb{Z}_2$ spin flip 
symmetry of the tricritical-critical massless flow, and we apply
the bootstrap approach to obtain analytic expressions for their two 
and four-particle form factors. Using them, we also investigate the 
corresponding $\Delta$-function.

The paper is organised as follows. In Sec.~\ref{sec:flow} we introduce the system we 
are interested in, the massless renormalisation flow between tricritical and critical Ising theories. In Sec.~\ref{sec:twistfield}, we review 
how entanglement entropies can be calculated in QFT as correlation
functions of branch point twist fields and we discuss their spectral
expansion in terms of form factors. In Sec.~\ref{sec:twistFormFactors}, we obtain the bootstrap equations for the standard branch point twist field form factors, we 
propose a general ansatz for their solution, and we derive the explicit
expressions for the two and four particle form factors.
In Sec.~\ref{sec:compositeFormFactors}, we repeat the same reasoning for the form factors of the 
composite twist fields associated to the $\mathbb{Z}_2$ spin flip symmetry of the massless flow. In Sec.~\ref{sec:RoamingLimit}, we rederive the twist field
form factors by taking the roaming limit of those of the sinh-Gordon model. In Sec.~\ref{sec:entropy}, we first study the $\Delta$-function along the flow of the (composite) twist fields and we derive a cumulant expansion for the entanglement entropies. We end up in Sec.~\ref{sec:conclusions} with
some conclusions and future prospects. We also include two appendices where we discuss in detail the derivation of some of the results presented in the main text. 

\section{The massless RG flow from the tricritical to the critical Ising theory}\label{sec:flow}

In this paper, we investigate the ground state entanglement entropy along the renormalisation group flow that connects the tricritical and critical Ising CFTs, which are respectively the unitary minimal models $\Em_{4}$ and $\Em_3$~\cite{AnZamo123, AnZamo2, AnZamo3} with central charges~\cite{yellow-book, m-book}
\begin{equation}\label{eq:CIRUV}
    c_\text{UV} =\frac{7}{10}\,, \quad \text{and} \quad c_\text{IR} = \frac{1}{2}\,,
\end{equation}
and with Kac tables reported in Table~\ref{tab:Kac}.
This integrable QFT, usually denoted as $A_2$,
is the simplest member of the infinite family of massless theories $A_p$ that interpolate between two consecutive $A$-series diagonal conformal minimal models $\Em_{p+2}\rightarrow\Em_{p+1}$ with central charges
\begin{equation}
c_\text{UV} =1-\frac{6}{\left(p+2\right)\left(p+3\right)} \,, \quad \text{ and } \quad c_\text{IR} =1-\frac{6}{\left(p+1\right)\left(p+2\right)}\,.
\label{eq:CIRUVAn}
\end{equation}
This family of integrable RG trajectories is obtained by deforming the UV CFT $\Em_{p+2}$ with its relevant field $\phi_{1,3}$~\cite{Zamo-87-serie, cl-87-serie,AnZamo123,AnZamo2,AnZamo3}.
In particular, in the tricritical Ising CFT, $\phi_{1, 3}$ corresponds to the vacancy density field with conformal dimension $h_{1,3} = \frac{3}{5}$ (see Table~\ref{tab:KacTri})~\cite{AnZamo123,AnZamo2,AnZamo3}. 
In the Euclidean formalism, the action $\mathcal{A}_\text{flow}$ of this flow takes the form
\begin{equation}\label{eq:flow_action}
    \mathcal{A}_\text{flow} = \mathcal{A}_{\Em_{4}} + \lambda \int \dd^2 x \,\phi_{1,3}(\textbf{x})\,,
\end{equation}
where $\lambda$ is a dimensionful coupling and, importantly, $\lambda$ is positive, since for negative coupling a different massive integrable theory is obtained. Several other families of massless integrable flows have been identified as well~\cite{Lassig,RGFlow2,RGFlow3,RGFlow4,RGFlow5,homSinGordon,cafkm-00,CastroFring1}.

\begin{table}[t]
    \centering
    \begin{subfigure}[b]{0.48\textwidth}
        \centering
        \begin{tabular}{c|cccc|}
            \hline
            3&  \sfrac{3}{2}&   \sfrac{3}{5}&   \sfrac{1}{10}&  0 \\
            2&  \sfrac{7}{16}&  \sfrac{3}{80}&  \sfrac{3}{80}&  \sfrac{7}{16} \\
            1&  0&              \sfrac{1}{10}&  \sfrac{3}{5}&   \sfrac{3}{2} \\
            \hline
            $s$/$r$&   1&  2&  3&  4
        \end{tabular}
        \caption{$\Em_4$, $c = \frac{7}{10}$}
        \label{tab:KacTri}
    \end{subfigure}
    \begin{subfigure}[b]{0.48\textwidth}
        \centering
        \begin{tabular}{c|ccc|}
            \hline
            2&  \sfrac{1}{2}& \sfrac{1}{16}& 0\\
            1&  0&           \sfrac{1}{16}& \sfrac{1}{2}\\
            \hline
            $s$/$r$ &   1&  2&  3
        \end{tabular}        
        \caption{$\Em_3$, $c = \frac{1}{2}$}
        \label{tab:KacIsing}
    \end{subfigure}
    \caption{
    Kac tables of the tricritical (Table~\ref{tab:KacTri}) and critical (Table~\ref{tab:KacIsing}) Ising CFTs~\cite{yellow-book, m-book}. In each case, we report the conformal dimension of the primary fields $\phi_{r,s}$ of the theory. The vertical and horizontal axes correspond to the $s$ and $r$ indices respectively.}
    \label{tab:Kac}
\end{table}

The masslessness of the flow described by Eq.~\eqref{eq:flow_action} can be understood by recalling that the tricritical Ising CFT $\Em_{4}$ is one of the simplest examples of superconformal theory~\cite{fqs-85-superconf, q-86-superconf, bkt-85-superconf}.
The deformation with the vacancy field $\phi_{1,3}$ leads to a spontaneous supersymmetry breaking~\cite{KastorMartinecTTBar} which gives rise to right- and left-moving massless Goldstone fermions $\psi, \bar{\psi}$ and ensures that the theory has vanishing mass gap.

In Ref.~\cite{KastorMartinecTTBar}, it was shown that the low energy behaviour of the massless flow is described by the effective Lagrangian
\begin{equation}\label{eq:LEff}
    \mathcal{L}_\text{eff} = \frac{1}{2\pi} \left (  \psi\bar{\partial}\psi + \bar{\psi}\partial\bar{\psi} \right ) - \frac{1}{\pi^2 M^{2}}\left(\psi\partial\psi\right)\left(\bar{\psi}\bar{\partial}\bar{\psi}\right) + \ldots\, ,
\end{equation}
that is, the $T\bar{T}$ deformation of the critical Ising model.
Notice that the Majorana fermions $\psi, \bar{\psi}$ of the Ising model are now identified with the Goldstone fermions of the massless flow $A_2$, which are the only stable particles in this theory~\cite{ZamolodchikovTTBar}.
It is worth stressing that the massless flow at low-energies is described by a $T\bar{T}$-deformed CFT.
Such theories have been studied in great detail~\cite{ds-18, llst-20, cch-18, bbc-19, jkn-19, mnrs-19, gst-19, p-19, a-20, hs-20, ss-19, g-19, akf-20, cans-23} and hence they provide non-trivial benchmarking for some of our results.

The massless flow~\eqref{eq:flow_action} as well as the effective Lagrangian~\eqref{eq:LEff} possess a mass scale $M$, which plays the role of the momentum scale at which non-trivial scattering happens between the fermions.
We can parameterise the energy and momenta of the right- and left-moving Goldstone fermions in terms of a rapidity variable $\theta$ and of this mass scale $M$ as~\cite{ZamolodchikovTTBar}
\begin{equation}\label{eq:MasslessEnergyMomentum}
    \begin{aligned}
        E_R(\theta) &= \frac{M}{2} e^\theta\,,\\
        p_R(\theta) &= \frac{M}{2} e^\theta\,,  
    \end{aligned}\qquad
    \begin{aligned}
        E_L(\theta) &= \frac{M}{2} e^{-\theta}\,,\\
        p_L(\theta) &= -\frac{M}{2} e^{-\theta}\,.
    \end{aligned}
\end{equation}
Since the massless fermions are the only stable particles, they form a complete basis of asymptotic states, which in the rapidity parameterisation read as
\begin{equation}\label{eq:asymp_states}
\ket{\theta_1,\ldots\theta_r,\theta_1',\ldots\theta_l'}_{r,l}\, = \psi\!\left(\theta_1\right )\ldots\psi\!\left(\theta_r\right ) \bar{\psi}\!\left (\theta_1'\right )\ldots\bar{\psi}\!\left (\theta_l'\right )\ket{0},
\end{equation}
which contains $r$ right-moving and $l$ left-moving fermions.
If the rapidities are ordered as $\theta_1>\theta_2>\ldots>\theta_r$ and $\theta_l'>\ldots>\theta_2'>\theta_1'$, then the set of states~\eqref{eq:asymp_states} corresponds to in-states, whereas the opposite ordering results in out-states.
Different orderings are linked by scattering processes between the particles. Since the theory is integrable, the scattering of particles is completely elastic, preserves particle number and rapidity, and is fully characterised by the two-body $S$-matrices.
Since the scattering of the particles is diagonal, the $S$-matrices are scalars and functions of the rapidity difference of the particles.
In particular~\cite{ZamolodchikovTTBar}
\begin{equation}\label{eq:s_matrix_flow}
    \begin{gathered}
        S_\text{RR} = S_\text{LL} = -1\,,\\
        S_\text{RL}\!\left(\theta \right ) = S^{-1}_\text{LR}\!\left (-\theta \right ) = \tanh\! \left(\frac{\theta }{2}-\frac{\ii \pi }{4}\right),
    \end{gathered}
\end{equation}
that is, only the scattering between left- and right-movers is non-trivial.

The massless flow~\eqref{eq:flow_action} has been the subject of numerous studies.
These involve its description in terms of the thermodynamic Bethe ansatz~\cite{AnZamo123,AnZamo2,AnZamo3}, or the determination of form factors, i.e., the matrix elements of the off-critical versions of the UV scaling fields, as well as certain correlation functions~\cite{mds-95}.
At the level of the free energy and form factors~\cite{RefRoaming, RoamingFF}, the massless flow can also be recovered from the staircase model~\cite{RefZamo, Lassig}, which we shall introduce in Section~\ref{sec:RoamingLimit}.
The model also shows interesting properties in inhomogeneous out-of-equilibrium situations as studied in~\cite{MasslessHydro} via generalised hydrodynamics.

To complete the brief review of this massless flow, we discuss its symmetry properties.
Both the UV and IR limiting CFTs enjoy a spin-flip $\ZZ_2$ symmetry under which the perturbing field also transforms trivally.
Consequently, the massless flow inherits this symmetry as well. This fact can be made more transparent using the Landau-Ginzburg formalism, which allows the identification of the multicritical Ising CFTs with a Lagrangian~\cite{yellow-book, m-book}.
In particular, the tricritical and critical Ising models can be described by the following actions in terms of the bosonic field $\varphi$
\begin{equation}
\mathcal{A}_\text{tri} = \int \dd^2x\, \frac{1}{2} \left (\partial_\mu\varphi\, \partial^\mu \varphi \right ) + g  \nord{\varphi^6}\,, \qquad 
    \mathcal{A}_\text{crit} = \int \dd^2x\, \frac{1}{2} \left ( \partial_\mu\varphi\, \partial^\mu \varphi \right ) + g' \nord{\varphi^4}\,, 
    \label{TriCrit-CritAction}
\end{equation}
where $\nord{\;}$ denotes normal ordering of the fields. The perturbing field of the UV theory $\phi_{1,3}$ corresponds to $\nord{\varphi^4}$~\cite{yellow-book, m-book},
which means that the action of the massless flow can be equivalently written as
\begin{equation}
    \mathcal{A}_\text{flow}=\int \dd^2x\, \frac{1}{2} \left ( \partial_\mu\varphi\, \partial^\mu \varphi \right ) + \tilde{\lambda}\nord{\varphi^4} + g\nord{\varphi^6}\,,
\end{equation}
in which the invariance under the spin-flip $\ZZ_2$ symmetry, which maps $\varphi\rightarrow-\varphi$, is explicit. 

Given the presence of a global $\mathbb{Z}_2$ symmetry, a relevant question is whether the ground state entanglement entropy along the massless flow~\eqref{eq:flow_action} can be resolved with respect to it. It is not immediately obvious if a reduced density matrix of the ground state of the theory commutes with the charge operator associated with the $\mathbb{Z}_2$ symmetry.
While for a continuous symmetry this is ensured by Noether theorem, in the case of discrete symmetries, the existence of a local charge density is not guaranteed.
In order to justify the existence of such a local $\ZZ_2$ charge, we can appeal to the defect line formalism. 
As understood in recent years, global symmetries in QFT are implemented by topological defects \cite{ffrs-04, gksw}, which, in the case of CFT minimal models, correspond to the Verlinde lines operators. 
In particular, the spin-flip $\ZZ_2$ symmetry is implemented by the Verlinde line associated with the primary operator $\varepsilon$.
Such a defect line can be restricted to the subsystem $A$, with two disorder operators $\mu$ inserted at the end-points \cite{ffrs-04, gksw} (which we discuss in more detail in the next section).
This formalism has been very recently used to study the symmetry resolution of entanglement in CFTs with respect to both continuous and discrete finite groups in Ref.~\cite{kmop-23}. Since the operators $\varepsilon$ and $\mu$ exist along the entire massless flow, the previous construction may be extended outside the fixed points.

\section{Entanglement entropy and Branch Point Twist Fields in QFT}\label{sec:twistfield}

In this section, we review the computation of the entanglement entropies in QFT as correlators of branch point twist fields.
In QFT, the non-trivial task of computing entanglement entropies can be naturally formulated via the path integral approach.
The main idea is that the moments of the reduced density matrix $\Tr(\rho_{A}^{n})$ and the charged moments $\Tr (\rho_{A}^{n}e^{i\alpha Q_{A}})$ can be regarded as partition functions of the QFT on a Riemann surface consisting of $n$ replicas of the space-time that are sewed along the subsystem $A$ in a cyclical way~\cite{hlw-94, cc-04}.

Alternatively, one can take $n$ copies of the QFT under analysis and quotient them by the $\mathbb{Z}_n$ symmetry associated to the cyclic exchange of the copies.
In $(1+1)$-dimensional relativistic QFTs, there exist local fields in the $n$-replica theory, called branch points twist fields (BPTF), that implement the boundary conditions imposed on the fields in the path integral on the $n$-sheeted Riemann surface.
These twist fields can be generalised to cases in which the boundary conditions also involve additional phases, such as in the calculation of the charged moments $\Tr(\rho_A^n e^{i\alpha Q_A})$, in which an Aharonov-Bohm flux is introduced between the sheets of the Riemann surface.
In this setup, the corresponding twist fields are called composite BPTFs and they were originally introduced in other context~\cite{cdl-12,leviFFandVEV}.
Both types of fields are associated with particular symmetries of the replicated theory, which allows us to discuss them on the same ground.
Therefore, for our purposes it is useful to distinguish the following twist fields:

\begin{enumerate}[label=\arabic*)]
    \item the disorder field $\mu$ associated with the $\ZZ_2$ spin-flip symmetry of the massless flow;
    \item the standard BPTFs, $\twist_n$ and its conjugate $\antitwist_n$, which are associated with the cyclic and the inverse cyclic permutation symmetry $\ZZ_n$ among the copies in the $n$-replica massless flow.
    These fields play a central role in the computation of the entanglement entropy;
    \item the $\ZZ_2$-composite BPTFs, denoted as $\twist^{\mu}_n$ and $\antitwist^{\mu}_n$, which are the result of fusing the former fields    \begin{equation}\label{eq:twistcompFusion}
        \twist_n^\mu(\mathbf{x}) =\,  \nord{\twist_n\,\mu}\!(\mathbf{x})\,, \qquad \antitwist_n^\mu(\mathbf{x}) =\,  \nord{\antitwist_n\,\mu}\!(\mathbf{x}).
    \end{equation}
    Therefore, they are associated both with the $\mathbb{Z}_n$ symmetry under the cyclic permutation of the replicas and with the global $\ZZ_2$ spin-flip symmetry present in the massless flow.
    These composite fields play the analogous role of the BPTF in the computation of the symmetry resolved entanglement entropies~\cite{gs-18}. 
\end{enumerate}
These twist fields are typically non-local or semi-local with respect to other quantum fields of the theory, in particular with respect to the fundamental field or to the interpolating field, which is associated with particle creation/annihilation.
Non-locality can be formulated by non-trivial equal-time exchange relations between the two fields. Let us first consider the disorder operator $\mu$ and an operator $\mathcal{O}_i$ living in the copy $i$ of the replicated theory. Since the disorder field introduces an Aharonov-Bohm flux in the region $y^1>x^1$, the exchange relations of these two operators can be written as
\begin{equation} \label{CTFSpatialExchange}
        \mathcal{O}_{i}(\mathbf{y})\mu(\mathbf{x}) = 
    \begin{cases}
        e^{\ii\kappa_\mathcal{O} \pi}\, \mu(\mathbf{x})\, \mathcal{O}_{i}(\mathbf{y})\,,  &\text{for } y^{1}>x^{1},\\
        \mu(\mathbf{x})\, \mathcal{O}_{i}(\mathbf{y})\,,  &\text{otherwise.}
    \end{cases}
\end{equation} 
We refer to $\kappa_\mathcal{O}$ as the charge of the operator $\mathcal{O}$ with respect to the $\ZZ_2$ spin-flip symmetry. In particular, the Goldstone fermions $\psi, \bar{\psi}$ which generate the asymptotic states~\eqref{eq:asymp_states} have charge $\kappa_\psi = 1$, i.e., they are odd under the spin-flip transformation.

The action of the standard BPTFs when winding around a field is to cyclically map it from one replica to the next, as encoded in the equal time exchange relation
\begin{equation}\label{eq:ExchangeRelationTwist}
    \mathcal{O}_{i}(\mathbf{y})\twist_n(\mathbf{x}) =
    \begin{cases}
        \twist_n(\mathbf{x})\,\mathcal{O}_{i+1}(\mathbf{y})\,,  & \text{for } y^{1}>x^{1}\,,\\
        \twist_n(\mathbf{x})\, \mathcal{O}_{i}(\mathbf{y})\,,    & \text{otherwise.} 
    \end{cases}
\end{equation}
In the case of the composite BPTFs, the winding around them further adds a phase $e^{i\kappa_\mathcal{O}\pi}$. When considering discrete groups, as the $\mathbb{Z}_2$ spin-flip symmetry of the tricritical-critical massless flow, we must be careful on how we include this phase. Unlike the continuous $U(1)$ symmetry, discussed in Refs.~\cite{U(1)FreeFF, SGSRE}, here we cannot distribute the flux uniformly in all the copies by inserting a phase $e^{i\kappa_\mathcal{O}\pi/n}$ when moving between replicas since this operation in not compatible with the properties of the $\mathbb{Z}_2$ field $\mu$. This issue can be 
addressed in two different ways. The first possibility is to insert a phase $e^{i\kappa_\mathcal{O}\pi}$ between all the copies; this corresponds to consider the exchange relation
\begin{equation}\label{eq:ExchangeRelationComp1}
    \mathcal{O}_{i}(\mathbf{y})\twist_n^{\mu}(\mathbf{x}) =
    \begin{cases}
        e^{\ii\kappa_\mathcal{O} \pi}\, \twist_n^{\mu}(\mathbf{x})\,\mathcal{O}_{i+1}(\mathbf{y})\,,  & \text{for } y^{1}>x^{1}\,,\\
        \twist_n^{\mu}(\mathbf{x})\, \mathcal{O}_{i}(\mathbf{y})\,,    & \text{otherwise}. 
    \end{cases}
\end{equation}
 This approach was applied in Ref.~\cite{Z2IsingShg}, but it is only legitimate when we take an odd number of replicas $n=1,3,5,7,\ldots$ in which the identity $e^{i\pi n}=-1$ clearly holds.
The other approach consists of introducing the flux only between the last and the first replicas, in such a way that the phase $e^{i\pi\kappa_\mathcal{O}}$ only appears when a particles moves from the $n$-th copy to the $1$-st one, that is
\begin{equation}\label{eq:ExchangeRelationComp2}
    \mathcal{O}_{i}(\mathbf{y})\twist_n^{\mu}(\mathbf{x}) =
    \begin{cases}
        \twist_n^{\mu}(\mathbf{x})\,\mathcal{O}_{i+1}(\mathbf{y})\,,  & \text{for } y^{1}>x^{1} \text{ and } i \neq n\,, \\
        e^{\ii\kappa_\mathcal{O} \pi}\, \twist_n^{\mu}(\mathbf{x})\,\mathcal{O}_{i+1}(\mathbf{y})\,,  & \text{for } y^{1}>x^{1} \text{ and } i=n\,, \\
        \twist_n^{\mu}(\mathbf{x})\, \mathcal{O}_{i}(\mathbf{y})\,,    & \text{otherwise.} 
    \end{cases}
\end{equation}
This choice introduces a slight asymmetry between the replicas, but it is applicable to any number of replicas $n$. In Sec.~\ref{sec:compositeFormFactors}, we discuss in more detail the effect of the two conventions~\eqref{eq:ExchangeRelationComp1} and \eqref{eq:ExchangeRelationComp2}, showing that they provide results for the correlation functions under analysis.

Analogous exchange relations can be formulated for the Hermitian conjugate fields $\antitwist$ and $\antitwist^\mu$, with the difference that they move the field from the replica $i$ to $i-1$. In the following discussion, whenever we wish to treat both the standard and the composite twist fields at the same time, we use the notation $\twist_n^\tau$, $\antitwist_n^\tau$, where $\tau$ refers either to `$\mu$' for the composite or to the identity for the standard BPTF.

Using the (composite) BPTFs, one can switch from a path-integral to an operator formulation of both the neutral and charged moments of $\rho_A$, which can be defined in terms of multi-point functions of the standard or the composite BPTFs in the replicated QFT inserted at the end-points of subsystem $A$. In particular, when $A$ consists of a single interval, $A = \left [ 0, \ell \right ]$, we have
\begin{equation}\label{eq:mom_twist_fields}
\Tr\!\left(\rho_A^n \right ) \sim \bra{0} \twist_n(0) \antitwist_n(\ell) \ket{0},
\end{equation}
and 
\begin{equation}\label{eq:charged_mom_twist_fields}
\Tr\!\left ( \rho_A^n\,  e^{\ii\pi Q_A} \right ) \sim \bra{0} \twist_n^\mu (0) \antitwist_n^\mu(\ell)\ket{0}.
\end{equation}

The twist field formalism is especially useful
at criticality, where conformal invariance fixes the properties of both the standard $\twist_n$ and the composite branch point twist field $\twist_n^\mu$.
In particular, in a unitary CFT with central charge $c$, the standard twist fields $\twist_n$, $\antitwist_n$ are known to be primary operators with conformal dimension~\cite{cc-04,cc-09}
\begin{equation}\label{eq:twistDim}
  h_\twist = \frac{c}{24} \left ( n - \frac{1}{n} \right ).
\end{equation}
We remind the reader that, in the $A$-diagonal unitary minimal models $\Em_p$, the central charge is given by Eq.~\eqref{eq:CIRUVAn}.

In order to identify the conformal dimension of the composite twist fields $\twist_n^\mu$, $\antitwist_n^\mu$, one can use the fact that they are the fusion of $\twist_n$, $\antitwist_n$ with the disorder field $\mu$ as shown in Eq.~\eqref{eq:twistcompFusion}.
In the tricritical and critical Ising models, the field $\mu$ is the Kramers-Wannier dual of the spin field $\sigma = \phi_{2,2}$ and has the same conformal dimension reported in the Kac table in Table~\ref{tab:Kac}~\cite{yellow-book, m-book}
\begin{equation}\label{eq:TricrDisorderDim}
    h_\mu^\text{UV} = h_\sigma^\text{UV} = \frac{3}{80}\,,\qquad
    h_\mu^\text{IR} = h_\sigma^\text{IR} = \frac{1}{16}\,,
\end{equation}
where we denote with UV the tricritical and with IR the critical Ising models respectively.
Knowing the dimension of the disorder field, the one of the composite twist fields $\twist_n^\mu$, $\antitwist_n^\mu$ is obtained as~\cite{gs-18}
\begin{equation}\label{eq:resolvedTwistDim}
  h_{\twist^\mu} = h_\twist + \frac{h_\mu}{n}\,.
\end{equation}
In particular, for the tricritical and critical Ising models, Eq.~\eqref{eq:resolvedTwistDim} gives respectively
\begin{equation}\label{eq:TricrResolvedTwistDim}
  h_{\twist^\mu}^\text{UV} = \frac{1}{240} \left ( 7 n + \frac{2}{n} \right ), \quad\text{and}\quad
  h_{\twist^\mu}^\text{IR} = \frac{1}{48} \left ( n + \frac{2}{n} \right ).
\end{equation}

The use of CFT techniques has provided exact results for $\Tr(\rho_A^n)$ in many different situations~\cite{cc-04,cc-09}.
On the other hand, away from criticality, the exact determination of the correlation functions of Eqs.~\eqref{eq:mom_twist_fields} and~\eqref{eq:charged_mom_twist_fields} is known to be an extremely difficult task, except in the case of free theories~\cite{Ola2,mdc-20}. 
In integrable QFTs, however, the form factor (FF) bootstrap approach provides a powerful tool to systematically investigate and construct (truncated) multi-point functions via form factors, namely matrix elements of generic local operators between the vacuum and the multi-particle states\cite{SmirnovBook,KarowskiU1}.
Although, in principle, all these matrix elements can be analytically computed, their resummation is an unsolved problem. 
Nevertheless, the multi-point correlation functions at large distances are generically dominated by the first few (lower-particle) form factors. 
For this reason this technique applies efficiently to the infrared properties of these theories as was first shown in~\cite{Ola} in the case of BPTFs and entanglement. 
As we shall see in this paper, the above considerations do not hold in massless theories, that is, when the IR limit of the QFT is described by a CFT as well. However, we show that it is possible to identify a subset of terms in the form factor expansion whose resummation reproduces the IR CFT results, while the remaining contributions yield non-trivial predictions for the behaviour of the entropies along the flow.

\subsection{Form factors and spectral representations of BPTF correlation functions}

From the knowledge of the exchange relations ~\eqref{eq:ExchangeRelationTwist} satisfied by the BPTFs, one can formulate bootstrap equations for their FF in integrable QFTs~\cite{Ola, OlaBen, cd-09}, generalising the standard form factor program for local fields~\cite{SmirnovBook,KarowskiU1}, which for the tricritical-critical Ising flow~\eqref{eq:flow_action} has been investigated in Ref.~\cite{mds-95}.

Let us consider the two-point correlation function of the (composite) BPTFs in the ground state of the theory and insert the set of asymptotic states~\eqref{eq:asymp_states}, which form a complete basis,
\begin{equation}
\begin{split}
&\bra{0} \twist^{\tau}_n(x)\antitwist^{\tau}_n(x')\ket{0} \\
&=\sum_{k=0}^{\infty}\sum_{\{\gamma\},\{\nu\}} \int \prod_{i=1}^k\dd\theta_i\, \bra{0}  \twist^{\tau}_n(x) \ket{\theta_{1},\ldots\theta_{k}}_{\gamma_{1}\ldots \gamma_{k}}^{\nu_{1}\ldots \nu_{k}}\times \phantom{}^{\nu_{1}\ldots \nu_{k}}_{\gamma_{1}\ldots \gamma_{k}}{\bra{\theta_{1},\ldots\theta_{k}} \antitwist^{\tau}_n(x')\ket{0}}\,,
\end{split}
\label{FFSeriesCF}
\end{equation}
where $\tau=0,\mu$ corresponds to the standard or the $\ZZ_2$-composite BPTF respectively. In the multi-particle states 
\begin{equation}\label{eq:basis}
\ket{\theta_1, \dots,\theta_k}_{\gamma_1\dots\gamma_k}^{\nu_1\dots\nu_k}\,,
\end{equation}
of the $n$-replica theory, the subindex $\gamma_i=R, L$ specifies if the particle with rapidity $\theta_i$ is a right- (R) or left-mover (L). Moreover, each particle is labelled by an extra index $\nu_{i}$ which indicates the copy where the particle lives; therefore, it takes
values from $1$ to $n$ and it is identified up to $\nu_i \sim \nu_i +n$. 

In the $n$-replica theory, the $S$-matrix connects non-trivially only particles living on the same replica,
while particles in different copies do not interact and no scattering events occur between them.
In light of this, the $S$-matrix of the replicated model takes the form \cite{Ola}
\begin{equation}
S_{\gamma_i,\gamma_j}^{\nu_i,\nu_j}(\theta) = \begin{cases}
        S_{\gamma_{i}\gamma_{j}}(\theta),   &\nu_{i}=\nu_{j},\\
        1,  &\nu_{i}\neq\nu_{j},
\end{cases}
\end{equation}
where $S_{\gamma_i,\gamma_j}(\theta)$ is the $S$ matrix of the original theory, which for the massless flow~\eqref{eq:flow_action} is reported in Eq.~\eqref{eq:s_matrix_flow}.
 
Since the vacuum of the theory is invariant under space and time translations, we can rewrite the spectral expansion in Eq.~\eqref{FFSeriesCF} as 
\begin{align}
    \begin{split}
        &\bra{0} \twist^{\tau}_n(x-x')\antitwist^{\tau}_n(0)\ket{0}\\
        &=\sum_{k=0}^{\infty}\sum_{\{\gamma\},\{\nu\}} \int \prod_{i=1}^k\dd\theta_i\, \bra{0} e^{\ii H(x_0-x_0')-\ii P(x_1-x_1')}\twist^{\tau}_n(0)e^{-\ii H(x_0-x_0')+\ii P(x_1-x_1')} \ket{\theta_{1},\ldots\theta_{k}}_{\gamma_{1}\ldots \gamma_{k}}^{\nu_{1}\ldots \nu_{k}} \times\\
        &\hspace{4cm} \times \phantom{}_{\gamma_{1}\ldots \gamma_{k}}^{\nu_{1}\ldots \nu_{k}}{\bra{\theta_{1},\ldots\theta_{k}} \antitwist^{\tau}_n(0)\ket{0}} 
    \end{split}\notag\\
    \begin{split}
        &=\sum_{k=0}^{\infty}\sum_{\{\gamma\},\{\nu\}} \int \prod_{i=1}^k \dd\theta_i\, \bra{0} \twist^{\tau}_n(0) \ket{\theta_{1},\ldots\theta_{k}}_{\gamma_{1}\ldots \gamma_{k}}^{\nu_{1}\ldots \nu_{n}}\, e^{-\ii \sum_i E_i(x_0-x_0')+\ii \sum_i p_i(x_1-x_1')}\\
        &\hspace{4cm} \times \phantom{}_{\gamma_{1}\ldots \gamma_{k}}^{\nu_{1}\ldots \nu_{k}}{\bra{\theta_{1},\ldots\theta_{k}} \antitwist^{\tau}_n(0) \ket{0}} ,
    \end{split} \label{FFSeriesCF2}
\end{align}
where $E_i$ and $p_i$ are the single particle energies and momenta reported in Eq.~\eqref{eq:MasslessEnergyMomentum}.
The elementary FFs of a generic (semi-)local operator $\mathcal{O}(x, t)$ are their matrix elements between the vacuum and the asymptotic multi-particle states~\eqref{eq:basis}, i.e.
\begin{equation}\label{eq:FF}
F_{\gamma_{1}\ldots \gamma_{k}}^{\mathcal{O}|\nu_1\ldots \nu_k}(\theta_{1},\ldots,\theta_{k}) = \bra{0} \mathcal{O}(0,0) \ket{\theta_{1},\ldots\theta_{k}}_{\gamma_{1}\ldots \gamma_{k}}^{\nu_{1}\ldots \nu_{k}}\,.
\end{equation}
Using the definition~\eqref{eq:FF} of the FFs in the spectral sum representation in Eq.~\eqref{FFSeriesCF2}, we finally obtain the following expansion of the twist field correlator
\begin{equation}
\begin{split}
\bra{0} \twist^{\tau}_n(\textbf{x})\antitwist^{\tau}_n(0) \ket{0} = \sum_{k=0}^{\infty}\sum_{\{\gamma\},\{\nu\}} \int \prod_{i=1}^k\dd\theta_i \abs{ F_{\gamma_{1}\ldots \gamma_{k}}^{\twist_{\tau}|\nu_1\ldots \nu_k}(\theta_{1},\ldots,\theta_{k};n)}^2\, \exp\!\left(-\ell \sum_i^k E_i \right ),
\end{split}
\label{FFSeriesCFFF}
\end{equation}
where we switched to the Euclidean formalism for simplicity and $\ell$ denotes the Euclidean distance.
We can see from the above formula that the computation of the twist field correlation functions can be naturally formulated by means of FFs via the insertion of a complete set of asymptotic multi-particle states.
Crucially, the form factors in integrable QFTs can often be determined exactly, giving access to the corresponding correlation functions.
In the following, we review some basic properties of the twist field FFs and present the bootstrap equations from which their analytic expressions can be obtained.

\section{Form factors of the branch point twist field in the massless flow}\label{sec:twistFormFactors}

Given the exchange properties of the standard BPTFs~\eqref{CTFSpatialExchange}, it is possible to write down the bootstrap equations for the form factors~\eqref{eq:FF} associated with these fields in integrable QFTs.
Relying on earlier works~\cite{Ola,OlaBen, cd-09}, we can immediately specify these equations for our massless theory~\eqref{eq:flow_action}.
If we denote the FFs of $\twist_n$ by $ F^{\twist|\underline{\nu}}_{\underline{\gamma}}(\underline{\theta};n)$, then their bootstrap equations can be written as \cite{Ola,OlaBen, cd-09}
\begin{align}
    &F^{\twist|\underline{\nu}}_{\underline{\gamma}}(\underline{\theta};n) = S_{\gamma_{i}\gamma_{i+1}}^{\nu_{i}\nu_{i+1}}(\theta_{i,i+1})\, F^{\twist|\ldots \nu_{i-1}\nu_{i+1}\nu_{i}\nu_{i+2}\ldots}_{\ldots \gamma_{i-1}\gamma_{i+1}\gamma_{i}\gamma_{i+2}\ldots}(\ldots\theta_{i+1},\theta_{i},\ldots;n),\label{eq:FFAxiom1}\\
&F^{\twist|\underline{\nu}}_{\underline{\gamma}}(\theta_{1}+2\pi \ii,\theta_{2},\ldots,\theta_{k};n)= F^{\twist|\nu_{2}\nu_{3}\ldots \nu_{k}\hat{\nu}_{1}}_{\gamma_{2}\gamma_{3}\ldots \gamma_{k}\gamma_{1}}(\theta_{2},\ldots,\theta_{k},\theta_{1};n),\label{eq:FFAxiom2}\\
    &-\ii\res_{\theta_{0}'=\theta_{0}+\ii\pi} F^{\twist|\nu_0\nu_0\underline{\nu}}_{\bar{\gamma}_{0}\gamma_{0}\underline{\gamma}}(\theta_{0}',\theta_{0},\underline{\theta};n)= F^{\twist|\underline{\nu}}_{\underline{\gamma}}(\underline{\theta};n),\label{eq:FFAxiom3}\\
    &-\ii\res_{\theta_{0}'=\theta_{0}+\ii\pi} F^{\twist|\nu_0\hat{\nu_0}\underline{\nu}}_{\bar{\gamma}_{0}\gamma_{0}\underline{\gamma}}(\theta_{0}',\theta_{0},\underline{\theta};n)=-\prod_{l=1}^k S_{\gamma_0\gamma_{l}}^{\hat{\nu}_{0}\nu_l}(\theta_{0l})\, F^{\twist|\underline{\nu}}_{\underline{\gamma}}(\underline{\theta};n),\label{eq:FFAxiom4}
\end{align}
where we introduced
\begin{equation}
    \bar{\gamma}_{i}=\gamma_{i},\qquad\hat{\nu}_{i}=\nu_{i}+1,
\end{equation}
and $\bar{\gamma}_{i}$ denotes the anti-particle of $\gamma_{i}$ (which coincides with the particle in the theory under consideration). Here $\underline{\theta}$ and $\underline{\gamma}$, $\underline{\nu}$  
are shorthands for $(\theta_{1},\theta_{2},\ldots,\theta_{k})$ and $(\gamma_{1},\gamma_{2},\ldots,\gamma_{k})$, $(\nu_{1},\nu_{2},\ldots,\nu_{k})$
respectively, with $\gamma=R, L$ and $\bar{R}=R$, $\bar{L}=L$. In the argument of the $S$-matrices, $\theta_{ij}=\theta_i-\theta_i$.

In the massless flow~\eqref{eq:flow_action}, two particles of any type cannot form a bound state.
It is also easy to see that the one-particle FFs of BPTF are vanishing. The reason for this is that these fields are neutral w.r.t. the $\ZZ_2$ charge.
This implies that only FFs with an even number of $R$ and an even number of $L$ particles are non-vanishing and, consequently, the odd-particle FFs are zero.

Moreover, relativistic invariance imposes that
\begin{equation}
 F^{\twist|\underline{\nu}}_{\underline{\gamma}}(\theta_{1}+\Lambda,\ldots,\theta_{k}+\Lambda;n)=e^{\Sigma\Lambda}F^{\twist|\underline{\nu}}_{\underline{\gamma}}(\underline{\theta};n)=F^{\twist|\underline{\nu}}_{\underline{\gamma}}(\underline{\theta};n)\,,\label{eq:RelInv}
\end{equation}
where $\Sigma$ is the Lorentz spin, which is $\Sigma = 0$ for the twist field.

Another important property of form factors which will be useful in our analysis is the cluster property, studied in detail in Ref.~\cite{delta_theorem} and recognised in different models, see e.g.~\cite{ShGFFZamo,s-90,km-93,caf-01,cl-11}.
In the limit in which the difference between the particle rapidities diverges, the form factors factorise in the product of form factors with a lower number of particles.
In our model, the clusterisation of the different particle species can be phrased as
\begin{equation}\label{eq:cluster}
\lim_{\Lambda \rightarrow \infty} F^{\twist|\underline{\nu},\underline{\nu'}}_{\underline{R},\underline{L}}(\underline{\theta}+\Lambda, \underline{\theta}'-\Lambda;n)=\avg{\twist_{n}}^{-1}F^{\twist|\underline{\nu}}_{\underline{R}}(\underline{\theta};n)F^{\twist|\underline{\nu}'}_{\underline{L}}(\underline{\theta'};n)\,,
\end{equation}
where $\underline{\theta}$ and $\underline{\nu}$ stand for the rapidities and replica indices of the 'R' particles, and $\underline{\theta}'$ and $\underline{\nu}'$ for the 'L' particles. The cluster property for particles of the same species is instead written as
\begin{equation}\label{eq:cluster2}
\lim_{\Lambda \rightarrow \infty} F^{\twist|\underline{\nu}_1+\underline{\nu}_2,\underline{\nu}'}_{\underline{R}_1+\underline{R}_2,\underline{L}}(\underline{\theta}_1+\Lambda,\underline{\theta}_2-\Lambda, \underline{\theta}'-\Lambda;n)=\avg{\twist_{n}}^{-1}F^{\twist|\underline{\nu}_1}_{\underline{R}_1}(\underline{\theta}_1;n)F^{\twist|\underline{\nu}_1,\underline{\nu}'}_{\underline{R}_2,\underline{L}}(\underline{\theta}_2,\underline{\theta'};n)\,,
\end{equation}
with an analogous expression for the clustering of 'L' particles.

Let us now use the previous axioms to construct a set of solutions of the bootstrap equations~\eqref{eq:FFAxiom1}-\eqref{eq:FFAxiom4} for the BPTF form factors. To fix the ideas, we first place every particle on the first replica $\nu_i = 1$.
A convenient ansatz for the form factors is~\cite{mds-95}
\begin{equation}
\begin{split}
F^{\twist}_{\underline{R},\underline{L}}(\underline{\theta},\underline{\theta'};n) &= H^{\twist}_{r,l}Q^{\twist}_{r,l}(\underline{x},\underline{y};n) \prod_{1\leqslant i<j\leqslant r} \frac{f_{RR}(\theta_{i}-\theta_{j};n)}{(x_{i}-\omega x_{j})(x_{j}-\omega x_{i})}\,\times\\
 & \hspace{1cm} \times \prod_{i=1}^{r} \prod_{j=1}^{l} f_{RL}(\theta_{i}-\theta'_{j};n)\prod_{1\leqslant i<j\leqslant l} \frac{f_{LL}(\theta'_{i}-\theta'_{j};n)}{(y_{i}-\omega y_{j})(y_{j}-\omega y_{ji})},
\end{split}\label{FFParametrizationFlow}
\end{equation}
where we have $r$ right-moving and $l$ left-moving particles and we have defined $x_i = e^{\theta_i / n}$, $y_i = e^{-\theta_i'/n}$ and $\omega=e^{\ii\pi/n}$.
Notice that we simplified our notation by omitting the reference to the replica indices.
In the ansatz~\eqref{FFParametrizationFlow}, $Q^{\twist}_{r,l}$ are polynomials of their variables and $f_{RR}=f_{LL}$ and $f_{RL}$ are the minimal form factors. 
In Eq.~\eqref{FFParametrizationFlow}, the kinematical singularity of the FF (see Eq.~\eqref{eq:FFAxiom3}) comes entirely from the denominators and therefore the cyclic permutation and the exchange axioms, Eqs.~\eqref{eq:FFAxiom1} and \eqref{eq:FFAxiom2}, are automatically satisfied requiring the following identities for the minimal form factors:
\begin{equation}\label{eq:FminRRexchange}
\begin{gathered}
f_{RR}(\theta;n) = -f_{RR}(-\theta;n)\,,\\
f_{RR}(2\pi n \ii+\theta;n) = f_{RR}(-\theta;n)\,.
\end{gathered}
\end{equation}
By prescribing that the minimal form factor $f_{RR}$ has no poles and has the mildest asymptotic behaviour, we end up with the unique solution
\begin{equation}
    f_{RR}(\theta;n)=\sinh\left(\frac{\theta}{2n}\right),
\label{FMinRR}
\end{equation}
and $f_{RR}=f_{LL}$, which is identical to the minimal form factor of the massive Ising theory~\cite{Ola}.

For $f_{RL}$, the defining equations are
\begin{equation}\label{eq:FminRLexchange}
\begin{gathered}
    f_{RL}(\theta;n) = S_{RL}(\theta)f_{LR}(-\theta;n)\,,\\
    f_{RL}(2\pi n \ii+\theta;n) = f_{LR}(-\theta;n)\,,
\end{gathered}
\end{equation}
whose solution can be explicitly given based on the knowledge of the Fourier representation of the non-trivial $S$-matrix $S_{RL}$ in Eq.~\eqref{eq:s_matrix_flow}.
In particular, we can write the solution as
\begin{equation}
\label{FMinRL_exponential}
  f_{RL}\!\left ( \theta; n \right ) = \exp\!\left [ \frac{\theta}{4n} - \int_0^\infty \frac{dt}{t} \frac{\sin^2\!\left ( \frac{(\ii \pi n- \theta) t}{2\pi} \right )}{\sinh(n t) \cosh \frac{t}{2}} \right ],
\end{equation}
using an integral representation, or, alternatively, in terms of a mixed product integral representation
\begin{equation}
\begin{split}
  f_{RL}\!\left ( \theta; n \right ) = \exp\!\left(\frac{\theta }{4 n}\right)&\left[\prod _{k=0}^m \frac{\Gamma \left(\frac{2 k+n+\frac{3}{2}}{2 n}\right)^2 \Gamma\! \left(\frac{\frac{\ii \theta }{\pi }+2 k+\frac{1}{2}+2 n}{2 n}\right) \Gamma\!\left(\frac{-\frac{\ii \theta }{\pi }+2 k+\frac{1}{2}}{2 n}\right)}{\Gamma\!\left(\frac{2 k+n+\frac{1}{2}}{2 n}\right)^2 \Gamma\!\left(\frac{\frac{\ii \theta }{\pi }+2 k+\frac{3}{2}+2 n}{2 n}\right) \Gamma\!\left(\frac{-\frac{\ii \theta }{\pi }+2 k+\frac{3}{2}}{2 n}\right)}\right] \times \\ 
  &\hspace{2cm}\times \exp\!\left(\int_0^\infty \dd t\, \frac{e^{-(2 m+2)t} \sinh ^2\left(\frac{1}{2} t \left(n+\frac{\ii \theta }{\pi }\right)\right)}{t \cosh \left(\frac{t}{2}\right) \sinh (n t)}\right),
\label{FMinRL}
\end{split}
\end{equation}
which is more convenient for numerical evaluation.
Notice that, for $n = 1$, the minimal form factor~\eqref{FMinRL_exponential} reduces to the known result for a single replica obtained in Ref.~\cite{mds-95}.
Moreover, it is bounded, that is,  $f_{RL}(\theta, n)\rightarrow 0$ when $\theta \rightarrow -\infty$, 
and $f_{RL}(\theta, n)\rightarrow \mathcal{N}_n^{-1}$ when $\theta \rightarrow \infty$.
Therefore, it is customary to normalise it such that 
\begin{equation}
  \lim_{\theta \rightarrow \infty} \tilde{f}_{RL}(\theta;n)=1\,.
\end{equation}
In order to fix this normalisation along the massless flow, we compute the value $\mathcal{N}_n^{-1}$ of $f_{RL}$ in the limit $\theta \rightarrow \infty$, which reads
\begin{equation}
\mathcal{N}_n^{-1}= e^{\ii \pi / 4}\, 2^{1/4} \exp\!\left(\frac{1}{8} \int_{-\infty}^{+\infty} \frac{\dd t}{t} \frac{1}{\sinh(nt)} \left[1 - \frac{1}{\cosh t \cosh(nt)} \right] \right) = e^{\ii \pi / 4}\, 2^{1/4}\, e^{G_n/\pi}\,,
\label{f_RLNormalisation}
\end{equation}
where we have defined the sequence
\begin{equation}
  G_n = \frac{\pi}{8} \int_{-\infty}^{+\infty} \frac{\dd t}{t} \frac{1}{\sinh(nt)} \left[1 - \frac{1}{\cosh t \cosh(nt)} \right ],
\end{equation}
which is equal to Catalan's constant $G$ for $n=1$, recovering the normalisation for $f_{RL}$ found in the non-replicated theory in Ref.~\cite{mds-95}. With the choice
\begin{equation}
  \tilde{f}_{RL}(\theta; n)=\mathcal{N}_n f_{RL}(\theta;n)\,,
\end{equation}
we fix all the constants in the form factors for the massless flow.
An important property that the $f_{RL}$ minimal form factor satisfies is
\begin{equation}
\mathcal{N}_n f_{RL}(\theta+\ii\pi;n)\, \mathcal{N}_n f_{RL}(\theta;n)=\left(1-e^{-\frac{\ii \pi }{2 n}} e^{-\frac{\theta }{ n}}\right)^{-1} ,
\label{FminRLFminRL}
\end{equation}
which shall be very useful in the rest of the section.

The ansatz~\eqref{FFParametrizationFlow}, with the definitions for the minimal FFs $f_{RR}$~\eqref{FMinRR} and $f_{RL}$~\eqref{FMinRL}, satisfies all the axioms for the 
BPTF FFs. The eventual determination of $F^{\twist}_{\underline{R},\underline{L}}(\underline{\theta},\underline{\theta'};n)$ can be done recursively.
In fact, by applying the residue axiom in Eq.~\eqref{eq:FFAxiom3} to the ansatz~\eqref{FFParametrizationFlow}, one can derive recursive equations for the unknown $Q^{\twist}_{r,l}(\underline{x},\underline{y};n)$ functions that relate $Q^{\twist}_{r+2,l}(\underline{x},\underline{y};n)$ or $Q^{\twist}_{r,l+2}(\underline{x},\underline{y};n)$ to $Q^{\twist}_{r,l}(\underline{x},\underline{y};n)$, that is, to $Q^{\twist}_{r,l}$ functions with fewer particles.
In the next subsections and in App.~\ref{app:FlowBootstrapTwist}, we explicitly demonstrate how the determination of higher-particle FFs is carried out by solving the recursive equations for the polynomials $Q_{r, l}^\twist$. 

\subsection{Two-particle form factors and form factors with only one species}
Since the Lorentz spin of the BPTFs is zero, their two-particle FFs only depend on one rapidity variable~\eqref{eq:RelInv}, that is, the rapidity difference $\theta_1-\theta_2$.
Recall that, because of the spin-flip symmetry, 
we can only have `RR' and `LL' form factors, which means that these quantities coincide with those of the massive Ising QFT (c.f. Eqs.~\eqref{FFParametrizationFlow} and~\eqref{FMinRR}) up to the vacuum expectation value $\langle \twist_n\rangle$.
These quantities, nevertheless, can also be easily obtained from the bootstrap equations~\eqref{eq:FFAxiom1}, \eqref{eq:FFAxiom2}.
For the two-particle form factors, they imply that 
\begin{equation}
F^{\twist|\nu_i\nu_j}_{\gamma_{i}\gamma_{j}}(\theta;n) = S_{\gamma_{i}\gamma_{j}}^{\nu_i\nu_j}(\theta)F^{\twist|\nu_j\nu_i}_{\gamma_{j}\gamma_{i}}(-\theta;n)=F^{\twist|\nu_j\nu_i}_{\gamma_{j}\gamma_{i}}(2\pi \ii n-\theta;n)\,.\label{fpm2U1}
\end{equation}
In this case, the kinematic residue equation~\eqref{eq:FFAxiom3},
\begin{equation} \label{kinU1}
    -\ii\, \res_{\theta=\ii\pi} F^{\twist|\nu\nu}_{\gamma\gamma}(\theta;n)=\avg{\twist_n},
\end{equation}
connects the two-particle FFs and the vacuum expectation value of the twist field. We can therefore write
\begin{equation}\label{eq:F2TwistField}
 F^{\twist|11}_{\gamma \gamma}(\theta;n)=\frac{\avg{\twist_{n}} \sin\frac{\pi}{n}}{2n\sinh\frac{\ii\pi+\theta}{2n} \sinh\frac{\ii\pi-\theta}{2n}}\, \frac{\sinh(\theta/(2n))}{\sinh(\ii\pi/(2n))}\,.
\end{equation}
If this formula is recast in the form of the ansatz~\eqref{FFParametrizationFlow}, then we have the equivalent expression
\begin{equation}\begin{split}
  F_{2,0}^\twist\!\left(\theta_1-\theta_2;n\right )
  &= -\ii\, \avg{\twist_n} \frac{4\,\omega}{n}\cos\left(\frac{\pi}{2 n}\right)\frac{ x_1\, x_2\, f_{RR}(\theta_1-\theta_2;n)}{(x_1-\omega x_2)(x_2-\omega x_1)} \\
  &= H^\twist_{2,0}\, Q^\twist_{2,0}\!\left ( x_1, x_2\right ) \frac{f_{RR}(\theta_{1}-\theta_{2};n)}{(x_{1}-\omega x_{2})(x_{2}-\omega x_{1})},
  \label{2pBPTFRecasted}
\end{split}\end{equation}
in which we identify
\begin{gather}
  H^\twist_{2,0} = -\ii \avg{\twist_n}\frac{4\,\omega}{n}\cos\left(\frac{\pi}{2 n}\right)\,,\\
  Q^\twist_{2,0}\!\left ( x_1, x_2\right ) = \sigma_2\!\left(x_1, x_2 \right ) = x_1 x_2,
\end{gather}
where $\sigma_j$ is the fully symmetric polynomial of degree $j$ in the variables $x_1$ and $x_2$.
Since in the formula above both particles live in the first replica, we slightly changed the notation, namely we denote the form factor corresponding to two right-moving particles living in the first replica $F_{R R}^{\twist|11}$ as $F_{2,0}^\twist$ . In the following, we shall use this convention whenever all the particles are on the first replica. The `LL' form factor can be obtained by replacing $x_1$ and $x_2$ by $y_1$ and $y_2$ in Eq.~\eqref{2pBPTFRecasted}.

From $F^{\twist|11}_{\gamma \gamma}(\theta;n)$, we can obtain the form factors $F^{\twist|jk}_{\gamma\gamma}(\theta;n)$ corresponding to particles in different replicas from~\cite{Ola}
\begin{equation}
 F^{\twist|jk}_{\gamma\gamma}(\theta;n)=\begin{cases}
 F^{\twist|11}_{\gamma\gamma}(2\pi -\ii(k-j)-\theta;n), & \text{if }k>j,\\
 F^{\twist|11}_{\gamma\gamma}(2\pi -\ii(j-k)+\theta;n), & \text{otherwise.}
\end{cases}\label{eq:FD2Full}
\end{equation}
The form factors $\tilde{F}$ of the antitwist field $\antitwist_n$ can be simply obtained from those of $\twist_n$ through the relation~\cite{Ola}
\begin{equation}
\tilde{F}^{\twist|jk}_{\gamma\gamma}(\theta;n)=F^{\twist|n-j,n-k}_{\gamma\gamma}(\theta;n)\,.
\end{equation}

As we already said, the only non-vanishing FFs with higher-particle number are those containing an even number of `R' and `L' particles.
It is easy to see that, in the particular case of form factors only containing an even number of particles of the same type, that is, the `RR...RR' and `LL...LL' form factors, they exactly coincide with the standard BPTF FFs of the massive Ising theory up to the vacuum expectation value $\langle\twist_n\rangle$, similarly to the two-particle case discussed above.
These form factors can be easily obtained from the two-particles ones. In particular, the form factor with $2k$ particles of the same type is given by
\begin{equation}
F_{\gamma\ldots \gamma}^{\twist|\nu_1\ldots\nu_{2k}}(\theta_{1},\ldots,\theta_{2k};n)=\avg{\twist_{n}}\Pf(W)\,,\label{eq:Pfaffain} 
\end{equation}
for $\nu_{1}\geq \nu_{2}\geq\ldots\geq \nu_{2k}$.\
Here $\Pf(W)$ is the Pfaffian of the $2k\times2k$ anti-symmetric matrix $W$ with entries 
\begin{equation}
    W_{lm} = \frac{1}{\avg{\twist_n}} \begin{cases}
        F_{\gamma\gamma}^{\twist|\nu_l\nu_m}\!\left ( \theta_{l}-\theta_{m};n\right ),  & m > l ,\\
        \left(-1\right)^{\delta_{\nu_{l},\nu_{m}}+1} F_{\gamma\gamma}^{\twist|\nu_l\nu_m}\!\left ( \theta_{l}-\theta_{m};n \right ),    & m < l .
    \end{cases}
\end{equation}
If the
ordering of the indices $\nu_{i}$ is not the canonical one, using the
exchange axiom~\eqref{eq:FFAxiom1} one can reshuffle the particles and their rapidities to satisfy $\nu_{1}\geq \nu_{2}\geq\ldots\geq \nu_{2k}$ and apply~\eqref{eq:Pfaffain}. 
In particular, for the `RRRR' or `LLLL' FFs with all the particles in the same replica, we have the simple formula
\begin{equation}
\begin{split}
F_{4,0}^{\twist}(\theta_{1},\theta_{2},\theta_{3},\theta_{4};n)&=\avg{\twist_{n}}^{-1} \left[F_{2,0}^{\twist}(\theta_{1}-\theta_{2};n) F_{2,0}^{\twist}(\theta_{3}-\theta_{4};n)\right.\\
&\hspace{.5cm}\left. -F_{2,0}^{\twist}(\theta_{1}-\theta_{3};n) F_{2,0}^{\twist}(\theta_{2}-\theta_{4};n)+F_{2,0}^{\twist}(\theta_{1}-\theta_{4};n) F_{2,0}^{\twist}(\theta_{2}-\theta_{3};n)\right].
\end{split}\label{FFBPTFRRRR}
\end{equation}

\subsection{Solution for the four particle `RRLL' form factor}\label{sec:RRLLtwistFF}

The first non-vanishing form factors that contain both `R' and `L' 
particles appear at the four-particle level: $F_{RRLL}^{\twist|\nu_1\nu_2\nu_3\nu_4}$ with any permutation of `R' and `L'.
Similarly to the other FFs previously discussed, it is sufficient to determine only the `RRLL' form factor with all the particles on the first replica.
Using then the exchange relation~\eqref{eq:FFAxiom1} we can readily obtain any other sequence of the particle species, and, applying the cyclic permutation axiom~\eqref{eq:FFAxiom2}, we can obtain FFs for particles living on different replicas.
Following the notation introduced for the form factors with all the particles on the first replica, we will denote $F_{RRLL}^{\mathcal{T}|1111}$ as $F_{2, 2}^\twist$.
In this case, the ansatz~\eqref{FFParametrizationFlow} takes the form
\begin{equation}
\begin{split}
F^{\twist}_{2,2}(\theta_1, \theta_2,\theta_1', \theta_2';n) =&\, H^{\twist}_{2,2}\, Q^{\twist}_{2,2}(x_1,x_2,y_1,y_2;n) \frac{f_{RR}(\theta_{1}-\theta_{2};n)}{(x_{1}-\omega x_{2})(x_{2}-\omega x_{1})}\times\\
 &\hspace{1cm} \times \prod_{i=1}^{2} \prod_{j=1}^{2} f_{RL}(\theta_{i}-\theta'_{j};n) \frac{f_{LL}(\theta'_{1}-\theta'_{2};n)}{(y_{1}-\omega y_{2})(y_{2}-\omega y_{1})}\:.
\end{split}
\label{FFParametrizationFlowRRLLComposite}
\end{equation}
Applying now the residue axiom~\eqref{eq:FFAxiom3} to Eq.~\eqref{FFParametrizationFlowRRLLComposite}, we can derive recursive equations for the $H^{\twist}_{2,2}$ normalisation factor and the $Q^{\twist}_{2,2}$ function. 
The detailed solution of this equation for the case of four-particles (RRLL) is presented in App.~\ref{app:FlowBootstrapTwist} and here we report the results of the calculations.

The normalisation factor reads
\begin{equation}
  H^{\twist}_{2,2} = -\avg{\twist_{n}}\mathcal{N}_n^4 \left [ \frac{4\,\omega}{n}\cos \left(\frac{\pi}{2 n}\right) \right ]^2,
\end{equation}
where $\mathcal{N}_n$ is given by Eq.~\eqref{f_RLNormalisation}, while for the polynomial $Q_{2, 2}^{\twist}$ we obtain
\begin{equation}\begin{split}
  Q^{\twist}_{2,2}\!\left (x_1, x_2, y_1, y_2 ; n \right) &= 1 -\frac{1}{2 \cos \frac{\pi}{2 n}} \sigma_1(x_1, x_2) \sigma_1(y_1, y_2) + \sigma_2(x_1, x_2)\sigma_2(y_1, y_2) \\
  &= 1 -\frac{1}{2 \cos \frac{\pi}{2 n}}\left ( x_1 + x_2 \right ) \left ( y_1 + y_2 \right ) + x_1 x_2 y_1 y_2,
\end{split}\end{equation}
where $\sigma_i, i = 1, 2$ denotes the completely symmetrical polynomial of degree $i$ in two variables. Using these results, the final solution for the full FF is
\begin{equation}\begin{split}
  F^\twist_{2,2}(\theta_{1},\theta_{2},\theta'_{1},\theta'_{2}; n) =& -\mathcal{N}_n^4 \left [ \frac{4\,\omega}{n}\cos \left(\frac{\pi}{2 n}\right) \right ]^2 \left [ 1 - \frac{1}{2 \cos \frac{\pi}{2 n}}\left ( x_1 + x_2 \right ) \left ( y_1 + y_2 \right ) + x_1 x_2 y_1 y_2 \right ] \times\\
  & \times \frac{f_{RR}(\theta_{1}-\theta_{2};n)}{(x_{1}-\omega x_{2})(x_{2}-\omega x_{1})} \prod_{i=1}^{2} \prod_{j=1}^{2} f_{RL}(\theta_{i}-\theta'_{j};n) \frac{f_{LL}(\theta'_{1}-\theta'_{2};n)}{(y_{1}-\omega y_{2})(y_{2}-\omega y_{1})}\,,
\end{split}
\label{FFRRLLFlow}
\end{equation}
which we can also rewrite as 
\begin{equation}\begin{split}
  &F^\twist_{2,2}(\theta_{1},\theta_{2},\theta'_{1},\theta'_{2}; n) =\\
  &=- 2\, \mathcal{N}_n^4\, e^{-\frac{\theta_1 + \theta_2 - \theta_1' - \theta_2'}{2n}} \left [ \cosh\!\left ( \frac{\theta_1 + \theta_2 - \theta_1' - \theta_2'}{2 n} \right ) - \frac{\cosh\!\left ( \frac{\theta_1 + \theta_2}{2 n} \right )\cosh\!\left ( \frac{ \theta_1' + \theta_2'}{2 n} \right )}{\cos \frac{\pi}{2 n}} \right ] \times \\
  &\hspace{5.5cm}\times F^\twist_{2,0}(\theta_{1},\theta_{2}; n)\prod_{i=1}^{2} \prod_{j=1}^{2} f_{RL}(\theta_{i}-\theta'_{j};n)\, F^\twist_{0,2}(\theta_{1}',\theta_{2}'; n)\,.
\label{FFRRLLFlow2}
\end{split}\end{equation}
We remark that the form factor in Eq.~\eqref{FFRRLLFlow} is one of the main results of this paper. As we will show in Sec.~\ref{sec:entropy}, it will provide the leading correction to the IR expressions for the entanglement entropy.

\section{Form factors of the \texorpdfstring{$\ZZ_2$}{Z2}-composite branch point twist field in the massless flow}\label{sec:compositeFormFactors}

In this section, we derive the bootstrap equations for the 
form factors of the $\mathbb{Z}_2$-composite BPTFs associated with the disorder field $\mu$ along the massless flow~\eqref{eq:flow_action} and we obtain their explicit solution for the two and four-particle cases. 
Similarly to the standard BPTFs discussed in Sec.~\ref{sec:twistFormFactors}, from the exchange properties of the $\mathbb{Z}_2$-composite twist fields~\eqref{eq:ExchangeRelationComp1} or~\eqref{eq:ExchangeRelationComp2}, we can easily write down their form factor bootstrap equations.
Importantly, these equations include the non-trivial phase $e^{i\pi\kappa_\mathcal{O}}$ in the monodromy properties due to the insertion of the disorder field $\mu$. The asymptotic states \eqref{eq:basis} that enter in the definition of the twist field FFs are constructed 
from the fields $\psi$, $\bar{\psi}$, which are odd under the $\mathbb{Z}_2$ transformation, i.e. $\kappa_\psi=1$, and therefore we must take into account a phase $e^{i\pi}$ when moving between replicas.
However, as we discussed around Eqs.~\eqref{eq:ExchangeRelationComp1} and \eqref{eq:ExchangeRelationComp2}, we have two different ways to introduce it, either as a whole phase $e^{i\pi}$ in each replica, which is valid only for odd $n$, or inserting it only in the last one.
These two approaches lead to slightly different form factor bootstrap equations.
In this section, we comment both choices. In particular, we will show that the two conventions give the same result for the form factors up to some $(-1)$ factors which do not influence the final physical result.

Let us denote as $F^{\twist^\mu|\underline{\nu}}_{\underline{\gamma}}(\underline{\theta},n)$ the form factors of the composite twist fields $\twist_n^\mu$. If we introduce the phase $e^{\ii\pi}$ on the last replica only, that is taking the exchange relations~\eqref{eq:ExchangeRelationComp2}, the bootstrap equations take the form
\begin{align}
    &F^{\twist^\mu|\underline{\nu}}_{\underline{\gamma}}(\underline{\theta};n) = S_{\gamma_{i}\gamma_{i+1}}^{\nu_{i}\nu_{i+1}}(\theta_{i,i+1})\, F^{\twist^\mu|\ldots \nu_{i-1}\nu_{i+1}\nu_{i}\nu_{i+2}\ldots}_{\ldots \gamma_{i-1}\gamma_{i+1}\gamma_{i}\gamma_{i+2}\ldots}(\ldots\theta_{i+1},\theta_{i},\ldots;n)\,,\label{eq:U1FFAxiom1}\\
    &F^{\twist^\mu|\underline{\nu}}_{\underline{\gamma}}(\theta_{1}+2\pi \ii,\theta_{2},\ldots,\theta_{k};n)=F^{\twist^\mu|\nu_{2}\nu_{3}\ldots \nu_{k}\hat{\nu}_{1}}_{\gamma_{2}\gamma_{3}\ldots \gamma_{k}\gamma_{1}}(\theta_{2},\ldots,\theta_{k},\theta_{1};n)\times \begin{cases} 
        -1,   &\nu_1=n, \\ 
        1,    &\text{otherwise,}
    \end{cases}\label{eq:U1FFAxiom2}\\
    &-\ii\res_{\theta_{0}'=\theta_{0}+\ii\pi} F^{\twist^\mu|\nu_0\nu_0\underline{\nu}}_{\bar{\gamma}_{0}\gamma_{0}\underline{\gamma}}(\theta_{0}',\theta_{0},\underline{\theta};n)= F^{\twist^\mu|\underline{\nu}}_{\underline{\gamma}}(\underline{\theta};n)\,,\label{eq:U1FFAxiom3}\\
    &-\ii\res_{\theta_{0}'=\theta_{0}+\ii\pi} F^{\twist^\mu|\nu_0\hat{\nu_0}\underline{\nu}}_{\bar{\gamma}_{0}\gamma_{0}\underline{\gamma}}(\theta_{0}',\theta_{0},\underline{\theta};n)=-\prod_{l=1}^k S_{\gamma_0\gamma_{l}}^{\hat{\nu}_{0}\nu_l}(\theta_{0l})\, F^{\twist^\mu|\underline{\nu}}_{\underline{\gamma}}(\underline{\theta};n)\times \begin{cases}
        -1, &\nu_0 = n, \\
        1,  &\text{otherwise.}
    \end{cases}\label{eq:U1FFAxiom4}
\end{align}
On the other hand, if we introduce the same flux between all the copies, we have
\begin{align}
     &F^{\twist^\mu|\underline{\nu}}_{\underline{\gamma}}(\underline{\theta};n) = S_{\gamma_{i}\gamma_{i+1}}^{\nu_{i}\nu_{i+1}}(\theta_{i,i+1})\, F^{\twist^\mu|\ldots \nu_{i-1}\nu_{i+1}\nu_{i}\nu_{i+2}\ldots}_{\ldots \gamma_{i-1}\gamma_{i+1}\gamma_{i}\gamma_{i+2}\ldots}(\ldots\theta_{i+1},\theta_{i},\ldots;n),\label{eq:U1FFAxiom1Sym}\\
     &F^{\twist^\mu|\underline{\nu}}_{\underline{\gamma}}(\theta_{1}+2\pi \ii,\theta_{2},\ldots,\theta_{k};n)= -F^{\twist^\mu|\nu_{2}\nu_{3}\ldots \nu_{k}\hat{\nu}_{1}}_{\gamma_{2}\gamma_{3}\ldots \gamma_{k}\gamma_{1}}(\theta_{2},\ldots,\theta_{k},\theta_{1};n) ,\label{eq:U1FFAxiom2Sym}\\
     &-\ii\res_{\theta_{0}'=\theta_{0}+\ii\pi} F^{\twist^\mu|\nu_0\nu_0\underline{\nu}}_{\bar{\gamma}_{0}\gamma_{0}\underline{\gamma}}(\theta_{0}',\theta_{0},\underline{\theta};n)= F^{\twist^\mu|\underline{\nu}}_{\underline{\gamma}}(\underline{\theta};n),\label{eq:U1FFAxiom3Sym}\\
     &-\ii\res_{\theta_{0}'=\theta_{0}+\ii\pi} F^{\twist^\mu|\nu_0\hat{\nu_0}\underline{\nu}}_{\bar{\gamma}_{0}\gamma_{0}\underline{\gamma}}(\theta_{0}',\theta_{0},\underline{\theta};n)=\prod_{l=1}^k S_{\gamma_0\gamma_{l}}^{\hat{\nu}_{0}\nu_l}(\theta_{0l})\, F^{\twist^\mu|\underline{\nu}}_{\underline{\gamma}}(\underline{\theta};n),\label{eq:U1FFAxiom4Sym}
\end{align}
where notations are the same as for the standard BPTF discussed in Sec.~\ref{sec:twistFormFactors}; in particular, we recall that $\gamma_i = R, L$. Both the Lorentz spin and the $\ZZ_2$ charge of the composite BPTFs are zero. Observe that the phase $(-1)$ in Eqs.~\eqref{eq:U1FFAxiom2Sym} and \eqref{eq:U1FFAxiom4Sym} as well as in Eqs.~\eqref{eq:U1FFAxiom2} and \eqref{eq:U1FFAxiom4} is due to the non-trivial monodromy of the fields $\psi$, $\bar{\psi}$  with $\mathcal{T}_n^\mu$ (compare with the analogous axioms for the standard BPTF in Eqs.~\eqref{eq:FFAxiom2} and \eqref{eq:FFAxiom4}).

Similarly to the standard BPTF, only FFs with an even number of `R' and `L' particles are non-vanishing and, consequently, the odd-particle FFs are zero. Additionally, the FFs of the composite BPTF satisfy the momentum space clustering property in the same form as the FFs of the standard BPTF in Eqs.~\eqref{eq:cluster} and \eqref{eq:cluster2}.

Analogously to what we have done in Sec.~\ref{sec:twistFormFactors}, let us assume the following ansatz for the composite twist field FFs in which, for simplicity, we place every particle in the first replica
\begin{equation}\label{FFParametrizationFlowComposite}
\begin{split}
F^{\twist^\mu}_{r,l}(\underline{\theta},\underline{\theta'};n) =&\, H^{\twist^\mu}_{r,l}\, Q^{\twist^\mu}_{r,l}(\underline{x},\underline{y};n)\prod_{1\leqslant i<j\leqslant r} \frac{f^\mu_{RR}(\theta_{i}-\theta_{j};n)}{(x_{i}-\omega x_{j})(x_{j}-\omega x_{i})}\times\\
 &\hspace{1cm} \times \prod_{i=1}^{r} \prod_{j=1}^{l} f^{\mu}_{RL}(\theta_{i}-\theta'_{j};n)\prod_{1\leqslant i<j\leqslant l} \frac{f^\mu_{LL}(\theta'_{i}-\theta'_{j};n)}{(y_{i}-\omega y_{j})(y_{j}-\omega y_{i})},
\end{split}
\end{equation}
where we have $r$ right-mover and $l$ left-mover particles, and $x_i=e^{\theta_i/n}$, $y_i=e^{-\theta'_i/n}$ and $\omega=e^{\ii\pi/n}$ as previously.
The cyclic permutation and the exchange axioms can automatically be satisfied if the equalities
\begin{equation}\label{eq:FMinFlowEquationComposite}
f^\mu_{\gamma\gamma}\!\left (2\pi \ii n-\theta;n \right ) = - f^\mu_{\gamma\gamma}\!\left (\theta;n \right ) = f^\mu_{\gamma\gamma}\!\left (-\theta;n \right ),
\end{equation}
are imposed, that is, the minimal form factors satisfy the non-trivial monodromy due to the insertion of the external flux. The solution of Eq.~\eqref{eq:FMinFlowEquationComposite} can be easily obtained from the standard minimal form factor in Eq.~\eqref{FMinRR} by simply introducing a factor $2 \cosh(\theta/2n)$ which changes the monodromy properties~\cite{Z2IsingShg} 
\begin{equation}\label{eq:fminRR^mu}
f^\mu_{\gamma\gamma}(\theta;n)=2 \cosh\!\left(\frac{\theta}{2n} \right )f_{\gamma\gamma}(\theta;n)=\sinh\left(\frac{\theta}{n}\right).
\end{equation}

For $f^{\mu}_{RL}(\theta;n)$ instead we have two possible choices. We might either choose the unaltered equation without the `$-1$' monodromy
\begin{equation}
f^{\mu}_{RL}\!\left ( \theta;n \right ) = S_{RL}\!\left (\theta\right ) f^{\mu}_{LR}\!\left ( - \theta;n \right ),
\label{F_RL^MuTriv}
\end{equation}
with the solution $f^{\mu}_{RL}=f_{RL}$, or we can also introduce the monodromy 
\begin{equation}
f^\mu_{RL}\!\left (2\pi \ii n-\theta;n\right ) = - f^\mu_{LR}\!\left ( \theta;n\right ) = - S_{LR}\!\left (\theta\right ) f^\mu_{LR}\!\left ( -\theta;n \right ),
\end{equation}
such that the solution becomes
\begin{equation}
f^\mu_{RL}\!\left ( \theta;n\right ) = e^{\theta/(2n)}f_{RL}\!\left (\theta;n\right ).
\label{F_RL^Mu}
\end{equation} 
As we will later see in Sec.~\ref{sec:RoamingLimit}, the exponential factor in Eq.~\eqref{F_RL^Mu} also appears in the roaming limit approach.
Importantly, the two choices for the minimal form factor $f^{\mu}_{RL}$ in Eqs.~\eqref{F_RL^MuTriv} and~\eqref{F_RL^Mu} are completely equivalent because for the composite BPTFs the number of `R' and `L' particles is always even. This implies that, in a FF, we always have the product of an even number of $f^\mu_{RL}$ terms, which implies that the $(-1)$ phases always mutually cancel. 
In order to connect in a clearer way with the roaming limit that we later discuss in Sec.~\ref{sec:RoamingLimit}, we choose Eq.~\eqref{F_RL^Mu} as the minimal form factors in the ansatz~\eqref{FFParametrizationFlowComposite} for the composite BPTF. If we had taken~\eqref{F_RL^MuTriv}, we would have got different expressions for the functions $Q^{\twist^\mu}_{r,l}$, which would differ only by products of $x_i$ and $y_j$ with the same integer powers.

We remark that, in contrast to what happened in Sec.~\ref{sec:twistFormFactors}, the ansatz~\eqref{FFParametrizationFlowComposite} does not guarantee that $Q^{\twist^\mu}_{r,l}$ is actually a polynomial.
In fact, as we will explicitly show, this function is in general a rational function.
The reason for this is the monodromy changing factor introduced in the minimal form factors in Eqs.~\eqref{eq:fminRR^mu} and \eqref{F_RL^Mu}. These terms possess additional zeros that cancel out with the denominator of the function $Q^{\twist^\mu}_{r,l}$, guaranteeing that the pole structure remains compatible with the bootstrap axioms.

\subsection{Two-particle form factors and form factors with only one species}

Similarly to the standard BPTFs, for the composite BPTFs the only non vanishing form factors at the two-particle level are those containing a pair of `R' or `L' particles, which coincide with the analogous expressions of the massive Ising QFT~\cite{Ola}. Alternatively, they can easily be obtained from the bootstrap equations, either from Eqs.~\eqref{eq:U1FFAxiom1},
\eqref{eq:U1FFAxiom2} or from Eqs.~\eqref{eq:U1FFAxiom1Sym},
\eqref{eq:U1FFAxiom2Sym}. For the two-particle form factors, the bootstrap equations imply that 
\begin{equation}\label{fpm2U1Comp}
F^{\twist^\mu|\nu_i\nu_j}_{\gamma_{i}\gamma_{j}}(\theta;n) = S_{\gamma_{i}\gamma_{j}}^{\nu_i\nu_j}(\theta)F^{\twist|\nu_j\nu_i}_{\gamma_{j}\gamma_{i}}(-\theta;n)=-F^{\twist^\mu|\nu_j\nu_i}_{\gamma_{j}\gamma_{i}}(2\pi \ii n-\theta;n)\,.
\end{equation}
The kinematic residue equations~\eqref{eq:U1FFAxiom3} or~\eqref{eq:U1FFAxiom3Sym} relate the FFs to the vacuum expectation value of $\twist_n^\mu$ as
\begin{equation}
 -\ii\, \res_{\theta=\ii\pi} F^{\twist^\mu|\nu\nu}_{\gamma\gamma}(\theta;n)=\avg{\twist_n^\mu}.
\end{equation}
The solution for the equations above can be immediately written by plugging in the two-particle FF of the standard twist field~\eqref{eq:F2TwistField} the minimal form factor of Eq.~\eqref{eq:fminRR^mu} that takes into account the non-trivial monodromy of $\twist_n^\mu$, obtaining 
\begin{equation}
F^{\twist^\mu | 11}_{\gamma \gamma}\!\left(\theta;n\right ) = \frac{\avg{\twist^{\mu}_{n}}\sin\frac{\pi}{n}}{2n\sinh\!\frac{\ii\pi+\theta}{2n}\, \sinh\!\frac{\ii\pi-\theta}{2n}}\,\frac{\sinh(\theta/n)}{\sinh(\ii\pi/n)}\,,
\label{eq:F2Z2TwistField}
\end{equation}
where, for simplicity, we have placed every particle on the first replica. 
Notice that Eq.~\eqref{eq:F2Z2TwistField} is not in the form of our ansatz~\eqref{FFParametrizationFlowComposite}, but it can be recast accordingly as
\begin{equation}
F^{\twist^\mu}_{2,0}(\theta_1-\theta_2;n)=\avg{\twist_{n}}\frac{\ii\, \omega }{n}\frac{2 x_1 x_2}{ (x_1-\omega x_2)(x_2-\omega x_1)}\sinh\left(\frac{\theta_1-\theta_2}{n}\right),
\label{FFCBFTRRRewritten}
\end{equation}
where an analogous expression for the `LL' form factor holds upon replacing $x_i$ with $y_i$. 
Since in the above formula each particle lives on the first replica, we again used the simplified notation to denote the FF, namely we write $F_{2,0}^{\twist^\mu}$ which indicates that we have two right-moving particles on the first replica. 
In the following, we shall use this convention whenever all the particles are on the first replica.

The two-particle FFs with arbitrary replica indices can be straightforwardly obtained from the result~\eqref{FFCBFTRRRewritten} with all the particles on the first replica. Importantly, the different flux convention in Eqs.~\eqref{eq:U1FFAxiom1}-\eqref{eq:U1FFAxiom4} or in Eqs.~\eqref{eq:U1FFAxiom1Sym}-\eqref{eq:U1FFAxiom4Sym} only differ in some $(-1)$ factors.
In particular, if the flux is only inserted on one replica, we have
\begin{equation}
    F^{\twist^\mu|jk}_{\gamma\gamma}\!\left (\theta;n \right ) = \begin{cases}
        F^{\twist^\mu|11}_{\gamma\gamma}\!\left (2\pi \ii(k-j)-\theta;n \right ),   & \text{ if }k>j,\\
        F^{\twist^\mu|11}_{\gamma\gamma}\!\left (2\pi \ii(j-k)+\theta;n \right ),   & \text{ otherwise.}
    \end{cases}
\label{eq:FCPTAAbarFull}
\end{equation}
The FFs of the anti-twist field $\antitwist^{\mu}_n$ denoted by $\widetilde{F}^{\twist^\mu}_{\underline{a}}(\underline{\theta},n)$ can be simply written as~\cite{U(1)FreeFF}
\begin{equation}
\widetilde{F}^{\twist^\mu|jk}_{\gamma\gamma}\!\left (\theta;n \right ) = F^{\twist^\mu|n-j, n-k}_{\gamma\gamma}\!\left (\theta;n \right ).
\label{FCBPT2ptInverse}
\end{equation}
If the flux is instead introduced on each replica, we have
\begin{equation}
    F^{\twist^\mu|jk}_{\gamma\gamma}\!\left (\theta;n \right ) = (-1)^{(k-j)} \begin{cases}
        F^{\twist^\mu|11}_{\gamma\gamma}\!\left (2\pi \ii(k-j)-\theta;n \right ),   & \text{ if }k>j,\\
        F^{\twist^\mu|11}_{\gamma\gamma}\!\left (2\pi \ii(j-k)+\theta;n \right ),   & \text{ otherwise,}
    \end{cases}
\label{eq:FCPTAAbarFull_eachReplica}
\end{equation}
while the FFs of the anti-twist field $\antitwist^{\mu}_n$ satisfy Eq.~\eqref{FCBPT2ptInverse}. In the computation of the symmetry resolved entropy, the additional factor $(-1)^{k-j}$ always cancels out, leading to the same value for both choices.

Similarly to the treatment of the standard BPTFs, it is easy to see that, in the particular case of form factors that only contain an even number of particles of the same type ---that is the `RR...RR' and `LL...LL' form factors---, they exactly coincide with the $\ZZ_2$-composite BPTF FFs of the massive Ising theory~\cite{Ola}, as occurs in the two-particle case discussed above.
Assuming that $\nu_{1}\geq \nu_{2}\geq\ldots\geq \nu_{2k}$, they can be written in terms of a Pfaffian involving the two-particle FFs as
\begin{equation}
\label{eq:Pfaffian2}
F^{\twist^\mu | \nu_1\ldots\nu_{2k}}_{\gamma,\ldots, \gamma}\!\left (\theta_{1},\ldots,\theta_{2k};n \right ) = \avg{\twist_{n}^\mu} \Pf\!\left ( W^{\mu} \right ) ,
\end{equation}
where $W^{\mu}$ is an anti-symmetric matrix with entries 
\begin{equation}
    W^{\mu}_{lm} = \frac{1}{\avg{\twist_n^\mu}} \begin{cases}
        F^{\twist^\mu|\nu_l \nu_m}_{\gamma \gamma}\!\left (\theta_{l}-\theta_{m};n\right ),    & m > l , \\
        \left(-1\right)^{\delta_{\nu_{l},\nu_{m}}+1} F^{\twist^\mu|\nu_l \nu_m}_{\gamma \gamma}\!\left (\theta_{l}-\theta_{m};n\right ),   & m < l.
\end{cases}
\end{equation}
For a different ordering of the replica indices $\nu_{i}$, we can apply the exchange axiom~\eqref{eq:U1FFAxiom1} to reorder them in the form $\nu_{1}\geq \nu_{2}\geq\ldots\geq \nu_{2k}$ and then use Eq.~\eqref{eq:Pfaffian2}.

In particular, for the four-particle `RRRR' or `LLLL' FF with all particles on the same replica, Eq.~\eqref{eq:Pfaffian2} takes the form
\begin{equation}
\begin{split}
F^{\twist^\mu}_{4,0}(\theta_{1},\theta_{2},\theta_{3},\theta_{4};n)&=\avg{\twist^{\mu}_{n}}^{-1} \left[F^{\twist^\mu}_{2,0}(\theta_{1}-\theta_{2};n)) F^{\twist^\mu}_{2,0}(\theta_{3}-\theta_{4};n)\right.\\
&\left. -F^{\twist^\mu}_{2,0}(\theta_{1}-\theta_{3};n) F^{\twist^\mu}_{2,0}(\theta_{2}-\theta_{4};n)+F^{\twist^\mu}_{2,0}(\theta_{1}-\theta_{4};n) F^{\twist^\mu}_{2,0}(\theta_{2}-\theta_{3};n)\right].
\end{split}\label{FFCBFTRRRR}
\end{equation}

\subsection{Solution for the four particle `RRLL' form factor}\label{sec:RRLLcompFF}

To obtain the first non-zero form factors that couple right- and left-moving particles, we have to move to the four-particle level, in which we find $F_{RRLL}^{\twist^\mu|\nu_1\nu_2\nu_3\nu_4}$ and all the possible permutations of `R' and `L'. As for the standard BPTFs, it is sufficient to determine only the `RRLL' form factor with all the particles on the first replica. In fact, using the exchange relation~\eqref{eq:U1FFAxiom1} we can directly get any other sequence of the particle species and, applying the cyclic permutation axiom~\eqref{eq:U1FFAxiom2}, we can find the FFs for particles living on different replicas. If we denote the form factor $F_{RRLL}^{\twist^\mu|1111}$ as $F^{\twist^\mu}_{2,2}$, then it reads 
\begin{equation}
\begin{split}
F^{\twist^\mu}_{2,2}(\theta_1, \theta_2,\theta_1', \theta_2';n) =&\, H^{\twist^\mu}_{2,2}\, Q^{\twist^\mu}_{2,2}(x_1,x_2,y_1,y_2;n) \frac{f_{RR}^\mu(\theta_{1}-\theta_{2};n)}{(x_{1}-\omega x_{2})(x_{2}-\omega x_{1})}\times\\
 &\hspace{1cm} \times \prod_{i=1}^{2} \prod_{j=1}^{2} f^{\mu}_{RL}(\theta_{i}-\theta'_{j};n) \frac{f_{LL}^\mu(\theta'_{1}-\theta'_{2};n)}{(y_{1}-\omega y_{2})(y_{2}-\omega y_{1})},
\label{FFParametrizationFlowCompositeRRLL}
\end{split}
\end{equation}
according to the ansatz~\eqref{FFParametrizationFlowComposite}.

Applying now the residue axiom to Eq.~\eqref{FFParametrizationFlowCompositeRRLL} we can derive recursive equations for the normalisation factors $H^{\twist^\mu}_{2,2}$ and the $Q^{\twist^\mu}_{2,2}$ functions in a similar way as we did for the standard BPTFs in Sec.~\ref{sec:RRLLtwistFF}. 
In App.~\ref{app:FlowBootstrapComp}, we find the solution for the functions $H^{\twist^\mu}_{2,2}$,
\begin{equation}\label{eq:HRRLLFlowComp}
H^{\twist^\mu}_{2,2}= -4\avg{\twist^{\mu}_{n}} \mathcal{N}_n^{4}\,,
\end{equation}
and $Q^{\twist^\mu}_{2,2}$,
\begin{equation}\label{eq:QRRLLFlowComp}
\begin{split}
&Q^{\twist^\mu}_{2,2}(x_1,x_2,y_1,y_2;n)=
 Q^{\twist^\mu}_{2,2}(x_1,x_2,y_1,y_2;n)^{(0)}+Q^{\twist^\mu}_{2,2}(x_1,x_2,y_1,y_2;n)^{(k)}\\
 =&\,\frac{\omega^2}{n^2}\frac{1+x_1 x_2 y_1 y_2}{x_1x_2y_1y_2}+\\
 &\hspace{2cm}-\frac{2 \omega^2 \cos \frac{\pi }{2 n}}{n^2} \frac{ \left(x_1 x_2 (y_1+y_2)^2+y_1 y_2 (x_1+x_2)^2-2 x_1 x_2 y_1 y_2 (\cos \left(\frac{\pi }{n}\right)+1)\right)}{(x_1x_2y_1y_2)(x_1+x_2) (y_1+y_2)}\,.
\end{split}
\end{equation}
As we explain in Appendix~\ref{app:FlowBootstrapComp}, the set of equations that allows to obtain $Q_{2,2}^{\twist^\mu}$ recursively is under-determined. This ambiguity in the solution can be fixed by requiring  that the form factor $F_{2, 2}^{\twist^\mu}$ reduces to 
the one of the disorder field $\mu$ in the single replica limit $n\to 1$. One can further check that the normalisation term $H_{2,2}^{\twist^\mu}$ also matches the one of $\mu$ in that limit. 
In Sec.~\ref{sec:entropy}, we use the $\Delta$-sum rule to provide an additional test of the validity of our solution.

\section{Roaming limit of twist field form factors}\label{sec:RoamingLimit}

In the previous sections, we computed the form factors of the twist fields along the tricritical Ising massless flow directly from the solution of their bootstrap equations. In this section, in order to provide a non-trivial check of our expressions, we present an alternative derivation based on the roaming limit of the sinh-Gordon model. After reviewing the general notions of this approach, we will then use it to recover the form factors in the massless flow as the limit of those in the sinh-Gordon theory.

Let us first briefly introduce the sinh-Gordon (ShG) model.
This theory is defined via the Euclidean action 
\begin{equation}\label{eq:ShGAction}
\mathcal{A}_\text{ShG} = \int \dd^{2}x\left\{ \frac{1}{2}\left[\partial\phi(x)\right]^{2}+\frac{\mu^{2}}{g^{2}}\nord{\cosh\!\left[g\phi(x)\right]}\right\}.
\end{equation}
This is the simplest interacting integrable relativistic QFT and has been the subject of an intense research activity since many decades, see, e.g.,~\cite{m-book,AFZ,FMS,km-93,ShG-Ising,ns-13, n-14,sgOverlaps, klm-20}.
The spectrum of the model consists of multi-particle states of a massive bosonic particle with the dispersion relation $E=m\cosh\theta, p=m\sinh\theta$, where $m$ is the particle mass.
The two-particle $S$-matrix is given by~\cite{AFZ}
\begin{equation}\label{eq:ShGSMatrix}
S_{\text{ShG}}\!\left (\theta\right ) = \frac{\tanh\!\frac{1}{2}\!\left(\theta-\ii\frac{\pi}{2} B \right)}{\tanh\!\frac{1}{2}\!\left(\theta+\ii\frac{\pi}{2} B \right)}\,,
\end{equation}
where $B$ is defined in terms of the coupling $g$ appearing in the action in Eq.~\eqref{eq:ShGAction} as 
\begin{equation}\label{eq:ShGB}
B(g)\,=\,\frac{2g^{2}}{8\pi+g^{2}}\,.
\end{equation}
For the ShG model, the form factors of various operators are known~\cite{ShGFFZamo,FMS,km-93}, including the standard and the $\ZZ_2$-composite BPTFs in the $n$-replica theory~\cite{Ola, cl-11, Z2IsingShg}.

It was observed in Ref.~\cite{RefZamo} that the $S$-matrix of the sinh-Gordon model can be analytically continued from the self-dual point $B=1$ to complex values 
\begin{equation}
B(\theta_0)=1+\ii\frac{2}{\pi}\theta_0\,,
\label{BTheta_0}
\end{equation}
and the resulting $S$-matrix is a new perfectly valid scattering theory, which has been called the staircase or roaming trajectories model.
Using Bethe ansatz, it was found that, as the real parameter $\theta_{0}$ increases,  the $c$-function shows a `staircase' of defined plateaux with values equal to the central charges of the $\Em_{p}$ unitary diagonal minimal models and, in the intervals between the plateaux, the flow was found to approximate the $A_p$ massless crossovers $\Em_{p+2}\rightarrow\Em_{p+1}$ generated by the perturbing field $\phi_{1,3}$ discussed in Sec.~\ref{sec:flow}.
Therefore, in the roaming limit $\theta_0\to\infty$, the staircase 
model describes a renormalisation group flow that passes by 
the successive minimal models $\mathcal{M}_p$. The final point of the flow is a massive Ising theory.

In another work~\cite{RefRoaming}, it was shown that the $c$-function defined by the $c$-theorem~\cite{ZamoRoaming,CardyCentralCharge} using a spectral series in terms of the form factors of the trace of the stress-energy tensor $\Theta$~\cite{FMS,ShGFFZamo} presents the same behaviour.
In addition, it was explicitly demonstrated that the FFs of the stress-energy tensor for the $A_p$ massless flows can be reconstructed from that of the ShG model.
Importantly, for this construction to work, the rapidities in the FFs have to be also shifted by $\pm k\theta_0/2$ with specific integers $k$.
A follow-up publication targeted specifically the $A_2$ tricritical-critical Ising flow~\eqref{eq:flow_action}, and showed that the form factors of the order and disorder operators along the flow can also be obtained via the roaming limit of the appropriate ShG FFs and, although not published, the correspondence holds for the $\varepsilon$ field of the flow as well.
As we have said, the staircase model also incorporates the massive Ising field theory, which is regarded in this context as a flow from the critical Ising fixed point to a massive one, and where the consecutive RG flows between the multicritical Ising CFTs terminate.
Accordingly, it was demonstrated in~\cite{ShG-Ising} that the FFs of the massive Ising theory can be obtained from the ShG FFs by merely taking the limit of Eq.~\eqref{BTheta_0}, i.e., scaling the rapidity variables within the FFs. 
In contrast, for other than the $A_2$ flow and massive Ising QFT, only the FFs of the field $\Theta$ were found to be reproduced by the roaming limiting procedure, and hence the validity of this approach is not a priori obvious and well understood.

Regarding the replicated staircase model, in Ref.~\cite{cl-11} the form factors of the standard BPTFs in the sinh-Gordon have been computed up to the four-particle order.
While the explicit roaming limit of these form factors was not carried out, they were used in the computation of the conformal dimension of the BPTFs applying the $\Delta$-sum rule \cite{delta_theorem}, which we discuss in more detail in Sec.~\ref{sec:entropy}.
In particular, it was found that the two-particle contribution correctly reproduces the first `step' of the staircase, from the critical Ising CFT to the massive theory, while the four-particle one gives the result for the massless flow $A_2$ from tricritical to critical Ising~\cite{cl-11}.
This result reveals that the roaming limit also holds for the branch point twist fields of the replicated theory.
In the following, we make a step further, by explicitly performing the roaming limit of the form factors of both the standard and the composite BPTFs up to the four-particle order, showing that they reduce to the exact expressions in the $A_2$ flow~\eqref{eq:flow_action} obtained via the bootstrap program in Secs.~\ref{sec:twistFormFactors} and \ref{sec:compositeFormFactors}.
Proving the correspondence in the first few non-trivial particle levels provides strong evidence that the roaming limit for standard and composite BPTFs is valid for any (composite) BPTF form factors in the $A_2$ massless flow.

In the ShG model, the $k$-particle form factors of the BPTFs can be parameterised in the usual fashion, that is~\cite{cl-11}, 
\begin{equation}
F^{\twist^\tau}_{k}(\underline{\theta},B;n)=H^{\twist^\tau}_{k}Q^{\twist^\tau}_{k}(\underline{x},B;n)\prod_{1\leqslant i<j\leqslant k} \frac{f_\text{ShG}^\tau(\theta_{i}-\theta_{j},B;n)}{(x_{i}-\omega x_{j})(x_{j}-\omega x_{i})}\,,
\label{FFParametrizationShG}
\end{equation}
where each particle is put on the first replica and the superscript $\tau=0,\mu$ denotes the standard or the composite BPTF respectively.
The minimal form factor for the standard twist field $f_\text{ShG}(\theta;n)$ is given by
\begin{equation}\label{eq:fminShG}
f_\text{ShG}(\theta,B;n) = \exp\!\left[- 2 \int_0^\infty dt \frac{\sinh\!\left(\frac{t B}{4}\right) \sinh\!\left(\frac{t (2-B)}{4}\right)}{t \sinh(nt) \cosh\!\left(\frac{t}{2}\right )} \cosh\!\left(\frac{\ii t \left ( \theta - \ii \pi n \right )}{\pi} \right ) \right ],
\end{equation}
while the one for the composite field is obtained by including an appropriate monodromy changing factor
\begin{equation}\label{eq:fminShGComp}
f^\mu_\text{ShG}\!\left (\theta,B;n \right ) = 2 \cosh\!\left( \frac{\theta}{2n}\right ) f_\text{ShG}\!\left (\theta,B;n\right ),
\end{equation}
analogously to what we have done in Eq.~\eqref{eq:fminRR^mu} for the massless flow.
The minimal FF $f_\text{ShG}$ in Eq.~\eqref{eq:fminShG} is normalised in such a way that $f_\text{ShG}(\pm \infty,B;n)=1$.
The roaming limit construction of the twist field FFs in the massless flow can then be formulated as
\begin{equation}
    \frac{1}{\langle\twist^\tau_{n,\text{flow}}\rangle}\, F^{\twist^\tau}_{r,l}\!\left (\underline{\theta},\underline{\theta'};n\right ) = \lim_{\theta_0\rightarrow\infty} \frac{1}{\langle\twist^\tau_{n,\text{ShG}}\rangle}\, F^{\twist^\tau}_{r+l}\!\left ( \underline{\theta}+\theta_0/2,\underline{\theta'}-\theta_0/2,B(\theta_0);n \right ),
\label{roaming}
\end{equation}
where we split the rapidities in the sinh-Gordon FF into $r$ right-moving ($\underline{\theta}$) and $l$ left-moving ($\underline{\theta'}$) ones, which we shift by $\theta_0$ and $-\theta_0$ respectively. The function $B\!\left ( \theta_0\right )$ is defined in Eq.~\eqref{BTheta_0}. In the rest of this section, we explicitly demonstrate the validity of the limit in Eq.~\eqref{roaming} up to the four-particle level. 

Let us first focus on the ShG minimal FFs. Based on Ref.~\cite{RefRoaming}, it can be shown that, in the roaming limit~\eqref{roaming}, the minimal form factor in Eq.~\eqref{eq:fminShG} reduces to 
\begin{equation}\label{eq:roamingFMin}
\begin{gathered}
f_\text{ShG}\!\left (\theta, B(\theta_0); n \right ) \longrightarrow e^{-\frac{\theta_0}{2 n}} \left [ - 2 \ii \sinh\!\left(\frac{\theta}{2 n}\right) \right ],\\
f_\text{ShG}\!\left (\ii \pi, B(\theta_0); n \right ) \longrightarrow e^{-\frac{\theta_0}{2 n}} \, 2 \sin\!\left(\frac{\pi}{2 n}\right) ,\\
f_\text{ShG}\!\left (\theta + \theta_0,B(\theta_0); n \right ) \longrightarrow \mathcal{N}_n \, f_{RL}\!\left (\theta; n \right ),
\end{gathered}
\end{equation}
where $f_{RL}$ is the minimal form factor~\eqref{FMinRL} in the massless flow and the normalisation constant $\mathcal{N}_n$ was found in Eq.~\eqref{f_RLNormalisation}. Similarly, for the composite twist field one has
\begin{equation}\label{eq:roamingFMinComp}
\begin{gathered}
f^\mu_\text{ShG}\!\left (\theta, B(\theta_0); n \right ) \longrightarrow e^{-\frac{\theta_0}{2n}} \left [ - 2 \ii \sinh\!\left(\frac{\theta}{n}\right) \right ],\\
f^\mu_\text{ShG}\!\left (\ii \pi, B(\theta_0); n \right ) \longrightarrow e^{-\frac{\theta_0}{2n}} \, 2 \sin\!\left(\frac{\pi}{n}\right) ,\\
f^\mu_\text{ShG}\!\left (\theta + \theta_0,B(\theta_0); n \right ) \longrightarrow e^{\frac{\theta_0}{2n}}\mathcal{N}_n \, f^\mu_{RL}\!\left (\theta; n \right ) = e^{\frac{\theta_0 + \theta}{2n}}\mathcal{N}_n\,  f_{RL}\!\left (\theta;n\right ),
\end{gathered}
\end{equation}
with $f^\mu_{RL}$ given by Eq.~\eqref{F_RL^Mu}.
Note that, according to the definition~\eqref{roaming}, only the above cases are the relevant limits for the minimal form factor. Some of them involve an exponential factor $e^{\pm\theta_0/(2n)}$ but we anticipate that similar factors originate from other terms of the entire FF and they eventually cancel. 

From Eqs.~\eqref{eq:roamingFMin} and \eqref{eq:roamingFMinComp}, it is easy to see that the roaming limit in the two-particle case correctly reproduces the form factors of the massless flow. This is clearer when the two-particle ShG FF is rewritten  as a function of the rapidity difference as
\begin{equation}
 F^{\twist^\tau}_{2}(\theta,B;n)=\frac{\langle\twist^\tau_{n, \text{ShG}}\rangle\sin\frac{\pi}{n}}{2n\sinh\frac{i\pi+\theta}{2n}\sinh\frac{i\pi-\theta}{2n}}\frac{f^\tau_\text{ShG}(\theta,B;n)}{f^\tau_\text{ShG}(i\pi,B;n)}\,.\label{eq:F2TwistFieldShG}
\end{equation}
In the limit~\eqref{roaming} this expression reproduces either Eq.~\eqref{eq:F2TwistField} (for the standard BPTF) or Eq.~\eqref{eq:F2Z2TwistField} (for the composite BPTF) for the `RR' and `LL' cases, while it vanishes in the `RL' case because of the diverging denominator.

\subsection{Roaming limit of the four-particle FFs of the standard BPTF}\label{sec:RoamingLimitStandard}

It is also not difficult to show that the four-particle `RRRR' and `LLLL' FFs are provided by the roaming limit~\eqref{roaming}.
If we consider the standard BPTFs, using as $Q^{\twist}_4$ the polynomial determined in~\cite{cl-11} and reviewed in App.~\ref{app:standardShG4ptBPTF}, we can proceed in the following way.
According to Eq.~\eqref{roaming}, the `RRRR' or `LLLL' form factors that only contain right or left movers are given by  
\begin{equation}\label{eq:RoamingRRRR}
\begin{split}
  &\frac{1}{\left < \twist_{n,\text{flow}} \right >} F_{4, 0\text{ tri}}^{\twist}\!\left ( \theta_1, \theta_2, \theta_3, \theta_4;n \right )=\\ 
&=\lim_{\theta_0 \to \infty} \frac{1}{\left < \twist_{n,\text{ShG}} \right >} F_{4\text{ ShG}}^{\twist}\!\left ( \theta_1+\theta_0/2, \theta_2+\theta_0/2, \theta_3+\theta_0/2, \theta_4+\theta_0/2 ,B(\theta_0);n\right ). \end{split}
\end{equation}
In this limit, the denominator of the ShG FF~\eqref{FFParametrizationShG} does not change but acquires the diverging factor $e^{6 \theta_0/n}$,
\begin{equation}
  \frac{1}{ \prod_{i<j} (x_j-\omega x_i)(x_i-\omega x_j)} 
  \longrightarrow e^{-6 \theta_0/n}\frac{1}{\prod_{i<j} (x_j-\omega x_i)(x_i-\omega x_j)}\,,
\label{ShG4ptDenomRRRRLimit}  
\end{equation}
whereas for the polynomial $Q_4^{\twist}$ we obtain the following lengthy expression
\begin{equation}
\begin{split}
&Q_4^{\twist}\!\left(x_1,x_2,x_3,x_4,B;n\right ) 
\longrightarrow \\
& e^{8 \theta_0/n} \times 2 e^{\frac{3 (\theta_1+\theta_2+\theta_3+\theta_4)}{n}} \left\{\cosh \left(\frac{\theta_1+\theta_2-\theta_3-\theta_4}{n}\right)+2 \cosh \left(\frac{\theta_1-\theta_2}{n}\right) \cosh \left(\frac{\theta_3-\theta_4}{n}\right) +\right.\\
&\left.+\left(1-2 \cos \frac{\pi }{n}\right)\left[\cosh \left(\frac{\theta_1-\theta_2}{n}\right)+\cosh \left(\frac{\theta_1-\theta_3}{n}\right)+\cosh \left(\frac{\theta_1-\theta_4}{n}\right)+\cosh \left(\frac{\theta_2-\theta_3}{n}\right)\right.\right.\\
&\left.\left.+\cosh \left(\frac{\theta_2-\theta_4}{n}\right)+\cosh \left(\frac{\theta_3-\theta_4}{n}\right)\right]+2-\cos \frac{\pi }{n}+\cos \frac{2 \pi }{n}+\cos \frac{3 \pi }{n}\right\}, 
\end{split}
\label{ShG4ptQ4RRRRLimit}
\end{equation}
which diverges exponentially as $e^{8 \theta_0/n}$ when $\theta_0\to \infty$. In addition, taking into account the limit of the minimal form factor reported in Eq.~\eqref{eq:roamingFMin}, we have
\begin{equation}
  \prod_{i<j} f_\text{ShG}\!(\theta_i - \theta_j,B;n) 
  \longrightarrow e^{-3 \theta_0/n} \left [ - 2^6 \prod_{i<j} \sinh\!\left( \frac{\theta_i-\theta_j}{2 n} \right ) \right ] ,
\label{ShG4ptfminRRRRLimit}
\end{equation}
and for the normalisation factor $H_n^\twist$, we find
\begin{equation}
  H_n^{\twist} = \left ( \frac{2 \sin(\pi/n) \omega^2}{n f^\text{ShG}\!(i \pi, B;n)} \right)^2 \omega^2 \longrightarrow e^{\theta_0/n} \left ( \frac{\sin(\pi/n) \omega^2}{n \sin(\pi/2n)} \right)^2 \omega^2 = e^{\theta_0/n} \left ( \frac{4\, e^{\ii \frac{6 \pi}{n}}}{n^2} \cos^2\!\left( \frac{\pi}{2n}\right ) \right ).
\label{ShG4ptNormRRRRLimit}
\end{equation}
Counting the divergent factors $e^{\theta_0/n}$ in the final expressions of Eqs.~\eqref{ShG4ptDenomRRRRLimit}, \eqref{ShG4ptQ4RRRRLimit}, \eqref{ShG4ptfminRRRRLimit} and~\eqref{ShG4ptNormRRRRLimit}, we can conlude that the `RRRR' (`LLLL') roaming limit form factor of the ShG twist field is finite.
In fact, putting all the above results together it is straightforward to check that the limit~\eqref{eq:RoamingRRRR} works and Eq.~\eqref{FFBPTFRRRR} is exactly reproduced.

Turning to the case of the `RRLL' form factor, we have to consider
\begin{equation}\label{eq:RoamingRRLL}
\begin{split}
  &\frac{1}{\left < \twist_{n,\text{flow}} \right >} F_{2, 2\text{ tri}}^{\twist}\!\left ( \theta_1, \theta_2, \theta_1', \theta_2';n \right )=\\ 
&=\lim_{\theta_0 \to \infty} \frac{1}{\left < \twist_{n,\text{ShG}} \right >} F_{4\text{ ShG}}^{\twist}\!\left ( \theta_1+\theta_0/2, \theta_2+\theta_0/2, \theta_1'-\theta_0/2, \theta_2'-\theta_0/2 ,B(\theta_0);n\right ).
\end{split}\end{equation}
For the denominator, the limit gives
\begin{equation}
  \frac{1}{ \prod_{i<j} (x_j-\omega x_i)(x_i-\omega x_j)}
  \longrightarrow e^{-4\theta_0/n}\frac{ y_1^2y_2^2}{x_1^4 x_2^4\omega^4 (x_1-\omega x_2)(x_2-\omega x_1)(y_1-\omega y_2)(y_2-\omega y_1)},
\label{ShG4ptDenomRRLLLimit}  
\end{equation}
whereas for the polynomial $Q_4^{\twist}$ we obtain the following expression
\begin{equation}
\begin{split}
&Q_4^{\twist}\!\left(x_1,x_2,x_3,x_4,B;n\right )
\longrightarrow \\
& e^{4 \theta_0/n} \times e^{\frac{4 (\theta_1+\theta_2)}{n}} \left(-\frac{\left(e^{\theta_1/n}+e^{\theta_2/n}\right) \left(e^{\theta_1'/n}+e^{\theta_2'/n}\right) e^{\frac{2 (\theta_1'+\theta_2')+i \pi }{2 n}}}{1+\omega}+e^{\frac{\theta_1+\theta_2+\theta_1'+\theta_2'}{n}}+e^{\frac{2 (\theta_1'+\theta_2')}{n}}\right ) ,
\end{split}
\label{ShG4ptQ4RRLLLimit}
\end{equation}
which we can rewrite as
\begin{equation}Q_4^{\twist}\!\left(x_1,x_2,x_3,x_4,B;n\right )
\longrightarrow e^{4 \theta_0/n} \times x_1^4x_2^4 \left(-\frac{\left(x_1+x_2\right) \left(y_1+y_2\right) e^{\frac{i \pi }{2 n}}}{(1+\omega)y_1^2y_2^2}+\frac{x_1x_2}{y_1y_2}+\frac{1}{y_1^2y_2^2}\right).
\label{ShG4ptQ4RRLLLimitB}
\end{equation}
For the product of the minimal FFs, we find
\begin{equation}
   \prod_{i<j} f_\text{ShG}\!(\theta_i - \theta_j,B;n)
   \longrightarrow e^{- \theta_0/n} \left [ - 2^2 \mathcal{N}_n^4 \sinh\!\left( \frac{\theta_1-\theta_2}{2 n} \right )\sinh\!\left( \frac{\theta_1'-\theta_2'}{2 n} \right )\prod_{i<j}f_{RL}(\theta_i-\theta_j') \right ].
\label{ShG4ptfminRRLLLimit}
\end{equation}
The limit of the normalisation factor $H_n^\twist$ gives the same result as in Eq.~\eqref{ShG4ptNormRRRRLimit}. Combining~\eqref{ShG4ptDenomRRLLLimit}, \eqref{ShG4ptQ4RRLLLimitB}, \eqref{ShG4ptfminRRLLLimit}, and the normalisation~\eqref{ShG4ptNormRRRRLimit}, it is immediate to see that the divergent exponential factors $e^{\theta_0}$ mutually cancel and that the roaming limit yields Eq.~\eqref{FFRRLLFlow}, confirming the validity of Eq.~\eqref{eq:RoamingRRLL}.

\subsection{Roaming limit of the four-particle FFs of the composite BPTF}\label{sec:RoamingLimitComposite}

Unlike the four-particle form factor of the standard twist field, the one of the composite twist field was not previously known in the sinh-Gordon theory.
In App.~\ref{app:compositeShGbootstrap}, we compute this form factor by constructing and solving the bootstrap equations, starting from the usual ansatz in Eq.~\eqref{FFParametrizationShG}.
Notice that, as we discuss in App.~\ref{app:compositeShGbootstrap}, now the function $Q^{\twist^\mu}_k$ is not a polynomial but a rational function.
At the four-particle level, the explicit expressions of the normalisation $H^{\twist^\mu}_4$ and of the polynomial $Q^{\twist^\mu}_4$ are reported in App.~\ref{app:compositeShGbootstrap} in Eq.~\eqref{eq:H4ShGComp} and in Eqs.~\eqref{eq:Q4ShGComp}, \eqref{eq:coefficientsQ4ShGComp}, respectively.

Let us first consider the form factors `RRRR' and `LLLL', containing only either right- or left-moving particles. Following Eq.~\eqref{roaming}, we see that we need to compute the limit of $F^{\twist^\mu}_4(\underline{\theta} + \theta_0/2, B(\theta_0); n)$. As in Sec.~\ref{sec:RoamingLimitStandard} for the standard twist field, we study separately this limit for the different terms that constitute the composite ShG form factor in Eq.~\eqref{FFParametrizationShG}. Applying the limit of the minimal composite form factor reported in Eq.~\eqref{eq:roamingFMinComp}, we have 
\begin{equation}\begin{split}
    \prod_{1\leqslant i<j\leqslant 4} f^\mu_\text{ShG}\!\left (\theta_i - \theta_j, B(\theta_0); n \right ) \longrightarrow &\, e^{-3 \theta_0/n} \left [- 2^6 \prod_{1\leqslant i<j\leqslant 4}  \sinh\!\left(\frac{\theta_i-\theta_j}{n}\right) \right ]\\
    &= e^{-3 \theta_0/n} \left [- \prod_{1\leqslant i<j\leqslant 4}  \frac{\left ( x_i + x_j \right ) \left ( x_i -x_j\right ) }{x_i x_j} \right ],
\end{split}\end{equation}
where we have rewritten it in terms of $x_i = e^{\theta_i / n}$. It is convenient to take the limit of the function $Q^{\twist^\mu}_4$, reported in Eqs.~\eqref{eq:Q4ShGComp}, \eqref{eq:coefficientsQ4ShGComp}, and of the denominator of the ansatz~\eqref{FFParametrizationShG} together with the one of the minimal form factor. We find that this limit reproduces the form factor in Eq.~\eqref{FFCBFTRRRR} up to a normalisation with an exponential $e^{-\theta_0/n}$
\begin{equation}\label{eq:roamingShGCompAllRRRR}\begin{split}
    &Q^{\twist^\mu}_4\!\left (x_1,x_2, x_3, x_4, B\!\left ( \theta_0 \right ); n \right )\prod_{1\leqslant i<j\leqslant 4}\frac{f^\mu_\text{ShG}\!\left (\theta_i - \theta_j, B(\theta_0); n \right )}{\left ( x_{i}-\omega x_{j} \right ) \left (x_{j}-\omega x_{i}\right)} \longrightarrow \\
    &\longrightarrow \frac{e^{-\theta_0/n}}{\omega^6 \left ( 1 + \omega \right )^2} \bigg [ - \left(\frac{x_1}{x_1 -  \omega x_4 }+\frac{x_4}{\omega x_1 -x_4}\right) \left(\frac{x_2}{x_2-\omega x_3 }+\frac{x_3}{ \omega x_2 -x_3}\right)  + \\
    &\hspace{.8cm} + \left(\frac{x_1}{x_1- \omega x_3}+\frac{x_3}{ \omega x_1-x_3}\right) \left(\frac{x_2}{x_2- \omega x_4}+\frac{x_4}{ \omega x_2- x_4}\right)  + \\
    &\hspace{.8cm} - \left(\frac{x_1}{x_1- \omega x_2}+ \frac{x_2}{ \omega x_1-x_2}\right) \left(\frac{x_3}{x_3- \omega x_4}+\frac{x_4}{x_3- \omega x_4}\right) \bigg ] = \frac{e^{-\theta_0/n}\, n^2}{\omega^6 \left ( 1 + \omega \right )^2} F^{\twist^\mu}_{4,0}\!\left ( \theta_1, \theta_2, \theta_3 \theta_4 \right ) .
\end{split}\end{equation}
Finally, we see that the normalisation term $H_n^{\twist^\mu}$, whose explicit expression is given in Eq.~\eqref{eq:H4ShGComp}, becomes
\begin{equation}\label{eq:roamingH4Comp}
    \frac{H^{\twist^\mu}_4\!\left ( B(\theta_0)\right ) }{\avg{\twist^\mu_\text{ShG}}} = \left ( \frac{2 \left ( 1 + \omega\right ) \sin\!\frac{\pi}{n} }{n\, f^\mu_\text{ShG}\!\left( \ii \pi;n\right )} \right )^2 \omega^{6} \longrightarrow e^{\theta_0/n}\, \frac{\omega^6 \left ( 1 + \omega\right )^2}{n^2}\,,
\end{equation}
cancelling precisely the multiplicative factor in Eq.~\eqref{eq:roamingShGCompAllRRRR}, such that the roaming limit correctly reproduces the `RRRR' (or `LLLL') form factor in Eq.~\eqref{FFCBFTRRRR}, as expected.

Considering now the `RRLL' form factor,  we can see that it can be obtained from the limit of Eq.~\eqref{roaming} in the particular case
\begin{equation}
    \frac{F^{\twist^\mu}_{2,2\text{ flow}}\!\left ( \theta_1, \theta_2, \theta_1', \theta_2'; n\right )}{\avg{\twist^\mu_\text{flow}}} = \lim_{\theta_0 \to \infty} \frac{F^{\twist^\mu}_{4\text{ ShG}}\!\left (\theta_1 + \theta_0/2, \theta_2 + \theta_0/2, \theta_1' - \theta_0/2, \theta_2' - \theta_0/2, B(\theta_0); n \right )}{\avg{\twist^\mu_\text{ShG}}}\,.
\end{equation}
The joint limit of the denominator of the ansatz~\eqref{FFParametrizationShG} and of the polynomial $Q_{4}^{\twist^\mu}$ reported in Eqs.~\eqref{eq:Q4ShGComp}, \eqref{eq:coefficientsQ4ShGComp} gives
\begin{equation}\begin{split}
    &\frac{Q_{4\text{ ShG}}^{\twist^\mu}\!\left ( x_1, x_2, x_3, x_4 \right )}{\prod_{1\leqslant i<j\leqslant 4} \left ( x_{i}-\omega x_{j} \right ) \left (x_{j}-\omega x_{i}\right)} \longrightarrow \\
    &\longrightarrow e^{-2\theta_0/n} \frac{n^2}{\omega^6 \left ( 1+ \omega \right )^2}
     \frac{Q_{2,2\text{ flow}}^{\twist^\mu}\!\left ( x_1, x_2, y_1, y_2 \right )}{\left ( x_1 - \omega x_2 \right ) \left (x_2 - \omega x_1 \right) \left ( y_1 - \omega y_2 \right ) \left (y_2 - \omega y_1 \right)},
\end{split}\end{equation}
where $Q_{2,2\text{ flow}}^{\twist^\mu}$ is the polynomial in the massless flow of Eq.~\eqref{eq:QRRLLFlowComp}. The normalisation $H_{4}^{\twist^\mu}$ has the same limit as in Eq.~\eqref{eq:roamingH4Comp} and, recalling the limit of the composite minimal form factors in Eq.~\eqref{eq:roamingFMinComp}, we have
\begin{equation}\begin{split}
    &f^\mu_\text{ShG}\!\left (\theta_1 - \theta_2, B(\theta_0); n \right ) \prod_{i,j = 1,2} f^\mu_\text{ShG}\!\left (\theta_i - \theta_j', B(\theta_0); n \right ) f^\mu_\text{ShG}\!\left (\theta_1' - \theta_2', B(\theta_0); n \right ) \longrightarrow\\
    & \longrightarrow e^{\theta_0/n} \, \left [ - 4\,  \mathcal{N}_n^4\,  \sinh\!\left(\frac{\theta_1 - \theta_2}{n}\right) \prod_{i,j = 1,2} f^\mu_{RL}\!\left (\theta; n \right )\, \sinh\!\left(\frac{\theta_1' - \theta_2'}{n}\right) \right ].
\end{split}\end{equation}
Putting all together, we find that the limit of the `RRLL' form factor is again finite as expected,
\begin{equation}\begin{split}
     &\lim_{\theta_0 \to \infty} \frac{F^{\twist^\mu}_{4\text{ ShG}}\!\left (\theta_1 + \theta_0/2, \theta_2 + \theta_0/2, \theta_1' - \theta_0/2, \theta_2' - \theta_0/2, B(\theta_0); n \right )}{\avg{\twist^\mu_\text{ShG}}}\\
     =&\, - 4\,  \mathcal{N}_n^4\, Q_{2,2\text{ flow}}^{\twist^\mu}\!\left ( x_1, x_2, y_1, y_2 \right ) \frac{\sinh\!\frac{\theta_1 - \theta_2}{n}}{\left ( x_1 - \omega x_2 \right ) \left (x_2 - \omega x_1 \right) } \prod_{i,j = 1,2} f^\mu_{RL}\!\left (\theta; n \right ) \frac{\sinh\!\frac{\theta_1' - \theta_2'}{n} }{\left ( y_1 - \omega y_2 \right ) \left (y_2 - \omega y_1 \right)}\\
     =& \,\frac{F^{\twist^\mu}_{2,2\text{ flow}}\!\left ( \theta_1, \theta_2, \theta_1', \theta_2'; n\right )}{\avg{\twist^\mu_\text{flow}}}\,,
\end{split}\end{equation}
confirming the validity of the roaming limit also for the composite twist field $\twist_n^\mu$. 

\section{Standard and symmetry resolved entropies for the massless flow}\label{sec:entropy}

In this section, we use the form factors computed in the previous sections to study the behaviour of the correlation functions of the standard and composite twist fields. 
After calculating the running dimension of the field along the renormalisation flow, we investigate the entanglement entropy, comparing it with expected results.

\subsection{Running dimension from the \texorpdfstring{$\Delta$}{Delta}-sum rule}

As we discussed in Sec.~\ref{sec:flow}, the model under examination interpolates between the tricritical Ising CFT $\Em_{4}$ in the UV and the Ising CFT $\Em_{3}$ in the IR, providing the simplest example of a massless renormalisation flow between two $A$-series diagonal minimal models~\cite{AnZamo123, AnZamo2, AnZamo3}.
In both fixed points, the properties of  the standard twist field $\twist_n$ and the $\mathbb{Z}_2$ composite one $\twist_n^\mu$ are known from conformal invariance~\cite{cc-04}, as we reviewed in Sec.~\ref{sec:twistfield}. 
In particular, the conformal dimension of the standard twist field is given by Eq.~\eqref{eq:twistDim} while the dimension of the composite one is in Eqs.~\eqref{eq:resolvedTwistDim}, \eqref{eq:TricrResolvedTwistDim} for the  fixed points of interest.

The knowledge of the exact conformal dimensions of the fields in the IR and the UV fixed points of the massless flow provides a non-trivial check of the correctness of the form factors via the $\Delta$-sum rule~\cite{delta_theorem}.
Let us start by considering the twist field $\twist_n$.
Along a renormalisation group flow, the difference of conformal dimensions of the field $\twist_n$ in the IR and in the UV is given by an integral of the two-point function between $\twist_n$ and the trace of the stress-energy tensor $\Theta$~\cite{delta_theorem}
\begin{equation}\label{eq:deltaThUV}
  h_\twist^\text{UV} - h_\twist^\text{IR} = - \frac{1}{2\avg{\twist_n}} \int_{0}^{+\infty} \dd t\, t\, \big \langle \Theta\!\left (0 \right ) \twist_n\!\left ( t \right )  \big \rangle \, .
\end{equation}
In order to compute the $\Delta$-sum rule~\eqref{eq:deltaThUV}, we expand the two-point function in form factors, analogously to what we did in Eq.~\eqref{FFSeriesCF2} for the correlator of twist fields.
The resulting spectral expansion of Eq.~\eqref{eq:deltaThUV} involves the form factors of the twist field and the ones of the trace $\Theta$, which in the case of the (non-replicated) massless tricritical flow have been  obtained in~\cite{mds-95}.
In particular, in a massless model, all the form factors of $\Theta$ containing either only left- (`L') or right-movers (`R') identically vanish.
When considering the replicated theory, we have to take the sum of $\Theta$ in each the copy. Therefore, the only non-vanishing form factors are the ones with identical replica indices $F_{r,l}^{\Theta | 11\ldots1} = F_{r,l}^{\Theta}$. After integrating out the distance $t$ in the spectral expansion of the $\Delta$-sum rule~\eqref{eq:deltaThUV}, we finally find~\cite{delta_theorem, Ola}
\begin{equation}\begin{split}\label{eq:ExpansionDeltaThUV}
  h^\text{UV} - h^\text{IR} &= -\frac{n}{2\avg{\twist_n}} \sum_{r, l \text{ even}} \int_{-\infty}^{+\infty} \frac{\prod_{i=1}^r \dd \theta_i \prod_{j=1}^l \dd \theta_j'}{r!\, l!\left ( 2\pi\right)^{r+l}} \, \frac{1}{2\, E^2}\,  F^{\Theta}_{r,l}\!\left( \theta_1,\ldots \right) \left ( F^{\twist|11\ldots1}_{r,l}\!\left( \theta_1,\ldots \right) \right )^*\!,
\end{split}\end{equation}
where $E$ is the energy (reported in Eq.~\eqref{eq:MasslessEnergyMomentum} for a massless model).

The leading non-trivial form factor of $\Theta$ is the four-particle `RRLL' one, coupling two right- and two left-movers~\cite{mds-95}
\begin{equation}\label{eq:FRRLLTheta}
    F_{2,2}^{\Theta}\!\left ( \theta_1, \theta_2; \theta_1', \theta_2' \right ) = \frac{4\pi  M^2}{\gamma^2} \sinh\!\frac{\theta_1 - \theta_2}{2} \prod_{i,j = 1,2} f_\text{RL}\!\left ( \theta_i - \theta_j' \right )\, \sinh\!\frac{\theta_1' - \theta_2'}{2}\,,
\end{equation}
where $\gamma$ is Euler-Mascheroni's constant and $f_\text{RL}(\theta) = f_\text{RL}(\theta; n = 1)$ is the minimal form factor in Eq.~\eqref{FMinRL} for a single replica $n=1$. Since all form factors have an even number of left- and right-moving particles, Eq.~\eqref{eq:FRRLLTheta} is the only contribution at the four-particle level~\cite{mds-95}. We can then consider the approximation
\begin{equation}\label{eq:4partDeltaTh}\begin{split}
  h^\text{UV} - h^\text{IR} &\approx -\frac{n}{2\avg{\twist_n}} \int_{-\infty}^{+\infty} \frac{\dd \theta_1 \dd \theta_2 \dd \theta_1' \dd \theta_2'}{2\times2 \left ( 2\pi\right)^4}\, \frac{1}{2\, E^2} \, F^{\Theta}_{2,2}\!\left( \theta_1,\theta_2,\theta_1',\theta_2'\right) \left ( F^{\twist|1111}_{2,2}\left( \theta_1,\theta_2,\theta_1',\theta_2';n\right) \right )^*\!,
\end{split}\end{equation}
where $F^{\Theta}_{2,2}$ is given in Eq.~\eqref{eq:FRRLLTheta} and $F^{\twist|1111}_{2,2}$ is the twist field FF that we obtained in Eqs.~\eqref{FFRRLLFlow}, \eqref{FFRRLLFlow2}.
Analogous expressions hold for the $\Delta$-sum rule of the composite twist field $\twist_n^\mu$, replacing $F^{\twist|1111}_{2, 2}$ with the form factor $F^{\twist^\mu|1111}_{2,2}$ of the composite field reported in Eqs.~\eqref{FFParametrizationFlowCompositeRRLL}, \eqref{eq:HRRLLFlowComp}, \eqref{eq:QRRLLFlowComp}.

\begin{table}[t]
  \centering
  \begin{tabular}{ccccc}
    \toprule
    &    \multicolumn{2}{c}{BPTF $h_{\twist}^\text{UV} - h_{\twist}^\text{IR}$}&   \multicolumn{2}{c}{composite TF $h_{\twist^\mu}^\text{UV} - h_{\twist^\mu}^\text{IR}$} \\ 
        \cmidrule(lr){2-3}           \cmidrule(lr){4-5}
    $n$&  CFT&        $\Delta$-sum rule&  CFT&          $\Delta$-sum rule \\
    \midrule
    2&   $0.0125$&      $0.0125$&      $0$&          $0.0002$ \\
    3&   $0.0\overline{2}$& $0.0223$&      $0.013\overline{8}$&  $0.0138$ \\
    4&   $0.03125$&     $0.0313$&      $0.025$&        $0.0249$ \\
    \bottomrule
  \end{tabular}
  \caption{Comparison of the difference of the conformal dimensions in the UV and IR fixed points $h^\text{UV} - h^\text{IR}$ with the results of the $\Delta$-sum rule, for both the standard twist field $\twist_n$ and the composite one $\twist^\mu_n$. The `CFT' columns collect the exact result fixed by conformal invariance in Eqs.~\eqref{eq:twistDim}, \eqref{eq:TricrResolvedTwistDim}, while `$\Delta$-sum rule' is the result of the $\Delta$-sum rule truncated at four-particle order, reported in Eq.~\eqref{eq:4partDeltaTh}. The column `$n$' indicates the number of replicas. We can see that at the four-particle order we already find good agreement for all the number of replicas considered. }
  \label{tab:deltaUV}
\end{table}

In Table~\ref{tab:deltaUV}, we compare the exact difference of conformal dimensions of both the standard and the composite twist fields with the result of the $\Delta$-sum rule at the four-particle order~\eqref{eq:4partDeltaTh} for $n = 2, 3, 4$ replicas.
The integral in Eq.~\eqref{eq:4partDeltaTh} has been computed numerically using the Divonne routine of the library \textsc{Cuba}~\cite{cuba} for the software \emph{Mathematica}, using a cut-off $\theta_j\in[-60, 60]$ for the rapidities.
Already at the four-particle order we find a good agreement between the exact CFT result and the $\Delta$-sum rule, confirming the correctness of the form factors computed in Secs.~\ref{sec:twistFormFactors} and \ref{sec:compositeFormFactors} and the relatively small weight carried by the higher order FFs.
This is consistent with  Ref.~\cite{cl-11}, where it was found that, for the staircase model (reviewed in Sec.~\ref{sec:RoamingLimit}), the four-particle contribution obtained in the roaming limit reproduces the difference in conformal dimensions of the standard twist field along the massless flow~\eqref{eq:flow_action}.

The $\Delta$-sum rule~\eqref{eq:deltaThUV} can be modified to give a \textit{running} dimension of the (composite) twist fields along the flow~\cite{CastroFring1, cl-11}
\begin{equation}\label{eq:deltaThFlow}
  h\!\left ( \ell \right ) - h^\text{IR} = - \frac{1}{2\avg{\twist_n}} \int_{\ell}^{\infty} \dd t\, t\, \big \langle \Theta\! \left(0 \right ) \twist_n\!\left (t \right ) \big \rangle\, ,
\end{equation}
where now the integral over the distance $t$ starts from a finite length $\ell$.
As we did before in Eq.~\eqref{eq:ExpansionDeltaThUV}, we expand the two-point function in form factors and we integrate over the distance $t$, obtaining~\cite{CastroFring1, cl-11}
\begin{equation}\begin{split}\label{eq:ExpansionDeltaThFlow}
  h\!\left ( \ell \right ) - h^\text{IR} &= -\frac{n}{2\avg{\twist_n}} \sum_{r, l\text{ even}} \int_{-\infty}^{+\infty} \frac{\prod_{i=1}^r \dd \theta_i \prod_{j=1}^l \dd \theta_j'}{r!\, l!\left ( 2\pi\right)^{r+l}} \, \frac{\left ( 1+\ell\, E \right )e^{-\ell\,E}}{2\, E^2} \times \\
  &\hspace{5cm} \times F^{\Theta}_{r,l}\!\left( \theta_1,\ldots \right) \left ( F^{\twist|11\ldots1}_{r,l}\!\left( \theta_1,\ldots \right) \right )^*\!,
\end{split}\end{equation}
where again, in the massless flow, the leading contribution is given by the `RRLL' form factors,
\begin{equation}\label{eq:4partDeltaThFlow}\begin{split}
  h\!\left ( \ell \right ) - h^\text{IR} &\approx -\frac{n}{2\avg{\twist_n}} \int_{-\infty}^{+\infty} \frac{\dd \theta_1 \dd \theta_2 \dd \theta_1' \dd \theta_2'}{2\times2 \left ( 2\pi\right)^4}\, \frac{\left ( 1+\ell\, E \right )e^{-\ell\,E}}{2\, E^2} \times\\
  &\hspace{3cm}\times F^{\Theta}_{2,2}\!\left( \theta_1,\theta_2,\theta_1',\theta_2'\right) \left ( F^{\twist|1111}_{2,2}\left( \theta_1,\theta_2,\theta_1',\theta_2';n\right) \right )^*\!.
\end{split}\end{equation}
A running $\Delta$-theorem~\eqref{eq:deltaThFlow} can also be formulated for the composite twist field by considering the appropriate form factor $F^{\twist^\mu|1111}_{2,2}$ of that operator obtained in Eqs.~\eqref{FFParametrizationFlowCompositeRRLL}-\eqref{eq:QRRLLFlowComp}.

\begin{figure}[t]
    \centering
    \begin{subfigure}[b]{0.48\textwidth}
        \centering
        \includegraphics[width=\textwidth]{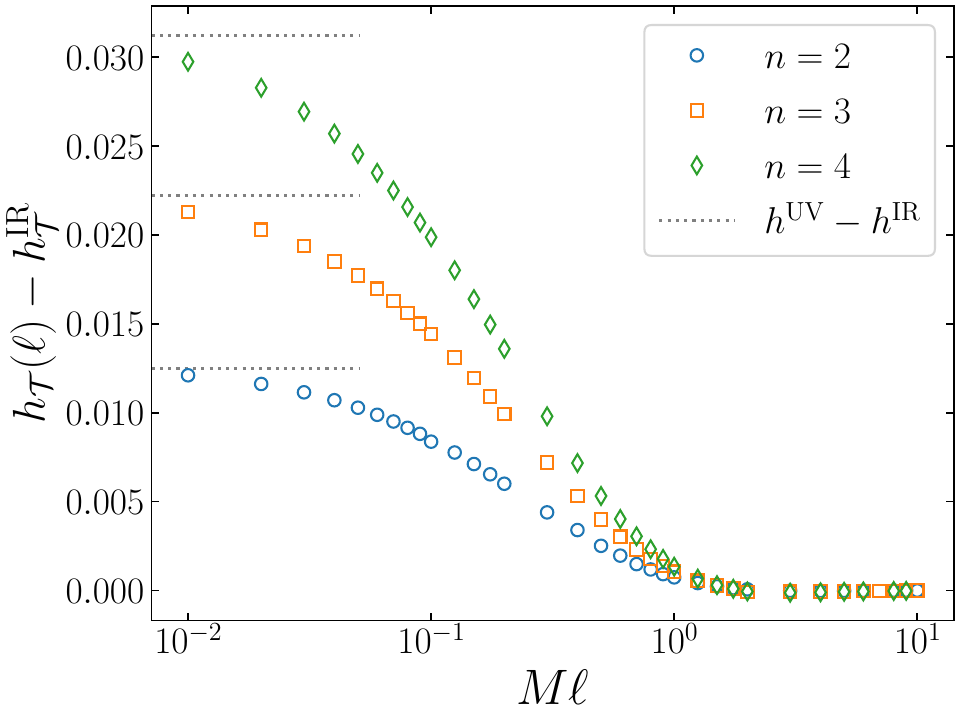}
        \caption{Branch point twist field $\twist_n$}
        \label{fig:deltaTwist}
    \end{subfigure}
    \hfill
    \begin{subfigure}[b]{0.48\textwidth}
        \centering
        \includegraphics[width=\textwidth]{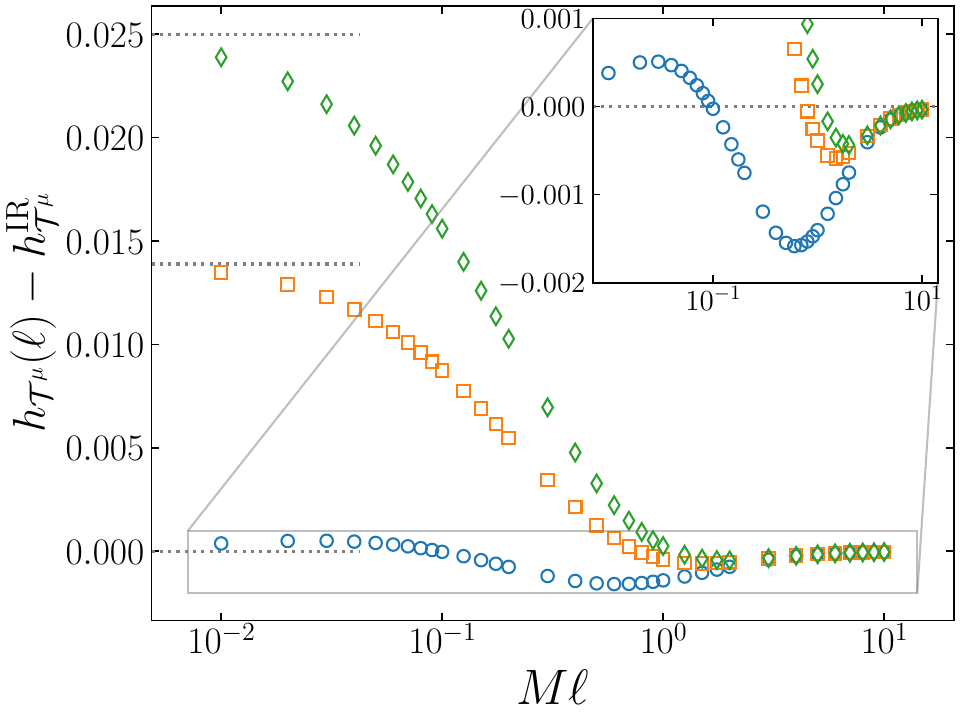}
        \caption{Composite twist field $\twist_n^\mu$}
        \label{fig:deltaComp}
    \end{subfigure}
    \caption{Semi-logarithmic plot of the running conformal dimension $h_{\mathcal{T}^\tau}$ obtained using the $\Delta$-sum rule at four-particle order reported in Eq.~\eqref{eq:4partDeltaThFlow}. In the plot on the left, we show the result for the standard twist field $\twist_n$, while on the right for the composite one $\twist_n^\mu$. The dotted gray lines indicate the exact difference between the UV and IR conformal dimension obtained from Eq.~\eqref{eq:twistDim} in the plot on the left and from Eqs.~\eqref{eq:TricrResolvedTwistDim} in the plot on the right. As expected, for small distances $\ell$, the running conformal dimension approaches the exact UV results, as also reported in Table~\ref{tab:deltaUV}. On the left, we see that the one of the standard twist field decreases monotonically, consistently with its behaviour as an entropic $c$-function. On the right, in the inset we zoom on the running dimension of the composite twist field (for $n = 2$ we have $h_{\twist^\mu}^\text{UV} = h_{\twist^\mu}^\text{IR}$). The running dimension of the composite field is not monotonic along the flow.}
    \label{fig:delta}
\end{figure}

In Ref.~\cite{cdl-12}, it was argued that the running dimension $h(\ell)$ of the branch point twist field $\twist_n$ provides an entropic $c$-function which is monotonically decreasing along the flow.
In Fig.~\ref{fig:deltaTwist} we report the result of the numerical integration of the running Delta theorem in Eq.~\eqref{eq:4partDeltaThFlow} for the standard branch point twist field $\twist_n$ taking $n = 2, 3, 4$ replicas.
We observe that, already at the four-particle order, the running dimension monotonically decreases with $\ell$ for all the number of replicas considered. In Fig.~\ref{fig:deltaComp}, we plot the running dimension of the composite twist field $\twist_n^\mu$ at four-particle order.
In particular for $n = 2$ replicas, the dimensions of the twist fields and of the charge operator conspire to give the same ultraviolet and infrared conformal dimensions for the composite twist field. Remarkably, we see that along the flow the running dimension varies and is not monotonic in $\ell$, differently from the standard twist field. 
In the inset, we zoom in the region of small running dimension which shows that also
for larger number of replicas $n$ the behaviour is non-monotonic.

\subsection{Cumulant expansion of the entanglement entropy}

As a main result of this paper, in this section we discuss the form factor expansion of the entanglement entropy along the massless renormalisation group flow. 
As we will show, the formal expressions require a suitable regularisation, after which the form factors containing particles with the same chirality reproduce the logarithmic entanglement entropy of the infrared Ising CFT, while those that include particles of different chirality provide the corrections along the flow.

Instead of studying the correlator of the twist field, we find more convenient to directly apply its form factor expansion to the R\'enyi entanglement entropies defined in Eq.~\eqref{eq:renyi_ent}.
Plugging in Eq.~\eqref{eq:mom_twist_fields} the spectral series~\eqref{FFSeriesCFFF} of the twist field correlator and expanding the logarithm for the R\'enyi entropy order by order in the number of particles, we obtain the cumulant expansion~\cite{bk-03, cd-09b}
\begin{equation}\label{eq:cumulantExpansion}
  S_n\!\left ( M \ell \right ) = \frac{1}{1-n}\log\big \langle \twist_n\!\left(0\right ) \antitwist_n\!\left (\ell\right ) \big \rangle \approx \frac{2}{1-n}\log\avg{\twist_n} + \frac{1}{1-n}\sum_{r, l \text{ even}} c_{r, l}^\twist\!\left ( M\ell; n \right ),
\end{equation}
where, in analogy with Ref.~\cite{bk-03}, we have introduced the cumulants
\begin{equation}\label{eq:cumulant}
  c^\twist_{r, l}\!\left ( M\ell; n \right ) = \sum_{j, j'}  \int_{-\infty}^{+\infty} \frac{\prod_{i=1}^r \dd \theta_i \prod_{k=1}^l \dd \theta_k'}{r!\, l!\left ( 2\pi\right)^{r+l}} \,\, f_{r,l}^{\twist | j_1\ldots j_1' \ldots}\!\left (\theta_1, \ldots, \theta_1', \ldots; n\right ) e^{- \frac{M\ell}{2}\left (\sum_i e^{\theta_i} + \sum_k e^{-\theta_k'} \right )}.
\end{equation}
In the integral in Eq.~\eqref{eq:cumulant}, $f_{r,l}^{\twist | j_1\ldots}$ denotes the connected part of the square of the form factor $\abs{F_{r,l}^{\twist | j_1\ldots}}^2$, obtained by subtracting all the possible clusterisations for large rapidities~\cite{bk-03, cd-09b}.
Recall from the discussion around Eqs.~\eqref{eq:cluster}, \eqref{eq:cluster2} that, due to the clustering property, at large rapidity differences between the particles, the form factor factorises in the product of form factors with less particles.
For example, up to the four-particle level, the connected components take the form
\begin{equation}\label{eq:connectedRR}
  f_{2,0}^{\twist | j_1 j_2}\!\left (\theta_1, \theta_2; n\right ) = \frac{1}{\avg{\twist_n}^2} \abs{F_{2,0}^{\twist | j_1 j_2}\!\left (\theta_1, \theta_2; n\right )}^2,
\end{equation}
\begin{equation}\label{eq:connectedRRRR}\begin{split}
  &f_{4,0}^{\twist | j_1 j_2 j_3 j_4}\!\left (\theta_1, \theta_2, \theta_3, \theta_4; n\right ) = \frac{1}{\avg{\twist_n}^2} \abs{F_{4,0}^{\twist | j_1 j_2 j_3 j_4}\!\left (\theta_1, \theta_2, \theta_3, \theta_4; n\right )}^2 + \\
  &\hspace{2cm} - f_{2,0}^{\twist | j_1 j_2}\!\left (\theta_1, \theta_2; n\right )\, f_{2,0}^{\twist | j_3 j_4}\!\left (\theta_3, \theta_4; n\right )  - f_{2,0}^{\twist | j_1 j_3}\!\left (\theta_1, \theta_3; n\right )\, f_{2,0}^{\twist | j_2 j_4}\!\left (\theta_2, \theta_4; n\right ) +\\
  &\hspace{2cm} - f_{2,0}^{\twist | j_1 j_4}\!\left (\theta_1, \theta_4; n\right )\, f_{2,0}^{\twist | j_2 j_3}\!\left (\theta_2, \theta_3; n\right ) ,
\end{split}\end{equation}
\begin{equation}\label{eq:connectedRRLL}\begin{split}
  f_{2,2}^{\twist | j_1 j_2 j_1' j_2'}\!\left (\theta_1, \theta_2, \theta_1', \theta_2'; n\right ) =&\, \frac{1}{\avg{\twist_n}^2} \abs{F_{2,2}^{\twist | j_1 j_2 j_1' j_2'}\!\left (\theta_1, \theta_2, \theta_1', \theta_2'; n\right )}^2 + \\
  & - f_{2,0}^{\twist | j_1 j_2}\!\left (\theta_1, \theta_2; n\right )\, f_{0,2}^{\twist | j_1' j_2'}\!\left (\theta_1', \theta_2'; n\right ) .
\end{split}\end{equation}
By definition, the connected form factors $f_{r, l}^{\twist | j_1\ldots}$ vanish for large rapidities. As we will see, this improves the convergence of the integral in Eq.~\eqref{eq:cumulant}.

In the expansion~\eqref{eq:cumulantExpansion}, we recognise two different kinds of cumulants. Those containing only form factors diagonal in the chiralities, $c^\twist_{r,0}$, $c^\twist_{0,l}$, which we will call non-interacting cumulants, and the ones that couple left- and right-movers, which we will call interacting. In the rest of the section, we treat the two kinds of terms separately since, as we will see, they give different contributions to the entanglement entropy~\eqref{eq:cumulantExpansion}.

\subsubsection{Non-interacting cumulants}\label{sec:noninteractingCumulants}

Let us first focus on the non-interacting cumulants.
As we saw in Sec.~\ref{sec:twistFormFactors}, in the massless flow, the form factors containing either only right- or only left-movers are identical to those of the massive Ising theory except for the vacuum expectation value $\langle \twist_n\rangle$, implying that their connected components are identical $f_{k, 0}^{\twist | j_1\ldots j_k} = f_{0, k}^{\twist | j_1\ldots j_k} = f_{k,\text{ Ising}}^{\twist | j_1\ldots j_k} $.
Given this identity, we can analyse them by applying the same strategy as in Ref.~\cite{cd-09b} for the massive Ising theory, which we also report in App.~\ref{app:cumulantIsing}.

Using the Pfaffian structure of the form factors in Eq.~\eqref{eq:Pfaffain}, it was shown that the non-interacting cumulants $c_{r, 0}^\twist$ have the general expression~\cite{cd-09,cd-09b,Ola2,cm-23}
\begin{equation}\label{eq:cumulantOlalla}\begin{split}
    &c_{k,0}^{\twist}\!\left ( M\ell; n \right ) = \frac{n}{k \left ( 2 \pi\right )^{k}} \sum_{j_2, \ldots, j_{k} = 1}^n \int_{-\infty}^{+\infty} \prod_{i = 1}^{k} \dd \theta_i\, e^{-\ell\, E\left (\theta_1, \ldots \right ) } \Big [  w\!\left(-\theta_{12} + 2\pi\ii j_2 \right )  w\!\left(\theta_{1, k} + 2\pi\ii j_{k} \right ) \times\\
    &\hspace{3cm}\times \prod_{l = 1}^{k/2 - 1} w\!\left ( -\theta_{2 l, 2l+1} + 2\pi\ii\left ( j_{2l} - j_{2l+1}\right )  \right ) w\!\left ( \theta_{2 l+1, 2l+2} + 2\pi\ii\left ( j_{2l+1} - j_{2l + 2}\right )  \right ) \Big ],
\end{split}\end{equation}
where $\theta_{ij} = \theta_i - \theta_j$, we have summed over $j_1$, and we have introduced the notation
\begin{equation}\label{eq:smallw}
    w\!\left ( \theta \right ) = \frac{1}{\avg{\twist_n}} F^{\twist|11}_{2,0}\!\left ( \theta; n \right ).
\end{equation}
From the form of Eq.~\eqref{eq:cumulantOlalla}, with all terms cyclically connected~\cite{cd-09b,cm-23}, it is clear why they are  known as fully connected.
Importantly,  Eq.~\eqref{eq:cumulantOlalla} holds for both the massless tricritical-critical flow and for the massive Ising theory.
The only difference between the cumulants in these two models is the form of the energy $E$ appearing in the exponential factor. 
This difference has however a major effect in the integral in Eq.~\eqref{eq:cumulantOlalla}.

For simplicity, we can start by analysing the two-right-mover cumulant $c_{2,0}^\twist$. The generalisation to higher particles will be straightforward. In our massless flow, as already recognised in Ref.~\cite{mds-95} for a different correlation function, the two-fold integral in Eq.~\eqref{eq:cumulantOlalla} is IR divergent due to the absence of a mass gap. In fact, in the relative and center-of-mass coordinates, $\theta_{12}=\theta_1-\theta_2$ and $A=(\theta_1+\theta_2)/2$, the energy~\eqref{eq:MasslessEnergyMomentum} of two right-moving particles takes the form 
\begin{equation}
  E\!\left ( \theta_1, \theta_2 \right ) = \frac{M}{2} \left ( e^{\theta_1} + e^{\theta_2} \right ) = M e^{A} \cosh\!\frac{\theta_{12}}{2}\,.
\end{equation}
For right-movers, the IR region $E \to 0$ corresponds to large and negative center-of-mass rapidity $A \to -\infty$. Since the form factors do not depend on $A$, we see that the integrand of Eq.~\eqref{eq:cumulantOlalla} tends to a non-zero constant for $A \to -\infty$, leading to a divergence when the integral in $A$ is performed.

In order to cure this IR divergence, we introduce a cut-off $\Lambda$ in the center-of-mass rapidity $A$. Since the form factors do not depend on $A$, the resulting integral can be cast in terms of the exponential integral function $\Ei(x)$, 
\begin{equation}\label{eq:cumulant2FlowCutoff}\begin{split}
  c^\twist_{2,0}\!\left ( M\ell, \Lambda; n\right ) =&\, \frac{n}{2 \left ( 2\pi \right )^2} \sum_{j_2}\int_{-\infty}^{+\infty} \dd \theta_{12}\, f_2^{\twist|1 j_2}\!\left ( \theta_{12}; n\right ) \int_{-\log \Lambda}^{+\infty} \dd A\, e^{- M \ell\, e^{A} \cosh\!\frac{\theta_{12}}{2}}\\
  =&\, \frac{n}{2 \left ( 2\pi \right )^2} \sum_{j_2}\int_{-\infty}^{+\infty} \dd \theta_{12}\, f_2^{\twist|1 j_2}\!\left ( \theta_{12}; n\right ) (-) \Ei\!\left (-\frac{M\ell}{\Lambda} \cosh\!\frac{\theta_{12}}{2} \right ) .
\end{split}\end{equation}
We see that $\Lambda$ plays the role of a cut-off at large distances with $M\ell \ll \Lambda$. In this limit, using the expansion 
\begin{equation}\label{eq:eiexpansion}
  \Ei\!\left( - x \right ) \underset{x\ll 1}{\approx} \log x + \gamma + \mathcal{O}\!\left ( x \right ), 
\end{equation}
we obtain a logarithmic dependence in the interval length $\ell$,
\begin{equation}\begin{split}
  c^\twist_{2,0}\!\left ( M\ell, \Lambda; n\right ) \approx - \frac{z_2\!\left ( n \right )}{2}\, \log\! \left ( \frac{M \ell}{\Lambda} \right ) + \text{const},
\end{split}\end{equation}
where $z_2$ is the function 
\begin{gather}
  z_{2}\!\left( n \right ) = \frac{n}{\left ( 2\pi \right )^2} \sum_{j_2} \int_{-\infty}^{+\infty} \dd \theta_{12}\, f_{2}^{\twist | 1 j_2}\!\left (\theta_{12}; n\right ). \label{eq:zeta2}
\end{gather}

Remarkably, up to an additive constant and the large distance cut-off $\Lambda$, the sum of the left- and right-moving two-particle cumulants in our massless flow, $c_{2,0}^\twist+c_{0, 2}^\twist$, is equal to the UV limit of the two-particle cumulant of the massive Ising model (cf. Eq.~\eqref{eq:c2_ising} in App.~\ref{app:cumulantIsing}). 
This is consistent with the expectation that, in the IR, the contributions of the interacting cumulants vanish because the flow leads to the critical Ising fixed point. As such, we expect that for large distances the non-interacting cumulants completely reproduce the logarithmic entanglement entropy of the Ising CFT.

Moving to higher particle cumulants $c_{r, 0}^\twist$, we expect a similar structure.
In the presence of more than two particles, a convenient set of coordinates is again provided by the center-of-mass rapidity $A = \frac{1}{k} \sum_{j=1}^k \theta_j$ and the difference between the rapidities $\theta_{j,j+1} = \theta_{j} - \theta_{j+1}$, with Jacobian equal to one.
For convenience, we further define the rapidities in the center-of-mass frame of reference
\begin{equation}\label{eq:comRapidity}
    \xi_j = \theta_j - A, \qquad \xi_k = - \sum_{j=1}^{k-1} \xi_j\,, 
\end{equation}
which can be shown to depend only on the rapidity differences.
In the massless flow, the $r$-fold integral in Eq.~\eqref{eq:cumulantOlalla} is divergent in the large negative center-of-mass rapidity region $A \to -\infty$. It is important to stress a subtle point. Due to the clustering property (see Eqs.~\eqref{eq:cluster} and \eqref{eq:cluster2}), the integral of the form factor is divergent in the direction of the sum of any two rapidities $\theta_j$. However, in the cumulants, the non-connected factorised component is subtracted as in, e.g., Eqs.~\eqref{eq:connectedRRRR} and \eqref{eq:connectedRRLL}, guaranteeing that the integral of the connected part converges in those directions. The only remaining divergence is the one in the direction of large negative center-of-mass $A$, as it happens for the two-particle cumulant~\eqref{eq:cumulant2FlowCutoff}.

In the center-of-mass coordinates defined before Eq.~\eqref{eq:comRapidity}, the energy of $r$ right-moving particles in the massless flow takes the form
\begin{equation}\begin{split}
  E\!\left ( \theta_1, \ldots, \theta_r \right ) =&\, \frac{M}{2} \sum_j e^{\theta_j} = \frac{M}{2}\, e^A \sum_j e^{\xi_j \left( \theta_{12}, \ldots \right )}\,.
\end{split}\end{equation}
As already done in Eq.~\eqref{eq:cumulant2FlowCutoff} for the two-particle case, we again introduce a cut-off $\Lambda$ on the large negative center-of-mass rapidity $A$ and we write the integral over it in terms of the exponential integral function $\Ei(x)$,
\begin{equation}\begin{split}
  c_{r,0}^\twist\!\left( M\ell, \Lambda;n \right ) =&\, \frac{n}{r! \left ( 2\pi\right )^r} \sum_j \int_{-\infty}^{+\infty} \prod_{j=1}^{r-1} \dd\theta_{j, j+1}\,  f_r^\twist\!\left (\theta_{12}, \ldots; n\right ) \int_{-\infty}^{+\infty} \dd A\, e^{-\frac{M}{2}\ell\, e^A \sum_j e^{\xi_j}} \\
  =&\, \frac{n}{r! \left ( 2\pi\right )^r} \sum_j \int_{-\infty}^{+\infty} \prod_{j=1}^{r-1} \dd\theta_{j, j+1}\,  f_r^\twist\!\left (\theta_{12}, \ldots; n\right ) \,\left ( - \right ) \Ei\!\left (\frac{1}{2} \frac{M\ell}{\Lambda} \sum_{j} e^{\xi_j} \right ).
\end{split}\end{equation}
In the large cut-off limit $\Lambda \gg M\ell$, we can approximate the cumulant using the expansion of the exponential integral in Eq.~\eqref{eq:eiexpansion}, obtaining the expected logarithmic behaviour
\begin{equation}\label{eq:cr0_higher}
  c_{r,0}^\twist\!\left ( M\ell,\Lambda; n \right ) \approx - \frac{z_r(n)}{2} \log\!\frac{M\ell}{\Lambda} + \text{const}\,,
\end{equation}
with $z_k(n)$ equal to 
\begin{align}
  z_k(n) &= \frac{2 n}{k! \left ( 2\pi\right )^k} \sum_j \int_{-\infty}^{+\infty} \prod_{j=1}^{k-1} \dd\theta_{j,j+1}\,  f_k^\twist\!\left (\theta_{12}, \ldots; n \right ) \notag \\
  &= \frac{2 n}{k \left ( 2 \pi\right )^{k}} \sum_{j_2, \ldots, j_{k} = 1}^n \int_{-\infty}^{+\infty} \prod_{i = 1}^{k} \dd \theta_i\, \Big [  w\!\left(-\theta_{12} + 2\pi\ii j_2 \right )  w\!\left(\theta_{1, k} + 2\pi\ii j_{k} \right ) \times \label{eq:zetaN}\\
    &\hspace{2cm}\times \prod_{l = 1}^{k/2 - 1} w\!\left ( -\theta_{2 l, 2l+1} + 2\pi\ii\left ( j_{2l} - j_{2l+1}\right )  \right ) w\!\left ( \theta_{2 l+1, 2l+2} + 2\pi\ii\left ( j_{2l+1} - j_{2l + 2}\right )  \right ) \Big ]. \notag
\end{align}

As happens with the two-particle cumulants, the sum $c_{r, 0}^\twist+c_{0,r}^\twist$ in the massless flow coincides with the UV limit of the $r$-particle cumulant of
the massive Ising theory up to additive constants (see Eq.~\eqref{eq:ckising} in App.~\ref{app:cumulantIsing}). In Ref.~\cite{cd-09b}, the resummation of the $z_r(n)$ terms is carried out. Taking Eq.~\eqref{eq:cr0_higher} and applying their result,
\begin{equation}\label{eq:isingCumulantResum}
    \sum_{k\text{ even}} z_k\!\left ( n \right ) = 4 h_{\twist}^\text{IR} = \frac{1}{12} \left ( n - \frac{1}{n} \right ),
\end{equation}
we find that
\begin{equation}
    \sum_{r\text{ even}} c_{r,0}^\twist\!\left ( M\ell,\Lambda; n \right ) + \sum_{l\text{ even}} c_{0,l}^\twist\!\left ( M\ell,\Lambda; n \right ) \approx \frac{1}{12} \left ( n - \frac{1}{n} \right )\log\!\frac{M\ell}{\Lambda} + \text{const}.
\end{equation}
 This shows that, up to additive constants, the sum of the non-interacting left- and right-movers contribution to the entanglement entropy~\eqref{eq:cumulantExpansion} in the massless flow gives the entropy of the Ising CFT in the IR fixed point.

\subsubsection{Interacting cumulants}\label{sec:interactingCumulants}

As shown in the previous discussion, the non-interacting cumulants contribute to the entropy of the IR Ising CFT; hence the corrections for smaller distances are provided by the interacting cumulants $c_{r,l}^\twist$, which couple left- and right-movers. In this section, we study the only interacting cumulant at four-particle level, namely $c_{2,2}^\twist$
\begin{equation}\label{eq:cumulantRRLLFlow}\begin{split}
  c_{2,2}^\twist\!\left(M\ell;n\right ) &= \frac{n}{2 \times 2 \left ( 2\pi \right )^4} \sum_{j_2, j_3, j_4} \int_{-\infty}^{+\infty} \dd \theta_1 \dd\theta_2 \dd\theta_3 \dd\theta_4 \, f_{2,2}^{\twist | 1 j_2 j_3 j_4}\!\left (\theta_1, \theta_2, \theta_1', \theta_2'; n \right ) \times\\
  &\hspace{8cm} \times e^{- \frac{M \ell}{2} \left ( e^{\theta_1} + e^{\theta_2} + e^{-\theta_1'} + e^{-\theta_2'} \right )} \\
  &= \frac{n}{2 \times 2 \left ( 2\pi \right )^4} \sum_{j_2, j_3, j_4} \int_{-\infty}^{+\infty} \dd \theta_1 \dd\theta_2 \dd\theta_3 \dd\theta_4 \bigg [ \frac{1}{\avg{\twist_n}^2} \abs{F_{2,2}^{\twist | 1 j_2 j_3 j_4}\!\left (\theta_1, \theta_2, \theta_1', \theta_2'; n \right )}^2 +\\ 
  &\hspace{1cm} - \frac{1}{\avg{\twist_n}^4}\abs{F_{2,0}^{\twist | 1 j_2}\!\left (\theta_1, \theta_2; n \right )}^2 \abs{F_{0,2}^{\twist | j_3 j_4}\!\left (\theta_1', \theta_2'; n \right )}^2 \bigg ] e^{- \frac{M \ell}{2} \left ( e^{\theta_1} + e^{\theta_2} + e^{-\theta_1'} + e^{-\theta_2'} \right )},
\end{split}\end{equation}
where the `RRLL' form factor $F_{2,2}^{\twist}$ is given in Eq.~\eqref{FFRRLLFlow}.
As for the non-interacting cumulants, the integral of $F_{2,2}^{\twist}$ is divergent in the IR limit. However, unlike the previous discussion, now the subtraction of the clusterisation in the connected component $f_{2,2}^{\twist | 1 j_2 j_3 j_4}$ ensures that the cumulant is convergent and a regularisation is not needed.

\begin{figure}[t]
    \centering
    \begin{subfigure}[b]{0.48\textwidth}
        \centering
        \includegraphics[width=\textwidth]{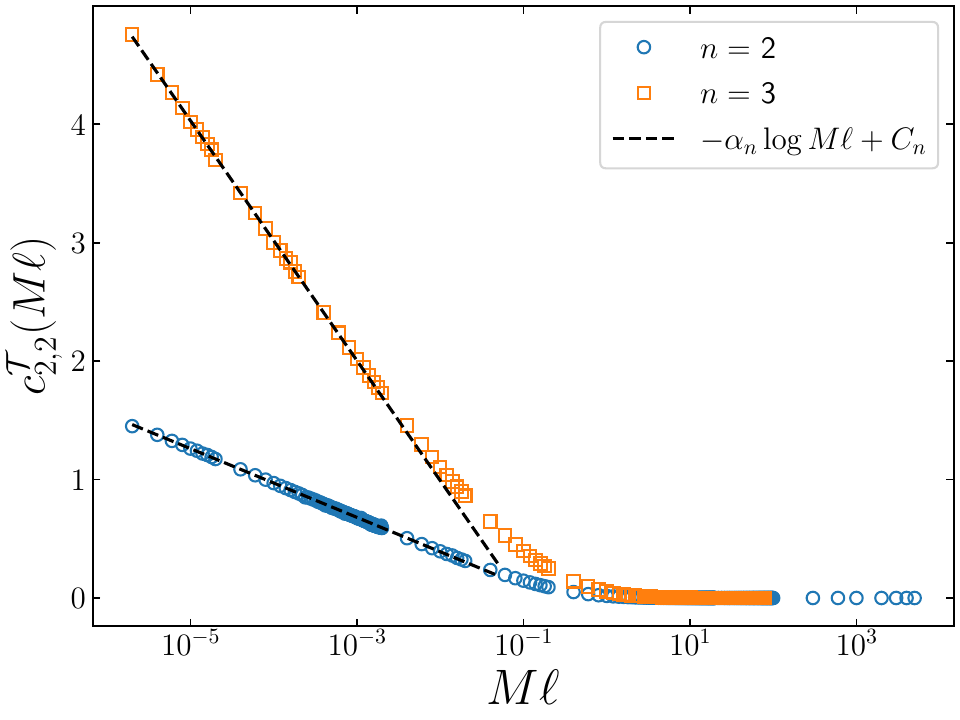}
        \caption{Cumulant $c_{2,2}^\twist$.}
        \label{fig:RRLLUV}
    \end{subfigure}
    \hfill
    \begin{subfigure}[b]{0.48\textwidth}
        \centering
        \includegraphics[width=\textwidth]{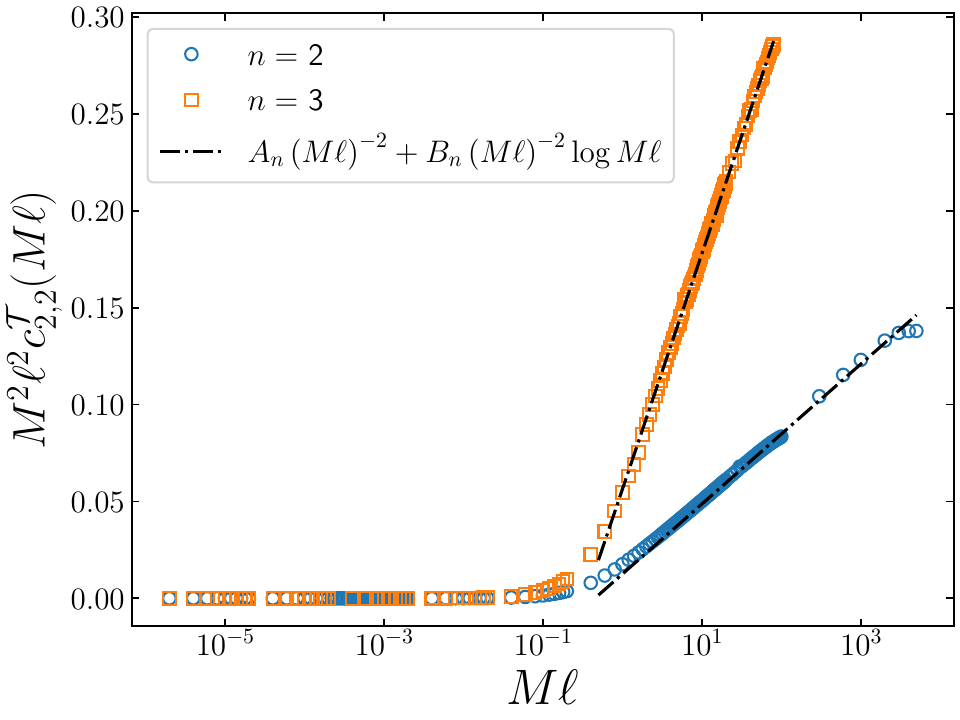}
        \caption{Cumulant $c_{2,2}^\twist$ times $M^2 \ell^2$.}
        \label{fig:RRLLIR}
    \end{subfigure}
    \caption{Semi-logarithmic plots of the cumulant $c_{2,2}^\twist$ in Eq.~\eqref{eq:cumulantRRLLFlow} for $n = 2, 3$ replicas as a function of $M\ell$. On the left, the dashed line is the curve $- \alpha_n \log\!\left (M \ell \right ) + C_n$, which provides the leading logarithmic behaviour in the UV regime $M\ell\ll 1$. The parameters $\alpha_n$, $C_n$ are obtained from numerical best fit of the points and are reported in Eqs.~\eqref{eq:fitLog2}, \eqref{eq:fitLog3}. On the right, we plot the product of the cumulant times $M^2\ell^2$. The dash-dotted line represents the leading IR ($M\ell\gg 1$) behaviour $c_{2,2}^\twist \simeq A_n\, \left( M \ell\right )^{-2} + B_n\, \left( M \ell\right )^{-2} \log\!\left ( M \ell\right )$, with parameters $A_n$, $B_n$ obtained from the best fit shown in Eqs.~\eqref{eq:fitTTbar2}, \eqref{eq:fitTTbar3}.}
    \label{fig:RRLLsemilog}
\end{figure}

In Fig.~\ref{fig:RRLLsemilog}, we report the result of the numerical integration of the cumulant $c_{2,2}^\twist$ for different values of $M\ell$ and for $n = 2, 3$ replicas, performed using the Divonne routine of the library \textsc{Cuba}~\cite{cuba}. In the UV region $M\ell \ll 1$, we expect a leading logarithmic behaviour, since the sum of the interacting and non-interacting form factors should reproduce the logarithmic entanglement entropy of the tricritical Ising UV fixed point.
In Fig.~\ref{fig:RRLLUV}, we plot the interacting cumulant $c_{2,2}^{\twist}$ in Eq.~\eqref{eq:cumulantRRLLFlow} for $n = 2, 3$ replicas and we compare it with the fit of the numerical points to a logarithmic function $- \alpha_n \log M\ell + C_n$.
We perform the fit for $M\ell \leqslant 2\times 10^{-4}$ when $n=2$ and for $M\ell \leqslant 6\times 10^{-4}$ when $n=3$, obtaining the parameters
\begin{alignat}{3}
    &\alpha_2 \approx 0.13, \quad   &&C_2 \approx - 0.19, \qquad    &&\text{for } n = 2,\, M\ell \leqslant 2\times 10^{-4}, \label{eq:fitLog2}\\
    &\alpha_3 \approx 0.44, \quad   &&C_3 \approx - 1.03, \qquad    &&\text{for } n = 3,\, M\ell \leqslant 6\times 10^{-4}. \label{eq:fitLog3}
\end{alignat}
From Fig.~\ref{fig:RRLLUV}, we see that for $M\ell \ll 1$ the cumulant is in good agreement with the expected logarithmic behaviour.

To understand the behaviour in the IR ($M\ell\gg 1$), recall from the general introduction of Sec.~\ref{sec:flow} that, near the IR fixed point, the effective theory describing the massless flow is the $T\bar{T}$ deformation of the critical Ising CFT, as shown in Eq.~\eqref{eq:LEff}.
The entanglement entropies of generic $T\bar{T}$-deformed CFTs have been heavily studied in recent years, see e.g.~\cite{ds-18, llst-20, cch-18, bbc-19, jkn-19, mnrs-19, gst-19, p-19, a-20, hs-20, ss-19, g-19, akf-20, cans-23}.
In particular, in Ref.~\cite{cch-18}, the entropy of an interval of length $\ell$ in a system of finite size $L$ has been computed perturbatively for a generic $T\bar{T}$-deformed CFT.
Let the deformed action be
\begin{equation}\label{eq:ttbarCFTaction}
  \mathcal{A}_{T\bar{T}} = \mathcal{A}_\text{CFT} + g \int \dd^2z \, T\bar{T}\, , 
\end{equation}
where the $T\bar{T}$ coupling $g$ has dimensions of an inverse mass squared, $g \propto M^{-2}$.
The first perturbative correction to the Rényi entanglement entropy of the IR CFT was found to be~\cite{cch-18}
\begin{equation}\label{eq:ttbarEEFiniteSize}
  (1-n)\, \delta S_n^{(1)}\!\left (\ell, L, g \right ) = - \frac{\pi c^2 g}{36} \frac{ \left (n^2-1 \right )^2}{n^3} \left [ \frac{1}{16\, \epsilon^2} - \frac{\left ( 11 \cos\!\frac{2\pi\ell}{L} +19 \right )}{24 \left ( \frac{L}{\pi} \sin\!\frac{\pi \ell}{L} \right )^2} + \frac{\cos\!\frac{\pi\ell}{L}\, \log\!\frac{L \sin\frac{\pi \ell}{L}}{2 \pi \epsilon}}{ \left ( \frac{L}{\pi} \sin\!\frac{\pi \ell}{L} \right )^2} \right ],
\end{equation}
where $\epsilon$ is a non-universal UV cut-off.
Here we are interested in the entanglement entropy in the thermodynamic limit $L \to \infty$ of Eq.~\eqref{eq:ttbarEEFiniteSize},
\begin{equation}\label{eq:ttbarEE}
  (1-n)\, \delta S_n^{(1)}\!\left (\ell, g \right ) = - \frac{\pi c^2 g}{36} \frac{\left ( n^2 - 1 \right )^2}{n^3} \left [ \frac{1}{16\, \epsilon^2} - \frac{5}{4} \frac{1}{\ell^2} + \frac{\log\frac{\ell}{2\epsilon}}{\ell^2}\right ]+ \mathcal{O}\!\left ( g^2 \ell^{-4} \right).
\end{equation}
Comparing the effective Lagrangian~\eqref{eq:LEff} with the generic one in Eq.~\eqref{eq:ttbarCFTaction} and taking into account that in our case $T = - \frac{1}{2} \psi \partial \psi$ and $\bar{T} = - \frac{1}{2} \bar{\psi} \bar{\partial} \bar{\psi}$,
we can conclude that $g = - \frac{4}{\pi^2 M^2}$.
Therefore, since the central charge of our IR point is $c = \frac{1}{2}$, Eq.~\eqref{eq:ttbarEE} specialised to our massless flow gives
\begin{equation}\label{eq:ttbarEETricr}
   ( 1 - n )\, \delta S_n^{(1)}\!\left (\ell, M \right ) = \frac{1}{36 \pi} \frac{\left ( n^2 - 1 \right )^2}{n^3} \left [\frac{1}{16 M^2 \epsilon^2} - \frac{5}{4} \frac{1}{M^2 \ell^2} + \frac{\log \frac{\ell}{2\epsilon}}{M^2 \ell^2}\right ] + \mathcal{O}\!\left ( M^{-4} \ell^{-4} \right).
\end{equation}
Observe that in the prediction of Eq.~\eqref{eq:ttbarEETricr} the leading correction is of the form $A_n\, \ell^{-2} + B_n\, \ell^{-2} \log \ell$. The coefficient $A_n$ is not universal due to the presence of the $UV$ cutoff $\epsilon$, while the factor $B_n$ is.
In particular, for $n = 2, 3$ replicas, its numerical value is
\begin{alignat}{2}
    &B_2 = \frac{1}{32\pi} = 0.00995\ldots, \qquad      &&\text{for } n = 2, \label{eq:predttbar2} \\
    &B_3 = \frac{16}{243 \pi} = 0.02096\ldots, \qquad   &&\text{for } n = 3. \label{eq:predttbar3} 
\end{alignat}
Note also that the leading correction in Eqs.~\eqref{eq:ttbarEEFiniteSize}, \eqref{eq:ttbarEE}, \eqref{eq:ttbarEETricr} is non-zero only for $n \geqslant 2$ R\'enyi entropies, while it vanishes in the replica limit $n \to 1$~\cite{cch-18}.
 
It is worthwhile to compare the first-order perturbative prediction in Eq.~\eqref{eq:ttbarEETricr} with the leading correction that we obtain here from the form factor cumulant expansion~\eqref{eq:cumulantExpansion}, which is given by the interacting cumulant $c_{2, 2}^\mathcal{T}$.
In Fig.~\ref{fig:RRLLIR} we study this cumulant for $n = 2, 3$ replicas as a function of $M\ell$ and we perform a best fit of the numerical points to a function $A_n\, \ell^{-2} + B_n\, \ell^{-2} \log \ell$, for $M \ell \geqslant 50$ when $n = 2$ and for $M \ell \geqslant 20$ when $n=3$. We obtain
\begin{alignat}{3}
    &A_2 \approx 0.013, \quad   &&B_2 \approx 0.016, \qquad &&\text{for } n = 2, \, M \ell \geqslant 50, \label{eq:fitTTbar2}\\
    &A_3 \approx 0.056, \quad   &&B_3 \approx 0.053, \qquad &&\text{for } n = 3, \, M \ell \geqslant 20. \label{eq:fitTTbar3}
\end{alignat}

For large distances, we find a good qualitative agreement with the functional form predicted in Eq.~\eqref{eq:ttbarEETricr}. However, while for $n=2$ replicas the numerical result of the fit for $B_2$ is close to the predicted value in Eq.~\eqref{eq:predttbar2}, this is not the case for $n = 3$. A possible explanation of this discrepancy is that
the higher-particle interacting cumulants $c_{r,l}^\twist$, which we are neglecting, also contribute to the term $\log\ell/\ell^2$ and their contribution depends on the number of replicas $n$.

\subsubsection{Entanglement entropy}\label{sec:entropyplot}
\begin{figure}[t]
    \centering
    \includegraphics[width=.7\textwidth]{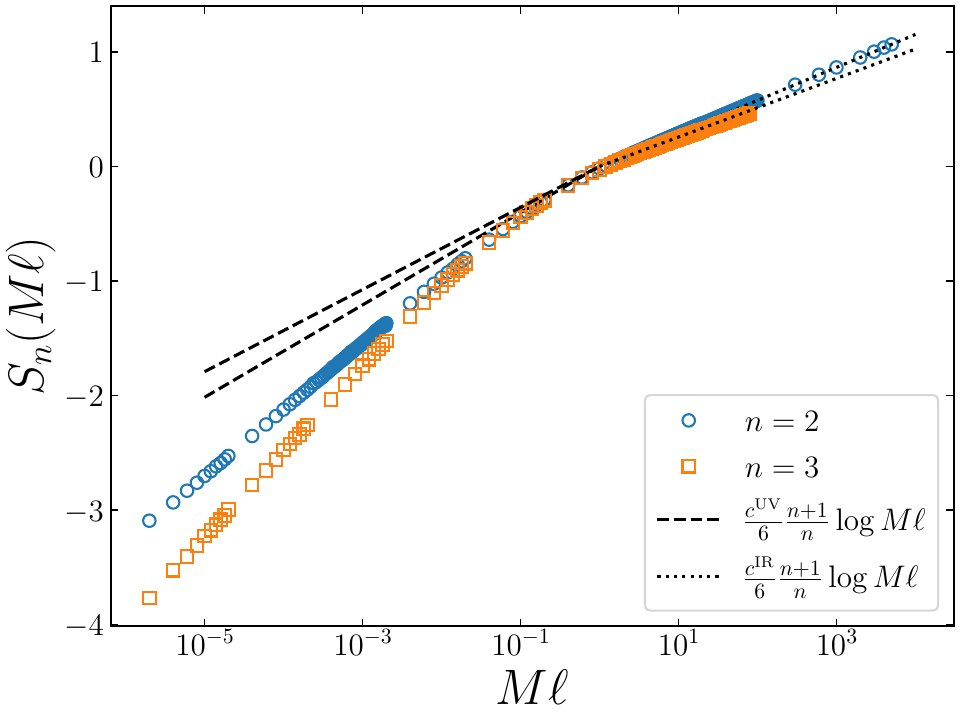}
    \caption{Semi-logarithmic plot of the Rényi entanglement entropy along the tricritical-critical Ising massless flow as a function of $M\ell$ for Rényi indices $n=2$ and $3$. The dots have been obtained using the truncated cumulant expansion~\eqref{eq:cumulantExpansion}, including the first $30$ non-interacting cumulants $c_{r, 0}^\twist, c_{0, l}^\twist$ and the leading interacting cumulant $c_{2, 2}^\twist$. The curves indicate the expected behaviour of the entropy when approaching the IR ($M\ell\gg 1$, dotted curves) and UV ($M\ell\ll 1$, dashed curves) regimes. In the IR, the truncated cumulant expansion agrees with the asymptotic behaviour, while in the UV it deviates. The reason of this mismatch is that we are only considering the leading interacting cumulant in the expansion~\eqref{eq:cumulantExpansion} of the entropy.}
    \label{fig:entropy}
\end{figure}

Finally, we can put together the results obtained in Secs.~\ref{sec:noninteractingCumulants} and \ref{sec:interactingCumulants} to get the total entanglement.
In Fig.~\ref{fig:entropy}, we consider the Rényi entanglement entropy in the massless flow as a function of $M\ell$ for $n=2$ and $3$. 
The results in this figure (represented as symbols) have been obtained with the cumulant expansion~\eqref{eq:cumulantExpansion}, including the first $30$ non-interacting cumulants $c_{r, 0}^\twist$ and $c_{0, l}^\twist$ and the leading interacting cumulant $c_{2, 2}^\twist$. 
In the plot, we also report the expected behaviour of the entanglement entropy when approaching the UV (dashed curves) and IR (dotted curves) fixed points.
When approaching the IR, $M\ell\gg 1$, we find a very good agreement between the truncated cumulant expansion and the expected IR asymptotics for both values of $n$.
In the UV, $M\ell\ll 1$, while the truncated expansion presents a behaviour compatible with a logarithmic divergence in $M\ell$, it does not quantitatively agree with the UV entropy.
This is expected, since the UV limit is the regime where the higher-particle interacting cumulants $c_{r, l}^\twist$ contribute more significantly and here we are only considering the four-particle one, $c_{2, 2}^\twist$.
It would be interesting to include higher order terms, but this is a challenging task due to the difficulty of computing higher-particle cumulants, which involve an increasing number of multidimensional integrals.
Nevertheless, in the light of Fig.~\ref{fig:entropy}, we can conclude that only including the leading interacting cumulant is enough to qualitatively observe the  crossover between the IR and UV regimes.

Before concluding this section, let us comment on the symmetry resolved entanglement entropy.
Also in this case, a cumulant expansion analogous to Eq.~\eqref{eq:cumulantExpansion} holds true, by replacing the form factors with the appropriate ones for the composite twist field that we have determined in Sec.~\ref{sec:compositeFormFactors}.
In particular, the expansion of the symmetry resolved entanglement entropy contains both the non-interacting cumulants reproducing the Ising CFT and the interacting ones providing the corrections, analogously to what happens for the standard entropy.
The symmetry resolved entropy in the massive Ising model has been recently studied in Ref.~\cite{cm-23} where it was found that (differently from what happens for the standard BPTF) the cumulants of the composite twist fields are divergent although the theory is massive; consequently a  regularisation was required.
This fact suggests that, also for the massless flow, the regularisation employed for the total entropy is not sufficient to find a finite result for the composite twist field.
Resolving such a regularisation is a problem that goes beyond the scope of this paper and we hope to return on the issue in the future.

\section{Conclusions}\label{sec:conclusions}

In this paper, we investigated the ground state Rényi entanglement entropies of a single interval in the massless QFT associated to the renormalisation group flow connecting the tricritical and critical Ising CFTs. 
The corresponding two-point correlation function of 
branch points twist fields admits a form factor expansion along the flow. 
We showed that these form factors can be calculated in two different and independent ways. 
On the one hand, we have directly applied the bootstrap approach of Ref.~\cite{Ola} for massive integrable QFTs: based on the symmetries of the theory and the exchange properties of the twist fields, we obtained a set of equations for the form factors and we found a general ansatz that solves them. Alternatively, we obtained the form factors using the Zamolodchikov's staircase model, an extension of the sinh-Gordon theory with complex couplings that includes the tricritical-critical renormalisation flow. 
In this framework, the form factors of several fields in this massless flow have been obtained as the roaming limit of those in the sinh-Gordon theory. We showed that the same strategy works for the branch points twist fields; we derived explicit expressions for the two and four-particle form factors, from which the higher-particle ones can be recursively derived. 
The two approaches gave identical results, confirming the validity of the roaming limit.

The form factor expansion of the entanglement entropy can be rearranged order by order in the number of particles in terms of
cumulants, which are given by the connected part of the form factors. In this cumulant expansion, we distinguished free and interacting cumulants. The former only contain particles of the same 
chirality and give the entropy of the IR Ising CFT. In fact, we found that, after a proper regularisation, they are equal to those that appear in the massive Ising theory. On the other hand, the interacting cumulants, which contain particles with different chiralities, describe the behaviour of the entanglement entropy along the flow. In particular, we checked that the lowest-particle interacting cumulant yields in the UV limit the expected logarithmic behaviour in the subsystem size. 
The IR limit can be described by a $T\bar{T}$ deformation of the Ising CFT. 
We showed that the lowest-particle interacting cumulant expansion approaching the IR point qualitatively reproduces the result for a generic $T\bar{T}$ perturbation at first order~\cite{cch-18}.
However, we could not obtain a quantitative agreement since the two expansions are organised in a different way. It would be interesting to compute higher particle form factors, to confirm that the agreement improves.

The massless flow~\eqref{eq:flow_action} that we studied here is also connected with the $SU(3)_2$-homogeneous sine-Gordon (HSG) model~\cite{homSinGordon}.
As shown in Refs.~\cite{cafkm-00, CastroFring1, cl-11}, along the renormalisation group flow, the central charge and the twist field dimension of this theory present two plateaux, analogously to the behaviour of the staircase model considered here.
For certain values of the parameters, it was shown that one of these plateaux corresponds to the massless flow from tricritical to critical Ising~\cite{cafkm-00, cl-11}.
Since the form factors of the standard twist field in the $SU(3)_2$-HSG model have been obtained in Ref.~\cite{cl-11} up to the four-particle order, it would be interesting to recover our results from an appropriate limit of the HSG expressions.

The massless flow connecting the tricritical and critical Ising CFTs enjoys a global $\mathbb{Z}_2$ symmetry. In this work, we  
also considered the composite branch point twist fields associated to this symmetry. Their two-point functions give the charged moments of the reduced density matrix from which one can determine the symmetry-resolved entanglement entropy. Similarly to the standard twist fields, we obtained their bootstrap equations, which now include the non-trivial monodromy due to the insertion of the charge, and we  found a general ansatz for their solution, which allows to obtain the higher-particle form factors recursively. We further derived them as the roaming limit of the composite twist field form factors of the sinh-Gordon theory. Remarkably, the latter were neither known in the literature, a gap which we also filled here.

Finally, we mention that, as we explained in Sec.~\ref{sec:flow}, the flow~\eqref{eq:flow_action} is the simplest example of the infinite family of massless renormalisation flows $A_p$, with $p \geqslant 2$, that interpolate between the unitary diagonal minimal models $\mathcal{M}_{p+2} \to \mathcal{M}_{p+1}$, all of which possess a global $\ZZ_2$ symmetry. A natural continuation of our work would be to study the form factor of both the standard and the composite twist fields in these flows and use them to study the (symmetry-resolved) entanglement entropies. The additional complication of these models is the presence of further magnonic excitations beside the fundamental ones. This fact makes more difficult the calculation of the twist field form factors compared to the $A_2$ case considered here.

\section*{Acknowledgements}
We thank Luca Capizzi, Olalla Castro-Alvaredo, Cecilia De Fazio, Michele Mazzoni and Stefano Negro for useful discussions.
We are particularly grateful to O.~Castro-Alvaredo for pointing out the results of Refs.~\cite{cd-09b, Ola2, cm-23} that we used in Sec.~\ref{sec:noninteractingCumulants} and App.~\ref{app:cumulantIsing}.
The authors acknowledge support from ERC under Consolidator grant number 771536 (NEMO).

\appendix

\section{Form factor bootstrap for Branch Point Twist Fields in the \texorpdfstring{$A_2$}{A2}~massless flow}
\label{app:FlowBootstrap}

In this appendix, we provide more information on the technical derivation of the four-particle form factors for the standard and the composite BPTFs in the massless flow~\eqref{eq:flow_action} from the tricritical to the critical Ising CFTs.

\subsection{Form factors of the standard BPTF}\label{app:FlowBootstrapTwist}
We begin by recalling the most general ansatz~\eqref{FFParametrizationFlow} for FFs at any particle level, which reads
\begin{equation}
\begin{split}
F^{\twist}_{\underline{R},\underline{L}}(\underline{\theta},\underline{\theta'};n) &=  H^{\twist}_{r,l}Q^{\twist}_{r,l}(\underline{x},\underline{y};n)\prod_{1\leqslant i<j\leqslant r} \frac{f_{RR}(\theta_{i}-\theta_{j};n)}{(x_{i}-\omega x_{j})(x_{j}-\omega x_{i})} \times\\
 &\hspace{1cm}\times \prod_{i=1}^{r} \prod_{j=1}^{l} f_{RL}(\theta_{i}-\theta'_{j};n)\prod_{1\leqslant i<j\leqslant l} \frac{f_{LL}(\theta'_{i}-\theta'_{j};n)}{(y_{i}-\omega y_{j})(y_{j}-\omega y_{ji})}\:.
\end{split}\label{FFParametrizationFlowRepeated}
\end{equation}
where we have $r$ right-mover and $l$ left-mover particles, $x_i=e^{\theta_i/n}$ and $y_i=e^{-\theta'_i/n}$. As discussed in the main text, the exchange~\eqref{eq:FFAxiom1} and the cyclic permutation axioms~\eqref{eq:FFAxiom2} are automatically satisfied by the above expression.

Our goal now is to identify the four-particle `RRLL' form factors using the well-known two-particle quantities. For simplicity we place every particle on the first replica and specify Eq.~\eqref{FFParametrizationFlowRepeated} to the case of interest
\begin{equation}
\begin{split}
&F^\twist_{2,2}(\theta_{1},\theta_{2},\theta'_{1},\theta'_{2}; n) = \\
=&\, H^{\twist}_{2,2}Q^{\twist}_{2,2}(x_1,x_2,y_1,y_2;n) \frac{f_{RR}(\theta_{1}-\theta_{2};n)}{(x_{1}-\omega x_{2})(x_{2}-\omega x_{1})} \prod_{i=1}^{2} \prod_{j=1}^{2} f_{RL}(\theta_{i}-\theta'_{j};n) \frac{f_{LL}(\theta'_{1}-\theta'_{2};n)}{(y_{1}-\omega y_{2})(y_{2}-\omega y_{1})}\,.
\end{split}
\label{FFParametrizationFlow4pt}
\end{equation}
Applying now the residue axiom in Eq.~\eqref{eq:FFAxiom3} to the ansatz~\eqref{FFParametrizationFlow4pt} we can derive recursive equations for the $H^{\twist}_{2,2}$ normalisation factors and the $Q^{\twist}_{2,2}$ functions. Let us first also recall that the minimal form factor $f_{RL}$ satisfies the identity 
\begin{equation}
\mathcal{N}_n f_{RL}(\theta+\ii\pi;n)\,  \mathcal{N}_n f_{RL}(\theta;n)=\left(1-e^{-\frac{\ii \pi }{2 n}} e^{-\frac{\theta }{ n}}\right)^{-1} \,.
\label{FMinRLFMinRLRepeated}
\end{equation}
The residue of the denominator of the ansatz~\eqref{FFParametrizationFlow4pt} takes the form
\begin{equation}
\begin{split}
&-\ii\res_{\theta_1=\ii\pi+\theta_2} \frac{1}{(x_1-\omega x_2)(x_2-\omega x_1)} \frac{1}{(y_1-\omega y_2)(y_2-\omega y_1)} =\\
&\hspace{5cm} = -\ii\,  \frac{1}{(y_1-\omega y_2)(y_2-\omega y_1)} \frac{n \left(\omega x_0^2 \right)^{-1}}{(e^{\frac{2 \ii \pi }{n}}-1)}\,, 
\label{4ptResidue}
\end{split}
\end{equation}
from which we can obtain the residue of the entire expression~\eqref{FFParametrizationFlow4pt} as
\begin{equation}
\begin{split}
& -\ii\res_{\theta_1=i\pi+\theta_2} F^\twist_{2,2}(\theta_{1},\theta_{2},\theta'_{1},\theta'_{2}; n) = \\
=&\, \ii\,  H^{\twist}_{2,2}Q^{\twist}_{2,2}(\omega x,x,y_1,y_2;n)f_{RR}(\ii\pi;n)\frac{n \left(x^2 \omega \right)^{-1}}{\omega^2-1}\times\\
 &\hspace{3cm}\times \prod_{j=1}^{2} f_{RL}(\theta+\ii\pi-\theta'_{j};n)f_{RL}(\theta-\theta'_{j};n)\frac{f_{LL}(\theta'_1-\theta'_2;n)}{(y_1-\omega y_2)(y_2-\omega y_1)}\\
 =&\, \ii\,  H^{\twist}_{2,2}Q^{\twist}_{2,2}(\omega x,x,y_1,y_2;n)f_{RR}(\ii\pi;n)\frac{n \left(x^2 \omega \right)^{-1}}{\omega^2-1}\times\\
 &\hspace{3cm}\times \mathcal{N}_n^{-4} \frac{1}{1 - \left (\omega^{1/2} x y_1 \right )^{-1}} \frac{1}{1 - \left (\omega^{1/2} x y_2 \right )^{-1}} \frac{f_{LL}(\theta'_1-\theta'_2;n)}{(y_1-\omega y_2)(y_2-\omega y_1)}\,,
\end{split}
\end{equation}
where we used Eqs.~\eqref{FMinRLFMinRLRepeated} and~\eqref{4ptResidue}. Via algebraic manipulations we can simplify the above formula to
\begin{equation}\label{eq:resF22flowTwist}
\begin{split}
& -\ii\res_{\theta_1=\ii\pi+\theta_2} F^\twist_{2,2}(\theta_{1},\theta_{2},\theta'_{1},\theta'_{2}; n) = \\
 =&\, \ii\, H^{\twist}_{2,2}Q^{\twist}_{2,2}(\omega x,x,y_1,y_2;n)\, \mathcal{N}_n^{-4} \frac{1}{(\omega^{1/2} x y_1 -1)(\omega^{1/2} x y_2 -1)} \left[ \frac{4\omega}{n} \cos\frac{\pi}{2n}  \right ]^{-1} \times\\
 & \hspace{2cm} \times  \frac{y_1 y_2\,f_{LL}(\theta'_1-\theta'_2;n)}{(y_1-\omega y_2)(y_2-\omega y_1)}\,.
\end{split}
\end{equation}

Following the residue axiom in Eq.~\eqref{eq:FFAxiom3}, the residue of the kinematical pole in Eq.~\eqref{eq:resF22flowTwist} has to reproduce the two-particle form factor in Eq.~\eqref{eq:F2TwistField}. We first recast it in the shape of our ansatz as
\begin{equation}\label{eq:FFRRflow_rewritten}\begin{split}
  F_{0,2}^\twist\!\left(\theta_1'-\theta_2';n\right ) &= \frac{ - \ii \cos\!\left (\frac{\pi}{2 n} \right )}{n \sinh\!\left ( \frac{\ii\pi + (\theta_1 - \theta_2)}{2n} \right ) \sinh\!\left ( \frac{\ii\pi - (\theta_1 - \theta_2)}{2n} \right )} \sinh\!\left( \frac{\theta_1'-\theta_2'}{2 n} \right )\\
  &= H^\twist_{0,2} Q^\twist_{0,2}\!\left ( y_1, y_2\right ) \frac{f_{LL}(\theta'_{1}-\theta'_{2};n)}{(y_{1}-\omega y_{2})(y_{2}-\omega y_{1})}\,,
\end{split}\end{equation}
where we have defined
\begin{gather}
  H^\twist_{0,2} = -\ii \frac{4\,\omega}{n}\cos\frac{\pi}{2 n}\,, \\
  Q^\twist_{0,2}\!\left ( y_1, y_2\right ) = \sigma_2\!\left(y_1, y_2 \right ) = y_1 y_2\,.
\end{gather}
Comparing the residue in Eq.~\eqref{eq:resF22flowTwist} with the two-particle form factor in Eq.~\eqref{eq:FFRRflow_rewritten} leads to the recursion equations for $H^{\twist}_{2,2}$ and for the polynomial $Q^{\twist}_{2,2}$
\begin{equation}
H^{\twist}_{2,2}Q^{\twist}_{2,2}(\omega x,x,y_1,y_2;n) = -\mathcal{N}_n^{4} \left [\frac{4\,\omega}{n}\cos\frac{\pi}{2 n}  \right ]^2(\omega^{1/2} x y_1 -1)(\omega^{1/2} x y_2 -1)\,,
\end{equation}
which we separate as 
\begin{gather}
  H^{\twist}_{2,2} = -\mathcal{N}_n^4 \left [ \frac{4\,\omega}{n}\cos \frac{\pi}{2 n}  \right ]^2,\\
  Q^{\twist}_{2,2}(\omega x,x,y_1,y_2;n) = (\omega^{1/2} x y_1 -1)(\omega^{1/2} x y_2 -1) = 1 - \omega^{1/2} x \left ( y_1 + y_2 \right ) + \omega x^2 y_1 y_2\,.\label{eq:Q22flowTwistEquation}
\end{gather}

We postulate a solution to Eq.~\eqref{eq:Q22flowTwistEquation} completely symmetrical in $x_1, x_2, y_1, y_2$ and hence writing
\begin{equation}\begin{split}
  Q^{\twist}_{2,2}\!\left (x_1, x_2, y_1, y_2 ; n \right) &= 1 + A \sigma_1(x_1, x_2) \sigma_1(y_1, y_2) + \sigma_2(x_1, x_2)\sigma_2(y_1, y_2) =\\
  &= 1 + A\left ( x_1 + x_2 \right ) \left ( y_1 + y_2 \right ) + x_1 x_2 y_1 y_2\,,
\end{split}\end{equation}
which is the most general expression compatible with the fact that the form factor has zero Lorentz spin, and that sending each rapidity to $\pm \infty$ the entire FF converges to zero.
Posing $x_1 = \omega x$, $x_2 = x$, we get a unique solution for the unknown constant $A$, namely
\begin{equation}
  A = - \frac{\omega^{1/2}}{1 + \omega} = -\frac{1}{\omega^{1/2} + \omega^{-1/2}} = -\frac{1}{2 \cos \frac{\pi}{2 n}}\,.
\end{equation}
This means that the entire solution can be written as
\begin{equation}\begin{split}
  &F^\twist_{2,2}(\theta_{1},\theta_{2},\theta'_{1},\theta'_{2}; n) = -\mathcal{N}_n^4 \left [ \frac{4\,\omega}{n}\cos \frac{\pi}{2 n}  \right ]^2 \left [ 1 - \frac{1}{2 \cos \frac{\pi}{2 n}}\left ( x_1 + x_2 \right ) \left ( y_1 + y_2 \right ) + x_1 x_2 y_1 y_2 \right ] \times\\
  &\hspace{4cm}\times \frac{f_{RR}(\theta_{1}-\theta_{2};n)}{(x_{1}-\omega x_{2})(x_{2}-\omega x_{1})} \prod_{i=1}^{2} \prod_{j=1}^{2} f_{RL}(\theta_{i}-\theta'_{j};n) \frac{f_{LL}(\theta'_{1}-\theta'_{2};n)}{(y_{1}-\omega y_{2})(y_{2}-\omega y_{1})},
\end{split}\end{equation}
which we can also rewrite as

\begin{equation}
\begin{split}
 &F^\twist_{2,2}(\theta_{1},\theta_{2},\theta'_{1},\theta'_{2}; n)= \\
  &= - 2 \mathcal{N}_n^4 e^{-\frac{\theta_1 + \theta_2 - \theta_1' - \theta_2'}{2n}} \left [ \cosh\!\left ( \frac{\theta_1 + \theta_2 - \theta_1' - \theta_2'}{2 n} \right ) - \frac{\cosh\!\left ( \frac{\theta_1 + \theta_2}{2 n} \right )\cosh\!\left ( \frac{ \theta_1' + \theta_2'}{2 n} \right )}{\cos\!\left ( \frac{\pi}{2 n}\right )} \right ] \times \\
  &\hspace{3cm}\times F^\twist_{2,0}(\theta_{1},\theta_{2}; n) \prod_{i=1}^{2} \prod_{j=1}^{2} f_{RL}(\theta_{i}-\theta'_{j};n) F^\twist_{0,2}(\theta_{1},\theta_{2}; n)\,.
\end{split} 
\end{equation}

\subsection{Form factors of the \texorpdfstring{$\ZZ_2$}{Z2}-composite BPTF}\label{app:FlowBootstrapComp}
Once again, we start our derivation by recalling and repeating the ansatz for $\ZZ_2$-composite BPTF
\begin{equation}
\begin{split}
F^{\twist^\mu}_{\underline{R},\underline{L}}(\underline{\theta},\underline{\theta'};n) &= H^{\twist^\mu}_{r,l}Q^{\twist^\mu}_{r,l}(\underline{x},,\underline{y};n)\prod_{1\leqslant i<j\leqslant r} \frac{f^\mu_{RR}(\theta_{i}-\theta_{j};n)}{(x_{i}-\omega x_{j})(x_{j}-\omega x_{i})}\times\\
 &\hspace{1cm} \times \prod_{i=1}^{r} \prod_{j=1}^{l} f^{\mu}_{RL}(\theta_{i}-\theta'_{j};n)\prod_{1\leqslant i<j\leqslant l} \frac{f^\mu_{LL}(\theta'_{i}-\theta'_{j};n)}{(y_{i}-\omega y_{j})(y_{j}-\omega y_{i})}\,,
\end{split}\label{FFParametrizationFlowCompositeRepeated}
\end{equation}
where we have $r$ right-mover and $l$ left-mover particles and again $x_i=e^{\theta_i/n}$ and $y_i=e^{-\theta'_i/n}$.
The cyclic permutation and the exchange axioms are already satisfied since
\begin{equation}
f^\mu_{\gamma\gamma}(2\pi \ii n-\theta;n)=-f^\mu_{\gamma\gamma}(\theta;n)=f^\mu_{\gamma\gamma}(-\theta;n)\,,
\end{equation}
is fulfilled via 
\begin{equation}
f^\mu_{\gamma\gamma}(\theta;n)=2 \cosh(\theta/(2n))f_{\gamma\gamma}(\theta;n)=\sinh(\theta/n)\,.
\end{equation}
Defining $f^\mu_{RL}(\theta;n)$ as
\begin{equation}
f^\mu_{RL}(\theta;n)=e^{\theta/(2n)}f_{RL}(\theta;n)\,,
\label{F_RL^MuApp}
\end{equation} for $\gamma'$ different from $\gamma$, we similarly satisfy
\begin{equation}
f^\mu_{\gamma\gamma'}(2\pi \ii n-\theta;n)=-f^\mu_{\gamma'\gamma}(\theta;n)=-S_{\gamma'\gamma}(\theta)f^\mu_{\gamma'\gamma}(-\theta;n)\,.
\end{equation}

In full analogy to what we have done in the previous section for the standard twist field, we apply the residue axiom Eq.~\eqref{eq:U1FFAxiom3} to the ansatz~\eqref{FFParametrizationFlowCompositeRepeated} in order to derive recursive equations for the normalisation factors $H^{\twist^\mu}_{r,l}$ and the $Q^{\twist^\mu}_{r,l}$ functions.
Since the denominator of the ansatz~\eqref{FFParametrizationFlowCompositeRepeated} is the same as the one for the standard twist fields in Eq.~\eqref{FFParametrizationFlowRepeated}, we can reuse the same residue we have computed in Eq.~\eqref{4ptResidue}.
Using again the property~\eqref{FMinRLFMinRLRepeated} of the minimal form factor, the residue of the ansatz with 4 particles yields
\begin{equation}
\begin{split}
&-\ii\res_{\theta_1=\ii\pi+\theta_0} F^{\twist^\mu}_{2,2}(\theta_1, \theta_2, \theta'_1,\theta'_2;n)=\\
=&\, \ii\,  H^{\twist^\mu}_{2,2}Q^{\twist^\mu}_{2,2}(\omega x,x,y_1,y_2;n)f^\mu_{RR}(\ii\pi;n)\frac{n \left(x^2 \omega \right)^{-1}}{\omega^2-1}\times\\
 & \hspace{3cm} \times\prod_{j=1}{2} f^{\mu}_{RL}(\theta+i\pi-\theta'_{j};n)f^{\mu}_{RL}(\theta-\theta'_{j};n)\frac{f^\mu_{LL}(\theta'_1-\theta'_2;n)}{(y_1-\omega y_2)(y_2-\omega y_1)} \\
=&\, \ii\, \frac{n \omega}{\omega(\omega-\omega^{-1})}H^{\twist^\mu}_{2,2}\mathcal{N}_n^{-4}\frac{x^2 (y_1y_2)^2 Q^{\twist^\mu}_{2,2}(\omega x,x,y_1,y_2;n)}{(\sqrt{\omega}xy_1-1)(\sqrt{\omega}xy_2-1)}\sinh(\ii\pi/n)\times\\
 &\hspace{8cm} \times  \frac{f^\mu_{LL}(\theta'_1-\theta'_2;n)}{(y_1-\omega y_2)(y_2-\omega y_1)}\,.
\end{split}\label{eq:resFF22flowComp}
\end{equation}
Again, from the residue axiom~\eqref{eq:U1FFAxiom3}, this expression must be compared with the two-particle FF, which we can rewrite as
\begin{equation}
\begin{split}
F^{\twist^\mu}_{2,0}\!\left (\theta'_1-\theta'_2;n\right ) &= \frac{\avg{\twist^{\mu}_{n}}\sin\frac{\pi}{n}}{2n\sinh\frac{\ii\pi+\theta'_1-\theta'_2}{2n}\sinh\frac{\ii \pi-(\theta'_1-\theta'_2)}{2n}}\frac{\sinh((\theta'_1-\theta'_2)/n)}{\sinh(\ii\pi/n)}\\
&= \avg{\twist_{n}}\frac{\ii \omega }{n}\frac{ \left(y_1^2-y_2^2\right)}{2 y_1 y_2}\frac{2 y_1 y_2}{ (y_1-\omega y_2)(y_2-\omega y_1)}\,,
\end{split}
\end{equation}
where $\left(y_2^2-y_1^2\right)/(2 y_1 y_2)=\sinh((\theta'_1-\theta'_2)/n)$.
We then end up with the equation for $Q^{\twist^\mu}_{2,2}$ as well as the normalisation
\begin{equation}
 H^{\twist^\mu}_{2,2}\mathcal{N}_n^{-4}Q^{\twist^\mu}_{2,2}(\omega x,x,y_1,y_2;n)=  -\avg{\twist^{\mu}_{n}} 4\frac{ \omega^2 }{n^2 \omega x^2 y_1 y_2} \left (\sqrt{\omega}xy_1-1 \right )\left (\sqrt{\omega}xy_2-1\right ),
\end{equation}
which we can separate as
\begin{gather}
 H^{\twist^\mu}_{2,2}=  -4\avg{\twist^{\mu}_{n}} \mathcal{N}_n^{4}\,, \\
 Q^{\twist^\mu}_{2,2}(\omega x,x,y_1,y_2;n)=\frac{ \omega^2 }{n^2 \omega x^2 y_1 y_2} (\sqrt{\omega}xy_1-1)(\sqrt{\omega}xy_2-1)\,.\label{eq:Q22flowCompEquation}
\end{gather}
Notice that, differently from what happened in Eq.~\eqref{eq:Q22flowTwistEquation} for the standard twist field, now the function $Q^{\twist^\mu}_{2,2}$ is not a polynomial but a rational function.

We write the solution to Eq.~\eqref{eq:Q22flowCompEquation} as
\begin{equation}
 Q^{\twist^\mu}_{2,2}( x_1,x_2,y_1,y_2;n)= \frac{\omega^2}{n^2 x_1x_2y_1y_2}\left(1+x_1x_2y_1y_2+A(x_1+x_2)(y_1+y_2)\right), 
\end{equation}
which is the most general expression compatible with (i) the form factor has zero Lorentz spin, and (ii) sending each rapidity to $\pm \infty$ the entire FF converges to a constant.
When setting $x_1=\omega x$ and $x_2=x$ we can obtain the same solution for $A$ as for the case of the standard BPTF, namely $A=-\frac{1}{2 \cos \frac{\pi }{2 n}}$ and hence
\begin{equation}
 Q^{\twist^\mu}_{2,2}( x_1,x_2,y_1,y_2;n)^{(0)}= \frac{\omega^2}{n^2 x_1x_2y_1y_2}\left(1-\frac{(x_1+x_2) (y_1+y_2)}{2 \cos \frac{\pi }{2 n}}+x_1 x_2 y_1 y_2\right).
\end{equation}
The ansatz with the above fraction of polynomial $Q^{\twist^\mu}_{2,2}(\omega x_1,x_2,y_1,y_2;n)^{(0)}$ satisfies all the FF axioms. Notice that while the $n\rightarrow 1$ limit of the standard BPTF is not well defined, the FFs of the composite twist field reduce to those of the disorder field
\begin{equation}
\begin{split}
F^{\mu}_{2,2}=H^{\mu}_{2,2} \frac{1+x_1 x_2 y_1 y_2}{x_1 x_2 y_1 y_2}\frac{\sinh(\theta_{1}-\theta_{2})}{(x_{1}+ x_{2})^2}\frac{\sinh(\theta'_{1}-\theta'_{2})}{(y_{1}+ y_{2})^2}  \prod_{i=1}^{2} \prod_{j=1}^{2} f^{\mu}_{RL}(\theta_{i}-\theta'_{j};1)\,.
\end{split}
\end{equation}

As pointed out in the main text, the solution of the bootstrap equation is in general not unique, since we can often add to our polynomial $Q$ also a (non-trivial) kernel solution, that is, another polynomial (or fraction of polynomials) $Q^{(k)}$ which satisfies the homogeneous equation
\begin{equation}
Q^{\twist^\mu}_{2,2}(\omega x,x,y_1,y_2;n)^{(k)}=0.
\end{equation}
Polynomial kernel solutions at the two- and four-particle level have been identified in~\cite{cl-11}. In particular, the two-particle kernel solution reads as
\begin{equation}
Q_{2,0}(x_1,x_2;n)^{(k)}=x_1 x_2-\left(\frac{x_1+x_2}{2 \cos \frac{\pi }{2 n}}\right)^2
\label{KerOlallaRepeated}
\end{equation}
from which the required four-particle kernel solution for the flow can be constructed by squaring the expression due to the anticipated symmetry between the variables of the RR and LL particles.
Based on the above consideration, we can write the eventual kernel as
\begin{equation}
\begin{split}
Q^{\twist^\mu}_{2,2}( x_1,x_2,y_1,y_2;n)^{(k)}=&\frac{\omega^2}{n^2}\frac{8 \cos ^3\left(\frac{\pi }{2 n}\right) \left(x_1 x_2-\frac{1}{4} \sec ^2\left(\frac{\pi }{2 n}\right) (x_1+x_2)^2\right) }{(x_1x_2)(x_1+x_2) (y_1+y_2)}\times\\
&\times \frac{\left(y_1 y_2-\frac{1}{4} \sec ^2\left(\frac{\pi }{2 n}\right) (y_1+y_2)^2\right)}{ y_1y_2(y_1+y_2)}
\end{split}
\end{equation}
that is, taking the product of~\eqref{KerOlallaRepeated} and additionally, by also renormalising the expression with $(x_1x_2y_1y_2)(x_1+x_2)(y_1+y_2)$ which does not spoil the kernel property. We chose the pre-factor in a way that the entire expression for the polynomial
\begin{equation}
\begin{split}
&Q^{\twist^\mu}_{2,2}(x_1,x_2,y_1,y_2;n) = 
 Q^{\twist^\mu}_{2,2}(x_1,x_2,y_1,y_2;n)^{(0)}+Q^{\twist^\mu}_{2,2}(x_1,x_2,y_1,y_2;n)^{(k)}\\
 =&\,\frac{\omega^2}{n^2}\frac{1+x_1 x_2 y_1 y_2}{x_1x_2y_1y_2} +\\
 &\hspace{2cm} -\frac{2 \omega^2 \cos \frac{\pi }{2 n}}{n^2} \frac{ \left(x_1 x_2 (y_1+y_2)^2+y_1 y_2 (x_1+x_2)^2-2 x_1 x_2 y_1 y_2 (\cos \left(\frac{\pi }{n}\right)+1)\right)}{(x_1x_2y_1y_2)(x_1+x_2) (y_1+y_2)}\,,
\end{split}
\end{equation}
gives $(1 + x_1 x_2 y_1 y_2)/(x_1x_2y_1y_2)$ in the $n\rightarrow 1$ limit, which reproduces 
$Q^{\mu}_{2,2}$. The normalisation factors match as well, since $H^\mu_{2,2}= -4 \mathcal{N}^{4}_1 = 2\, e^{-4 G / \pi}$.

\section{Form factor bootstrap for branch point twist fields in the sinh-Gordon model}\label{app:ShG4part}

In this appendix, we first report the known results for the four-particle form factor of the standard twist field in the sinh-Gordon model and we then derive the previously unknown form factor of the composite one.

\subsection{Form factors of the standard BPTF}\label{app:standardShG4ptBPTF}

In the sinh-Gordon model, the four-particle form factor of the standard branch point twist field has been computed in~\cite{cl-11} using the bootstrap program. In Sec.~\ref{sec:RoamingLimit} of the main text, this known result has been the starting point for the roaming limit. For completeness, in this appendix we report its explicit expression. 

The solution is written in the usual form, reported in Eq.~\eqref{FFParametrizationShG}
\begin{equation}
F^{\twist}_{4}(\theta_1, \theta_2,\theta_3, \theta_4;n) =\, H^{\twist}_{4}\, Q^{\twist}_{4}\!\left (x_1,x_2,x_3,x_4;n \right ) \prod_{1\leqslant i < j\leqslant 4}\frac{f_\text{ShG}(\theta_{i}-\theta_{j},B;n)}{(x_{i}-\omega x_{j})(x_{j}-\omega x_{i})}\,.
\label{FFBPFTShG4pt}
\end{equation}
where $f_\text{ShG}$ is the minimal form factor in the sinh-Gordon model, shown in the main text in Eq.~\eqref{eq:fminShG}.
The normalisation $H^{\twist}_{4}$ was found in~\cite{cl-11} to be
\begin{equation}
    H^{\twist}_{4} = \left ( \frac{2\, \sin\!\frac{\pi}{n}\, \omega^2}{n\, f_\text{ShG}\!\left ( \ii \pi \right )} \right )^2 \omega^2 \avg{\twist_n},
\end{equation}
while the polynomial $Q^{\twist}_{4}$ takes the form~\cite{cl-11}
\begin{equation}\begin{split}
    Q^{\twist}_{4}\!\left ( x_1,x_2,x_3,x_4;n\right ) = \frac{\sigma _4}{\beta^2\, \omega^4 \left ( \omega + 1 \right )} \Big [ & \sigma_1 \sigma_3 \left [ A_2 \sigma_1 \sigma_3 + A_6 \sigma_4 + A_7 \sigma _2^2 \right ] + \sigma_2^2 \left [ A_1 \sigma_2^2 + A_5 \sigma_4 \right ] + \\
    &+ A_3 \left(\sigma_1^2 \sigma_4 + \sigma_3^2 \right) \sigma_2 + A_4 \sigma_4^2  \Big ]\,,
\end{split}\end{equation}
with coefficients
\begin{align*}
    &A_1 = \beta ^2 \omega^4 \left ( \omega + 1 \right ),\\
    &A_2 = \beta\,  \omega ^3 \left ( \omega + 1 \right ) \left ( \beta +\omega + 1 \right ) \left ( \beta  \omega +\beta +\omega \right ),\\
    &A_3 = -\beta\, \omega^2 \left( \omega ^2 + \omega + 1 \right ) \left ( \beta  \omega - 1 \right ) \left( \beta - \omega^2 \right),\\
    &A_4 = \left ( \omega + 1 \right ) \left( \omega^2 + 1 \right)^2 \left(\beta^2 + \beta  \omega + \beta + \omega^2 + \omega + 1\right) \left(\left(\beta^2 + \beta + 1\right) \omega^2 + \beta^2 + (\beta +1) \beta  \omega \right),\\
    &A_5 = \omega^2 \left ( \omega + 1 \right ) \left( \beta^4 \omega + \beta^3 \left ( \omega + 1 \right )^3 - \beta^2 \left( \omega^2 - 4\,\omega + 1 \right ) \left( \omega^2 + \omega + 1 \right) + \beta  \omega \left ( \omega +1 \right )^3 + \omega^3 \right),\\
    &\begin{aligned}
        A_6 =&\, - \left( \left (\omega + 1 \right ) \omega ^2 \left( \beta^2 + \beta \omega + \beta + \omega^2 + \omega + 1\right) \left( \left( \beta^2 + \beta + 1 \right) \omega^2 + \beta^2 + \left ( \beta + 1 \right ) \beta  \omega \right) \right) +\\
        & - \beta \left ( \omega + 1 \right ) \left( \omega^2 + 1\right)^2 \omega \left ( \beta + \omega +1 \right ) \left (\beta  \omega +\beta +\omega \right ),
    \end{aligned}\\
    &A_7 = -\beta\, \omega^3 \left( \beta^2 \omega + \beta  \left ( \omega + 1 \right ) \left ( \omega^2 + 3\, \omega + 1 \right ) + \omega^2 \right), \stepcounter{equation}\tag{\theequation}
\end{align*}
where $\beta = e^{\frac{\ii\pi B}{2 n}}$ and $B$ is defined in terms of the sinh-Gordon coupling as in Eq.~\eqref{eq:ShGB}.

\subsection{Form factors of the \texorpdfstring{$\ZZ_2$}{Z2}-composite BPTF}
\label{app:compositeShGbootstrap}

As we mentioned in the main text, differently from the four-particle form factor of the standard twist field $\twist_n$, in the sinh-Gordon model the one of the composite field $\twist_n^\mu$ was not previously known in the literature.
In this appendix we compute this form factor by constructing and solving the bootstrap equations, in full analogy to what we have done in Sec.~\ref{sec:compositeFormFactors} and in App.~\ref{app:FlowBootstrapComp} in the case of the massless flow.

The minimal form factors $f^\mu_\text{ShG}$ of the composite twist field have been obtained in the main text in Eq.~\eqref{eq:fminShGComp} by multiplying the standard minimal form factor~\eqref{eq:fminShG} by the monodromy changing factor $2 \cosh(\theta/2n)$. 
Using this result, we parameterise the form factor in the usual way, as reported in Eq.~\eqref{FFParametrizationShG}
\begin{equation}\label{eq:ansatzCompNoDenomApp}
    F^{\twist^\mu|1\ldots 1}_{k}\!\left ( \theta_1, \ldots, \theta_k \right ) = H^{\twist^\mu}_k\, Q^{\twist^\mu}_k\!\left (x_1,\ldots,x_k\right ) \prod_{1\leqslant i< j\leqslant k} \frac{f^\mu_\text{ShG}\!\left( \theta_i - \theta_j;n \right )}{\left (x_i - \omega x_j \right ) \left (x_j - \omega x_i \right )}.
\end{equation}
Recall however that, as we discussed below Eq.~\eqref{F_RL^Mu}, for the composite twist field the function $Q^{\twist^\mu}_k$ appearing in the ansatz~\eqref{eq:ansatzCompNoDenomApp} is not guaranteed to be a polynomial but is in general a rational function. In the following, we will find it convenient to explicitly extract the denominator of the function $Q^{\twist^\mu}_k$ by defining 
\begin{equation}\label{eq:QNoDenom}
    \widetilde{Q}^{\twist^\mu}_k\!\left (x_1,\ldots,x_k\right ) = \frac{Q^{\twist^\mu}_k\!\left (x_1,\ldots,x_k\right )}{\prod_{1 \leqslant i < j \leqslant k} \left ( x_i + x_j \right )}\,.
\end{equation}
As we will see, extracting this factor is sufficient to guarantee that $\widetilde{Q}^{\twist^\mu}_k$ is indeed a polynomial. We stress that the denominator $\left ( x_i + x_j \right )$ does not introduce additional poles, as its zeroes exactly cancel out with those of the monodromy changing factor in $f^\mu_\text{ShG}$. Plugging the polynomial~\eqref{eq:QNoDenom} in the ansatz~\eqref{eq:ansatzCompNoDenomApp}, the form factor is alternatively parameterised as
\begin{equation}\label{eq:ansatzCompApp}
    F^{\twist^\mu|1\ldots 1}_{k}\!\left ( \theta_1, \ldots, \theta_k \right ) = H^{\twist^\mu}_k\, \widetilde{Q}^{\twist^\mu}_k\!\left (x_1,\ldots,x_k\right ) \prod_{1\leqslant i< j\leqslant k} \frac{f^\mu_\text{ShG}\!\left( \theta_i - \theta_j;n \right )}{\left (x_i + x_j \right ) \left (x_i - \omega x_j \right ) \left (x_j - \omega x_i \right )}\,.
\end{equation}

Before moving to the actual computation, we present a useful identity of the minimal form factor. It is known that the standard minimal form factor $f_\text{ShG}$ in the sinh-Gordon theory satisfies the identity~\cite{Ola}
\begin{equation}\label{eq:fminShGidentity}
    f_\text{ShG}\!\left( \theta_0 + \ii \pi - \theta_i;n \right ) f_\text{ShG} \!\left( \theta_0 - \theta_i;n \right ) = \frac{\left ( x_0 - x_i \right ) \left ( \omega x_0 - x_i \right )}{\left ( \beta x_0 - x_i \right ) \left ( \omega \beta^{-1} x_0 - x_i \right ) }\, ,
\end{equation}
where $\beta = e^{\frac{\ii \pi B}{2 n}}$ and $B$ is related to the sinh-Gordon coupling through Eq.~\eqref{eq:ShGB}.
In order to extend this relation to the composite minimal form factor, notice that for the monodromy changing factor we have
\begin{equation}
    2 \cosh\!\left( \frac{\theta_0 + \ii \pi - \theta_i}{2 n}\right ) 2 \cosh\!\left( \frac{\theta_0 - \theta_i}{2 n}\right ) = \frac{\omega x_0 + x_i}{\left ( \omega x_0 x_i \right )^{1/2}} \frac{x_0 + x_i}{\left (x_0 x_i \right )^{1/2}} = \frac{\left ( x_0 + x_i \right ) \left ( \omega x_0 + x_i \right )}{\omega^{1/2} x_0 x_i}\,,
\end{equation}\label{eq:fminShGCompidentity}
which from the definition of $f^\mu_\text{ShG}$ in Eq.~\eqref{eq:fminRR^mu} directly implies
\begin{equation}\label{eq:fminShGidentityCom}
    f^\mu_\text{ShG}\!\left( \theta_0 + \ii \pi - \theta_i;n \right ) f^\mu_\text{ShG}\!\left( \theta_0 - \theta_i;n \right ) = \frac{\left ( x_0^2 - x_i^2 \right ) \left ( \omega^2 x_0^2 - x_i^2 \right )}{\omega^{1/2} x_0 x_i \left ( \beta x_0 - x_i \right ) \left ( \omega \beta^{-1} x_0 - x_i \right ) }\, .
\end{equation}

We now have everything we need to write the bootstrap equation. For simplicity we consider all particles on the first replica and we apply the kinematic residue axiom of Eq.~\eqref{eq:U1FFAxiom3} to the modified ansatz~\eqref{eq:ansatzCompApp}.
Setting the first rapidity equal to $\theta_{-1} = \ii \pi +\theta_0$, the residue of the denominator (including the additional factor $(x_i + x_j)$) becomes
\begin{align}
\begin{split}
&-\ii \res_{\theta_{-1}=i\pi+\theta_0} \prod_{-1\leqslant i<j\leqslant k} \frac{1}{\left (x_{i}-\omega x_{j}\right )\left (x_{j}-\omega x_{i}\right )\left (x_{i} + x_{j}\right )} \\
=& -\ii\prod_{1\leqslant i<j\leqslant k} \frac{1}{\left (x_{i}-\omega x_{j}\right )\left (x_{j}-\omega x_{i}\right )\left (x_{i} + x_{j}\right )}\times\\
&\quad\times\left ( -\frac{n\, x_0^{-3}}{\omega (\omega^2-1) (\omega+1)} \right ) \left[\omega^{k}\prod_{i=1}^k \left (x_{0}-\omega x_{i}\right )\left (x_{i}^2-\omega^2 x_{0}^2\right )\left ( x_{0}^2 - x_{i}^2 \right )\left (x_{i}-\omega^2 x_{0}\right )\right]^{-1}
\end{split} \notag \\
\begin{split}
=&\prod_{1\leqslant i<j\leqslant k} \frac{1}{\left (x_{i}-\omega x_{j}\right )\left (x_{j}-\omega x_{i}\right )\left (x_{i} + x_{j}\right )}\times\\
&\quad\times\frac{n\, x_0^{-3}\, \omega^{- (k + 2)}}{2 \left ( \omega + 1 \right ) \sin\!\frac{\pi}{n} } \left[ \prod_{i=1}^k \left (x_{0}-\omega x_{i}\right )\left (x_{i}^2-\omega^2 x_{0}^2\right )\left ( x_{0}^2 - x_{i}^2 \right )\left (x_{i}-\omega^2 x_{0}\right )\right]^{-1}\!\!.
\end{split}\label{eq:resDenomShGComp}
\end{align}
Using this residue, the ansatz~\eqref{eq:ansatzCompApp} for the $(k+2)$-particle form factor reduces to 
\begin{align}
\begin{split}
    &-\ii\res_{\theta_{-1}=\ii\pi+\theta_0} F^{\twist^\mu | 1\ldots}_{k+2}\!\left ( \theta_{-1}, \theta_0, \theta_1, \ldots \right ) =\\
    &= H^{\twist^\mu}_{k+2} \widetilde{Q}^{\twist^\mu}_{k+2}\!\left ( \omega x_0, x_0, x_1, \ldots \right ) \prod_{1\leqslant i<j\leqslant k}\frac{f^\mu_\text{ShG}\!\left( \theta_i -\theta_j;n \right )}{\left (x_{i}-\omega x_{j}\right )\left (x_{j}-\omega x_{i}\right )\left (x_{i} + x_{j}\right )}\times\\
    &\quad\times \frac{n\, x_0^{-3}\, f^\mu_\text{ShG}\!\left( \ii \pi;n\right )}{2\,  \omega^{k+2} \left ( \omega + 1 \right ) \sin\!\frac{\pi}{n} } \left[\prod_{i=1}^k \frac{f^\mu_\text{ShG}\!\left( \theta_0+\ii \pi -\theta_i;n \right )f^\mu_\text{ShG}\!\left( \theta_0 -\theta_i;n \right )}{\left (x_{0}-\omega x_{i}\right )\left (x_{i}^2-\omega^2 x_{0}^2\right )\left ( x_{0}^2 - x_{i}^2 \right )\left (x_{i}-\omega^2 x_{0}\right )}\right]
\end{split}\notag\\
\begin{split}
    &= H^{\twist^\mu}_{k+2} \widetilde{Q}^{\twist^\mu}_{k+2}\!\left ( \omega x_0, x_0, x_1, \ldots\right ) \prod_{1\leqslant i<j\leqslant k}\frac{f^\mu_\text{ShG}\!\left( \theta_i -\theta_j;n \right )}{\left (x_{i}-\omega x_{j}\right )\left (x_{j}-\omega x_{i}\right )\left (x_{i} + x_{j}\right )}\times\\
    &\quad\times \frac{n\, f^\mu_\text{ShG}\!\left( \ii \pi;n\right )}{2\,  \omega^{k+2} \left ( \omega + 1 \right ) \sin\!\frac{\pi}{n} } \left[x_0^{k+3} \omega^{k/2} \prod_{i=1}^k  x_i \left (x_{i}-\omega^2 x_{0} \right ) \left (x_{0}-\omega x_{i} \right )   \left ( x_i - \omega \beta^{-1} x_0 \right ) \left ( \beta x_0 - x_i \right )\right]^{-1}\!\!.
\end{split}\label{eq:resFFShGComp}
\end{align}
where we have applied the identity in Eq.~\eqref{eq:fminShGidentityCom}.
Following the residue axiom in Eq.~\eqref{eq:U1FFAxiom3}, we compare the residue in Eq.~\eqref{eq:resFFShGComp} with the ansatz~\eqref{eq:ansatzCompApp} for $k$-particles, extracting immediately the recursion relation for the normalisation
\begin{equation}
    H^{\twist^\mu}_{k+2} = \frac{2\,  \omega^{k+2} \left ( \omega + 1 \right ) \sin\!\frac{\pi}{n} }{n\, f^\mu_\text{ShG}\!\left( \ii \pi;n\right )}\, H^{\twist^\mu}_{k}\,,
\end{equation}
which, using $H^{\twist^\mu}_0 = \avg{\twist_n^\mu}$, is solved as
\begin{equation}\label{eq:H4ShGComp}
    H^{\twist^\mu}_{k} = \left ( \frac{2 \left ( \omega + 1 \right ) \sin\!\frac{\pi}{n} }{n\, f^\mu_\text{ShG}\!\left( \ii \pi;n\right )} \right )^{\frac{k}{2}} \omega^{\frac{k}{2}\left ( \frac{k}{2} + 1 \right )} \, \avg{\twist_n^\mu}\,.
\end{equation}
The recursion relation for the polynomial $\widetilde{Q}^{\twist^\mu}_{k}$ instead takes the form 
\begin{equation}\label{eq:recursionpolyShGComp}
    \widetilde{Q}^{\twist^\mu}_{k+2}\!\left ( \omega x_0, x_0, x_1, \ldots, x_k\right ) = \widetilde{P}_k\!\left (x_0, x_1, \ldots, x_k \right ) \widetilde{Q}^{\twist^\mu}_k\!\left ( x_1, \ldots, x_k \right ),
\end{equation}
with the polynomial
\begin{equation}\begin{split}
    &\widetilde{P}_k\!\left (x_0, x_1, \ldots, x_k\right ) = x_0^{k+3} \omega^{k/2} \prod_{a,b,c,d, i=1}^k x_i \left (x_{a}-\omega^2 x_{0} \right ) \left (x_{0} - \omega x_{b} \right ) \left ( x_c - \omega \beta^{-1} x_0 \right ) \left ( \beta x_0 - x_d \right ) \\
    &= x_0^{k+3} \omega^{\frac{3}{2}k}\,  \sigma_k \sum_{a,b,c,d=1}^k \left (-\omega^2 x_{0} \right )^{k-a} \left (- \omega^{-1} x_{0}\right )^{k-b} \left ( - \omega \beta^{-1} x_0 \right )^{k-c} \left ( - \beta x_0 \right )^{k-d} \sigma_a\, \sigma_b\, \sigma_c\, \sigma_d\,,
\end{split}\end{equation}
where $\sigma_j$ are the fully symmetric polynomials of degree $j$ in $k$ variables and again we have $\beta = e^{\frac{\ii \pi B}{2 n}}$. 

In order to solve the recursion equation in Eq.~\eqref{eq:recursionpolyShGComp} for the four-particle form factor, we need to first rewrite the known two-particle form factor in Eq.~\eqref{eq:F2TwistFieldShG} in the form of our ansatz~\eqref{eq:ansatzCompNoDenomApp}, \eqref{eq:ansatzCompApp}
\begin{equation}\begin{split}
    F^{\twist^\mu | 11}_2\!\left ( \theta_i -\theta_j \right ) &= \frac{\avg{\twist^\mu} \sin\!\frac{\pi}{n}}{2 n \sinh\!\left ( \frac{\ii \pi + \theta_i -\theta_j}{2 n} \right ) \sinh\!\left(\frac{\ii \pi -\theta_i +\theta_j}{2 n}\right )} \frac{f^\mu_\text{ShG}\!\left( \theta_i -\theta_j;n \right )}{f^\mu_\text{ShG}\!\left( \ii \pi ;n \right )} \\
    &=\avg{\twist_n^\mu} \frac{ 2\, \omega^{2} \left( \omega + 1 \right ) \sin\!\frac{\pi}{n} }{n f^\mu_\text{ShG}\!\left( \ii \pi ;n \right )}\, \frac{ x_i x_j }{\omega \left( \omega + 1 \right ) } \frac{f^\mu_\text{ShG}\!\left( \theta_i -\theta_j;n \right )}{(x_{i}-\omega x_{j})(x_{j}-\omega x_{i})}\,,
\end{split}\end{equation}
which in agreement with Eq.~\eqref{eq:H4ShGComp} has to be divided as
\begin{gather}
    H^{\twist^\mu}_2 = \frac{ 2\, \omega^{2} \left( \omega + 1 \right ) \sin\!\frac{\pi}{n} }{n f^\mu_\text{ShG}\!\left( \ii \pi ;n \right )} \avg{\twist_n^\mu}, \\
    Q^{\twist^\mu}_2\!\left ( x_1, x_2 \right ) = \frac{ x_i x_j }{\omega \left( \omega + 1 \right ) } = \frac{\sigma_2}{\omega \left ( \omega + 1 \right )}\,, \quad
    \widetilde{Q}^{\twist^\mu}_2\!\left ( x_1, x_2 \right ) =  \left (x_{i}+x_{j}\right ) Q^{\twist^\mu}_2\!\left ( x_1, x_2 \right ) = \frac{\sigma_1\sigma_2}{\omega \left ( \omega + 1 \right )}\,,
\end{gather}
where $\sigma_1, \sigma_2$ are the fully symmetric polynomials in two variables.
Notice that the polynomial $\widetilde{Q}^{\twist^\mu}_2$ has total degree 3 and partial degree 2 in each variable.
Since the polynomial $\widetilde{P}_k$ in the recursion equation Eq.~\eqref{eq:recursionpolyShGComp} has partial degree 5 for any number of particles, this implies that $\widetilde{Q}^{\twist^\mu}_k$ at the $k$-particle order has partial degree $\frac{5}{2} k - 3$.

Assuming that the solution of Eq.~\eqref{eq:recursionpolyShGComp} is completely symmetrical in the variables $x_i$, it is in general not unique since one can always add a kernel solution, i.e., a solution of the homogeneous equation
\begin{equation}\label{eq:kernelEqShGComp}
    \widetilde{Q}^{\twist^\mu}_{k+2}\!\left ( \omega x_0, x_0, x_1, \ldots, x_k \right ) = 0.
\end{equation}
However, imposing that the polynomial has maximum partial degree $\frac{5}{2} k - 3$ in each variable, for $k = 2$ the kernel equation~\eqref{eq:kernelEqShGComp} has no solutions and the solution to the recursion equation Eq.~\eqref{eq:recursionpolyShGComp} is actually unique. We finally find the result 
\begin{gather} \begin{split}
        \widetilde{Q}^{\twist^\mu}_4\!\left( x_1, x_2, x_3, x_4 \right ) &= \frac{\sigma_4^2}{\beta ^2 \omega ^4 (\omega +1)^2} \Big [ \sigma_1 \sigma_3 \left [B_1 \sigma_2^3  + B_2 \sigma_1 \sigma_2 \sigma_3 + B_5 \left ( \sigma_1^2 \sigma_4 + \sigma_3^2 \right ) + B_8 \sigma_2 \sigma_4 \right ]   +\\
        &\qquad+ \sigma_2^2 \left [B_3 \sigma_2 \sigma_4 + B_4 \left ( \sigma_1^2 \sigma_4 + \sigma_3^2  \right ) \right ] + \sigma_4 \left [ B_6 \sigma_2 \sigma_4 + B_7 \left ( \sigma_1^2 \sigma_4 + \sigma_3^2 \right ) \right ]  \Big ],\label{eq:QNoDenom4ShGComp}
    \end{split}\\
    Q^{\twist^\mu}_4\!\left( x_1, x_2, x_3, x_4 \right ) = \frac{\widetilde{Q}^{\twist^\mu}_4\!\left( x_1, x_2, x_3, x_4 \right )}{\prod_{1\leqslant i < j \leqslant 4} \left ( x_i + x_j\right )}\,,\label{eq:Q4ShGComp}
\end{gather}
where the coefficients are
\begin{align*}
    &B_1 = \beta ^2 \omega ^4, \\
    &B_2 = -\beta ^2 \omega ^3 \left(\omega ^2+\omega +1\right),\\
    &B_3 = \beta \omega ^2 \left (\beta +1 \right ) \left (\omega +1 \right )^3 \left (\beta +\omega \right ),\\
    &B_4 = -\beta  \omega ^3 \left (\beta +\omega +1\right ) \left (\beta  \omega +\beta +\omega \right ),\\
    &B_5 = \beta  \omega ^2 \left(\omega ^2+\omega +1\right) \left (\beta +\omega +1 \right ) \left (\beta \omega +\beta +\omega \right ),\\
    &B_6 = \left (\beta +1 \right ) \left (\omega +1 \right )^3 \left(\omega ^2+1\right) \left (\beta +\omega \right ) \left(\beta^2 + (\beta +1)^2 \omega +\omega ^2\right), \stepcounter{equation} \label{eq:coefficientsQ4ShGComp} \tag{\theequation} \\
    &\begin{aligned}
        B_7 &= -\omega  \left(\omega ^2+\omega +1\right) \left(\beta ^2+\beta  \omega + \beta   +\omega ^2+\omega +1\right) \times\\
        &\hspace{4cm}\times\left(\left(\beta ^2+\beta +1\right) \omega^2 + \beta ^2+(\beta +1) \beta  \omega \right),
    \end{aligned}\\
    &\begin{aligned}
        B_8 &= -\omega\,  \Big(\beta ^3+\beta  \omega ^6 + \beta \omega^5 \left (\beta^2 + 5\beta + 5 \right ) + \omega^4 \left (5 \beta^3 + 9\beta^2 + 6\beta - 1 \right ) +\\
        &\hspace{1cm} + 2\, \beta \omega^3 \left ( 3 \beta^2 + 5 \beta + 3 \right ) + \beta \omega^2 \left (  9\beta - \beta^3 + 6\beta^2 +5 \right )
        +\beta \omega \left ( 5 \beta^2 + 5 \beta + 1 \right ) \Big)\,.
    \end{aligned}
\end{align*}
Plugging the function $Q^{\twist^\mu}_4$ in Eqs.~\eqref{eq:QNoDenom4ShGComp}-\eqref{eq:coefficientsQ4ShGComp} and the normalisation $H^{\twist^\mu}_4$ from Eq.~\eqref{eq:H4ShGComp} in the ansatz~\eqref{eq:ansatzCompNoDenomApp} we finally obtain the four-particle form factor for the composite twist field that we used in Sec.~\ref{sec:RoamingLimitComposite}.

\section{Cumulant expansion of the entanglement entropy in the massive Ising theory}\label{app:cumulantIsing}

In this appendix, we review the known results for the form factor expansion of the entanglement entropy in the massive Ising model, obtained in Refs.~\cite{cd-09b}.
In particular, we find a direct relation between UV limit of the cumulant expansion of the entropy in the massive Ising and the non-interacting part of the expansion in the massless flow, studied in Sec.~\ref{sec:noninteractingCumulants}.

In the massive Ising theory, if we denote by $m$ the mass gap, the ground state Rényi entanglement entropy admits the following cumulant 
expansion~\cite{cd-09b},
\begin{equation}
  S_n^\text{Ising}\!\left ( m \ell \right ) \approx \frac{1}{1-n}\sum_{k \text{ even}} c_{k,\text{ Ising}}^\twist\!\left ( m\ell; n \right ) + \text{const},
\end{equation}
where
\begin{equation}
  c^\twist_{k,\text{ Ising}}\!\left ( m\ell; n \right ) = \sum_{j_1, \ldots, j_k = 1}^n \int_{-\infty}^{+\infty} \frac{\prod_{i = 1}^{k} \dd \theta_i}{k! \left ( 2 \pi \right )^k}\, f_{k,\text{ Ising}}^{\twist | j_1\ldots j_k}\!\left (\theta_1, \ldots, \theta_k; n\right ) e^{-m \ell \sum_i \cosh \theta_i }. \label{eq:Isingcumulant}
\end{equation}
These cumulants can be reexpressed as in Eq.~\eqref{eq:cumulantOlalla} and, therefore, 
the $k$-particle cumulant $c_{k, \text{ Ising}}^\twist$ is similar to the $k$-right- or $k$-left-mover non-interacting cumulants $c_{k, 0}^\twist$, $c_{0, k}^\twist$ in the massless flow~\eqref{eq:cumulant}, differing only in the energy $E$ in the exponential factor. In the massive Ising theory, the energy of $k$ particles is
\begin{equation}\label{eq:massiveenergy}
    E\!\left ( \theta_1, \ldots, \theta_k \right ) = \sum_{i=1}^k m \cosh\!\left ( \theta_i \right ) .
\end{equation}
If we move to the center-of-mass coordinates, $A = \frac{1}{k}\sum_i \theta_i$ and $\theta_{ij} = \theta_i - \theta_j$, then the energy~\eqref{eq:massiveenergy} takes the form 
\begin{equation}\begin{split}
  &E(\theta_1, \ldots, \theta_k) = m \sum_{j = 1}^k \cosh \theta_j = m \sum_{j= 1}^{k} \cosh\!\left ( \theta_j + A - A \right ) =\\
  =&\, m \left [\cosh A \left ( \sum_{j= 1}^{k} \cosh \xi_j\!\left ( \theta_{12},\ldots, \theta_{k-1,k}\right ) \right ) + \sinh A \left (  \sum_{j= 1}^{k} \sinh \xi_j\!\left ( \theta_{12},\ldots, \theta_{k-1,k}\right ) \right ) \right ].
\end{split}\end{equation}

Let us first analyse the two-particle cumulant $c_{2, \text{ Ising}}^\twist$. In a massive theory, the exponential $e^{-\ell E}$ is responsible for a double exponential suppression in both the $\theta \to \infty$ and $\theta \to -\infty$ regimes, ensuring the convergence of the integrals.
For two particles, in particular, after changing coordinates to the relative $\theta_{12} = \theta_1 - \theta_2$ and center-of-mass rapidities $A = (\theta_1+\theta_2)/2$, we can integrate out the center-of-mass rapidity obtaining~\cite{Ola, cd-09,cd-09b, cl-11}
\begin{equation}\label{eq:cumulant2Ising}\begin{split}
  c^\twist_{2,\text{ Ising}}\!\left ( m\ell; n \right ) &= \frac{n}{2\left ( 2\pi \right )^2}\sum_{j_2} \int_{-\infty}^{+\infty} \dd \theta_{12}\, f_{2}^{\twist | 1 j_2}\!\left (\theta_{12}; n\right ) \int_{-\infty}^{+\infty} \dd A\, e^{- 2 m \ell \cosh A \cosh \frac{\theta_{12}}{2} }\\
  &= \frac{n}{2\left ( 2\pi \right )^2}\sum_{j_2} \int_{-\infty}^{+\infty} \dd \theta_{12}\, f_{2}^{\twist | 1 j_2}\!\left (\theta_{12}; n\right ) 2 K_0\!\left ( 2 m \ell \cosh\!\left( \theta_{12}/2\right ) \right ).
\end{split}\end{equation}
As shown in Ref.~\cite{Ola}, in the UV limit $m\ell \ll 1$, the expansion of the Bessel function $K_0$ 
\begin{equation}\label{eq:besselexpansion}
  K_0\!\left ( x \right) \underset{x \ll 1}{\approx} -\log\!\frac{x}{2} - \gamma + \mathcal{O}\!\left ( x^2 \right ) ,
\end{equation}
reproduces the expected UV logarithmic behaviour of the entanglement entropy up to an additive constant~\cite{Ola}
\begin{gather}
  c^\twist_{2,\text{ Ising}}\!\left ( m\ell; n \right ) \underset{m\ell \ll 1}{\approx} - z_{2}\!\left( n \right ) \log m \ell + \text{const},\label{eq:c2_ising}
\end{gather}
where the function $z_2(n)$ was introduced in Eq.~\eqref{eq:zeta2}.

We can now investigate the higher-particle cumulants $c_{k, \text{ Ising}}^\twist$. If we write them in terms of the center-of-mass coordinates, we can apply  the integral identity
\begin{equation}
  \int_{-\infty}^{+\infty} \dd t\, \exp\!\big \{ - C \cosh t - S \sinh t \big \} = 2 K_0\!\left ( \sqrt{C^2 - S^2} \right ),
\end{equation}
and the fact that the form factors only depend on the relative rapidities to integrate out the center-of-mass rapidity $A$. We then obtain
\begin{equation}\begin{split}
  c_{k,\text{ Ising}}^\twist\!\left( m\ell;n \right ) =&\, \frac{n}{k! \left ( 2\pi\right )^k} \sum_j \int_{-\infty}^{+\infty} \prod_{j=1}^{k-1} \dd\theta_{j,j+1}\,  f_k^\twist\!\left (\theta_{12}, \ldots;n \right ) \int_{-\infty}^{+\infty} \dd A\, e^{-m\ell\, ( C \cosh A + S \sinh A )} \\
  =&\, \frac{2 n}{k! \left ( 2\pi\right )^k} \sum_j \int_{-\infty}^{+\infty} \prod_{j=1}^{k-1} \dd\theta_{j,j+1}\,  f_k^\twist\!\left (\theta_{12}, \ldots; n \right ) \, K_0\!\left ( m\ell \sqrt{C^2 - S^2} \right ),
\end{split}\end{equation}
where
\begin{equation}
  C\!\left ( \theta_{12},\ldots\right ) = \sum_{j= 1}^{k} \cosh\!\left ( \xi_j\!\left ( \theta_{12},\ldots\right ) \right ), \quad 
  S\!\left ( \theta_{12},\ldots\right ) = \sum_{j= 1}^{k} \sinh\!\left ( \xi_j\!\left ( \theta_{12},\ldots\right ) \right ),
\end{equation}
and $\xi_j$ are defined in Eq.~\eqref{eq:comRapidity}. 
In the UV limit $m \ell \ll 1$,  by expanding the Bessel function using Eq.~\eqref{eq:besselexpansion}, we get at leading order the logarithmic behaviour of the entropy expected in the Ising CFT up to an additive constant
\begin{equation}\label{eq:ckising}
  c_{k,\text{ Ising}}^\twist\!\left(m\ell; n \right ) \underset{m\ell \ll 1}{\approx} - z_k(n)\, \log\!\left ( m\ell \right ) + \text{const},
\end{equation}
where the coefficient $z_k$ is the same as in Eq.~\eqref{eq:zetaN}.
Comparing Eq.~\eqref{eq:ckising} with the analogous formula in Eq.~\eqref{eq:cr0_higher}, we can immediately see that the UV limit of the $k$-particle massive Ising cumulants is twice the $k$-right-mover cumulants of our massless flow.
Notice that the factor 2 comes from the expansion of the Bessel function and ultimately its origin is the difference in the energy of the two models. 

Before concluding this appendix, let us make a remark on the computation of the coefficients $z_k(n)$.
The expression in Eq.~\eqref{eq:zetaN} contains $k-1$ integrals and, therefore, it is not practical for numerical calculations.
In Ref.~\cite{cd-09b}, the analytic continuation of Eq.~\eqref{eq:zetaN} was carried out for $n\geqslant 1$ replicas, writing $z_k(n)$ as a single integral for any $k$ (see also~\cite{Ola2, cm-23})
\begin{equation}\label{eq:olallaformulaFull}
    z_{k}(n) = \frac{2 n}{k \left (4\pi\right )^{k}} \int_{0}^{\infty} \dd x\, \mathcal{J}_{k}\!\left (x \right )^2 \mathcal{W}_k\!\left ( x; n \right ),
\end{equation}
where, for $k = 2 p$,
\begin{equation}\label{eq:olallaformulaJ}
    \mathcal{J}_{2p}\!\left ( x \right ) = \frac{(2\pi)^{p-1}}{(p-1)!} \begin{cases}
        \frac{x}{\pi} \frac{1}{\sinh\!\left ( \frac{x}{2}\right )} \prod_{j = 1}^{\frac{p}{2}-1} \left ( \frac{x^2}{\pi^2} + (2 j)^2 \right ), &\text{for $p$ even,}\\
        \frac{1}{\cosh\!\left ( \frac{x}{2} \right )} \prod_{j = 1}^{\frac{p-1}{2}} \left ( \frac{x^2}{\pi^2} + (2 j - 1)^2 \right ), &\text{for $p$ odd,}
    \end{cases}
\end{equation}
\begin{equation}\label{eq:olallaformulaW}
    \mathcal{W}_{2p}\!\left ( x; n \right ) = \left (-1 \right )^p \ii\, \sinh(x) \sum_{j=1}^{p} \binom{2 p - 1}{p - j} \left [ w\!\left ( 2 x + \left ( 2j-1 \right ) \ii\, \pi; n \right ) + w\!\left ( 2 x - \left ( 2j-1 \right ) \ii\, \pi; n \right ) \right ],
\end{equation}
and $w(\theta; n)$ is given in Eq.~\eqref{eq:smallw}.
Eq.~\eqref{eq:olallaformulaFull} is efficient for numerical calculations. We employed it to compute the first 30 non-interacting cumulants in the truncated expansion of the entropies of the massless flow plotted in Fig.~\ref{fig:entropy}.

\end{document}